\documentstyle[12pt]{article} \hoffset -0.5in \textwidth 6.5in \textheight
9.00in  \parskip 7pt \openup2.5\jot \parindent=0.5in
\newskip\humongous \humongous=0pt plus 1000pt minus 1000pt
\def\caja{\mathsurround=0pt}
\def\eqalign#1{\,\vcenter{\openup1\jot \caja
        \ialign{\strut \hfil$\displaystyle{##}$&$
        \displaystyle{{}##}$\hfil\crcr#1\crcr}}\,}
\newif\ifdtup
\def\panorama{\global\dtuptrue \openup1\jot \caja
        \everycr{\noalign{\ifdtup \global\dtupfalse
        \vskip-\lineskiplimit \vskip\normallineskiplimit
        \else \penalty\interdisplaylinepenalty \fi}}}
\def\eqalignno#1{\panorama \tabskip=\humongous
        \halign to\displaywidth{\hfil$\displaystyle{##}$
        \tabskip=0pt&$\displaystyle{{}##}$\hfil
        \tabskip=\humongous&\llap{$##$}\tabskip=0pt
        \crcr#1\crcr}}

\def\eqright #1\cr{\noalign{\hfill$\displaystyle{{}#1}$}}
\def\eqleft #1\cr{\noalign{\noindent$\displaystyle{{}#1}$\hfill}}

\def\oldreffmt#1{\rlap{[#1]} \hbox to 2\parindent{}}

\def\figfmt#1{\rlap{Figure {#1}} \hbox to 1in{}}

\def\etal{\hbox{\it et al.}}

\def\sectioneq{\def\theequation{\thesection.\arabic{equation}}{\let
\holdsection=\section\def\section{\setcounter{equation}{0}\holdsection}}}%
\def\sectiontab{\def\thetable{\thesection.\arabic{table}}{\let
\holdsection=\section\def\section{\setcounter{table}{0}\holdsection}}}%
\def\sectionfig{\def\thefigure{\thesection.\arabic{figure}}{\let
\holdsection=\section\def\section{\setcounter{figure}{0}\holdsection}}}%

\newcounter{holdequation}

\def\num{(\refstepcounter{equation}\theequation)}
\def\auto{\eqno(\refstepcounter{equation}\theequation)}
\def\begineq #1\endeq{$$ \refstepcounter{equation}\eqalign{#1}\eqno
        (\theequation) $$}
\def\contlimit{\,{\hbox{$\longrightarrow$}\kern-1.8em\lower1ex
\hbox{${\scriptstyle (a\rightarrow0)}$}}\,}
\def\centeron#1#2{{\setbox0=\hbox{#1}\setbox1=\hbox{#2}\ifdim
\wd1>\wd0\kern.5\wd1\kern-.5\wd0\fi
\copy0\kern-.5\wd0\kern-.5\wd1\copy1\ifdim\wd0>\wd1
\kern.5\wd0\kern-.5\wd1\fi}}
\def\centerover#1#2{\centeron{#1}{\setbox0=\hbox{#1}\setbox
1=\hbox{#2}\raise\ht0\hbox{\raise\dp1\hbox{\copy1}}}}
\def\centerunder#1#2{\centeron{#1}{\setbox0=\hbox{#1}\setbox
1=\hbox{#2}\lower\dp0\hbox{\lower\ht1\hbox{\copy1}}}}
\def\lsim{\;\centeron{\raise.35ex\hbox{$<$}}{\lower.65ex\hbox
{$\sim$}}\;}
\def\gsim{\;\centeron{\raise.35ex\hbox{$>$}}{\lower.65ex\hbox
{$\sim$}}\;}
\def\st#1{\centeron{$#1$}{$/$}}

\def\super#1{\ifmmode \hbox{\textsuper{#1}}\else\textsuper{#1}\fi}
\def\textsuper#1{\newcount\holdspacefactor\holdspacefactor=\spacefactor
$^{#1}$\spacefactor=\holdspacefactor}

\def\getcite#1,{\advance\citenumber by1
\ifnum\citenumber=1
\ref{#1}\let\next=\getcite\else\ifx#1@\let\next=\relax
\else ,\ref{#1}\let\next=\getcite\fi\fi\next}

\def\upon #1/#2 {{\textstyle{#1\over #2}}}
\relax
\renewcommand{\thefootnote}{\fnsymbol{footnote}}

\def\mainhead#1{\setcounter{equation}{0}\addtocounter{section}{1}
  \vbox{\begin{center}\large\bf #1\end{center}}\nobreak\par}
\sectioneq
\def\subhead#1{\bigskip\vbox{\noindent\bf #1}\nobreak\par}

\def\til#1{\centeron{\hbox{$#1$}}{\lower 2ex\hbox{$\char'176$}}}
\def\tild#1{\centeron{\hbox{$\,#1$}}{\lower 2.5ex\hbox{$\char'176$}}}
\def\sumtil{\centeron{\hbox{$\displaystyle\sum$}}{\lower
-1.5ex\hbox{$\widetilde{\phantom{xx}}$}}}

\def\kbar{\underline{k}}
\def\qbar{\underline{q}}
\def\kbarsl{\underline{\st k}}
\def\qbarsl{\underline{\st q}}
\def\pom{{\rm P\kern -0.53em\llap I\,}}
\def\spom{{\rm P\kern -0.36em\llap \small I\,}}
\def\sspom{{\rm P\kern -0.33em\llap \footnotesize I\,}}
\def\parens#1{\left(#1\right)}

\def\pbar{\underline{p}}
\def\pbarsl{\underline{\st p}}

\begin{document} \begin{titlepage} \rightline{\vbox{\halign{&#\hfil\cr
&\normalsize ANL-HEP-PR-93-16\cr
&\normalsize\today\cr}}}
\vspace{1in}
\begin{center}

\Large
{\bf ANALYTIC MULTI-REGGE THEORY AND THE POMERON IN QCD : II. ~GAUGE THEORY
ANALYSIS}\footnote{Work supported by the U.S. Department of Energy, Division of
High Energy Physics, Contract W-31-109-ENG-38.}
\medskip

\normalsize ALAN R. WHITE
\\ \smallskip
High Energy Physics Division\\Argonne National
Laboratory\\Argonne, IL 60439\\ \end{center}

\begin{abstract}

The high-energy Regge behavior of gauge theories is studied via the
formalism of Analytic Multi-Regge Theory. Perturbative results for
spontaneously-broken theories are first organised into reggeon diagrams.
Unbroken gauge theories are studied via a reggeon diagram infra-red analysis
of symmetry restoration. Massless fermions play a crucial role and
the case of $QCD$ involves the Super-Critical Pomeron as an essential
intermediate stage.

An introductory review of the build up of transverse momentum diagrams and
reggeon diagrams from leading log calculations in gauge theories is presented
first. It is then shown that the results closely reproduce the general
structure
for multi-regge amplitudes derived in Part I of the article, allowing the
construction of general reggeon diagrams for spontaneously-broken theories.
Next it is argued that, with a transverse-momentum cut-off, unbroken gauge
theories can be reached through an infra-red limiting process which
successively
decouples fundamental representation Higgs fields.

The first infra-red limit studied is the restoration of $SU(2)$ gauge symmetry.
The analysis is dominated by the exponentiation of divergences imposed by
Reggeon Unitarity and the contribution of massless quarks to reggeon
interactions. Massless quarks also produce ``triangle anomaly''
transverse-momentum divergences which {\em do not exponentiate} but instead
are absorbed into a {\em reggeon condensate} - which can be viewed as a
``generalised winding-number condensate''. The result is a reggeon spectrum
consistent with confinement and chiral symmetry breaking, but there is {\em
no Pomeron}. The analysis is valid when the gauge coupling does not grow in
the infra-red region, that is when a sufficient number of massless quarks is
present. An analogy is drawn between the confinement produced by the reggeon
condensate and that produced by regularisation of the fermion sea, in the
presence of the anomaly, in the two-dimensional Schwinger model.

When the analysis is extended to the case of $QCD$ with the gauge symmetry
restored to $SU(2)$, the Reggeon condensate can be identified with the Pomeron
condensate of Super-Critical Pomeron theory. In this case, the condensate
{\em converts an $SU(2)$ singlet reggeised gluon to a Pomeron Regge pole} -
which becomes an $SU(3)$ singlet when the full gauge symmetry is restored.
The condensate disappears as $SU(3)$ symmetry is recovered and, in general
this limit gives the Critical Pomeron at a particular value of the
transverse cut-off. If the maximal number of fermions consistent with
asymptotic freedom is present no transverse momentum cut-off is required.

For $SU(N)$ gauge theory it is argued that, when the theory contains many
fermions, there are $N-2$ Pomeron Regge poles of alternating signature. This
spectrum of Pomeron trajectories is in direct correspondence with the
topological properties of transverse flux tubes characterised by the center
$Z_N$ of the gauge group. The corresponding Reggeon Field Theory solution of
$s$-channel unitarity should include a representation of $Z_N$ in the cutting
rules.

Finally the implications of the results for the phenomenological study of the
Pomeron as well as the for the construction of $QCD$ with a small number of
flavors are discussed. Also discussed is the attractive possibility that a
flavor-doublet of color sextet quarks could both produce the Critical
Pomeron in QCD and also be responsible for electroweak dynamical symmetry
breaking.

\end{abstract}

\renewcommand{\thefootnote}{\arabic{footnote}} \end{titlepage}

\mainhead{1.   INTRODUCTION}

This is the second part of what we hope will be a definitive article,
presenting as well as we are able, the understanding of the Pomeron in $QCD$
that our long study of the problem has produced. The first part\cite{pt1} of
the
article was relatively straightforward in that it simply described all of the
general Analytic Multi-Regge Theory that we believe is necessary both
conceptually and technologically to attack the problem. This second part deals
directly with the analysis of $QCD$ and gauge theories in general and involves
a complex mixture of straightforward perturbative Regge limit calculations with
elaborate dynamical concepts and all-orders manipulations.

The problem of the Pomeron in $QCD$ necessarily couples the infra-red and
ultra-violet behavior of the theory in a ``vacuum'' sector and so
involves, either directly or indirectly, almost all dynamical properties of the
theory. As a consequence there are many points of principle and subtleties of
interpretation which we find it necessary to discuss at some length. While
this may make our presentation seem excessively complicated (and even
convoluted), we feel that the final outcome is both simple and attractive.
Indeed we hope that by the end of the article the reader will agree that we are
at a very exciting stage. While there remains an enormous amount of work to do
to completely construct the dynamical picture we formulate we do believe that,
at least in principle, we have solved the fundamental problem of how to extract
the physical Pomeron from the full dynamical complexity of $QCD$.

If we claim that we have a solution to a problem involving the infra-red
behavior of $QCD$, then clearly we must have something to say about
confinement.
Indeed we find that the nature of the Pomeron is intimately tied to the
question
of the true physical states of the theory. The core of our dynamical picture is
that, {\em in the Regge limit and with a sufficient number of massless quarks},
confinement can be understood via an analysis of transverse momentum infra-red
divergences. Quarks play a crucial role in our analysis and, in effect,
confinement emerges as an outcome of the regularisation of the massless quark
sea, in a manner that closely parallels the well-known solution of the
two-dimensional massless Schwinger model\cite{man}.

The Regge limit is essentially determined by the on mass-shell regions of
Feynman diagrams and, at the most elementary level, the technology we exploit
can be viewed as simply enforcing unitarity and analyticity on the organisation
and (infinite) summation of perturbative Regge region contributions to general
multiparticle scattering amplitudes. Although it would be an indirect way to
proceed, the asymptotic states of $QED$ (to the extent they are defined) can
actually be found by summing infra-red divergences in the Regge limit. In this
context, our arguments would confirm the result\cite{ce} that in $QED$, the
recently discovered\cite{gik} ``new anomaly'' due to the mass-shell region of a
massless electron loop decouples from the true asymptotic states.
However, as is well-known, the photon does not reggeize in $QED$ and so all
the infra-red divergences associated with reggeization are absent in this
case. Of course, $QCD$ has to be more complicated than $QED$ if the true
asymptotic states are to emerge as bound states as expected. Indeed in our
formalism it is the combination of the divergences of reggeization with the
infra-red ``anomalies'' of massless quarks which is the key to the emergence
of a confinement spectrum. (In this context an anomaly is the violation of
the consequences of a Ward identity that follow from the assumption of no
infra-red singularities.)

Our analysis suggests that the role of infra-red quark-loop anomalies is as
follows. First we note that the full array of multi-regge limits and multiple
discontinuities that we consider leads to a complex structure of ``off-shell''
reggeon amplitudes. In particular there exist ``unphysical'' reggeon amplitudes
which {\em are not related to on-shell particle amplitudes} but contribute
vitally to the process of summation of divergences. It is these amplitudes
which contain ``anomalous'' infra-red contributions from massless quark loops
and it is the resulting divergences which ultimately are responsible for
confinement. It seems to be a general property that the anomalies involved do
not destroy the reggeization properties of the theory (that are also a
manifestation of gauge invariance) just because they occur only in unphysical
amplitudes.

Our general understanding will enable us to arrive at a number of results and
conclusions. For example, we shall see that $QCD$ (that is SU(3) gauge theory)
is very special and uniquely able to produce the strong-interaction high-energy
asymptotic behavior seen in experiments. The associated Critical
Pomeron\cite{mpt} behavior occurs at a value of the transverse momentum
cut-off which increases with the number of flavors and occurs without such a
cut-off when the theory is ``saturated'' with quarks. We shall emphasise
that the Critical Pomeron is particularly important in the context of $QCD$
because of its ability to reconcile, at asymptotic energies, short distance
Parton Model properties with the confinement and chiral symmetry breaking
vacuum properties that are essential for the physical applicability of the
theory. The Super-Critical Pomeron\cite{pt1} emerges as describing a form of
gauge symmetry breaking and plays a vital role in relating quark and gluon
Regge behavior to that of the Pomeron and hadrons. The symmetry breaking is
associated with a ``reggeon condensate'' which, in a very particular sense,
can be thought of as a generalised ``winding number condensate''\cite{man}.

The ``Pomeranchon'' (which soon evolved into the Pomeron) was born\cite{cf} as
a simple Regge pole more than thirty years ago. The original reasons for
supposing its existence were entirely phenomenological, apart from the general
aesthetic appeal of the newly introduced concept of Regge behavior. Ten years
later a fundamental debate on whether such Regge behavior can occur in quantum
field theory was underway\cite{fgl}, based on its absence (due to the dominance
of large transverse momenta) in leading log {\em summations} in massive $QED$.
The field theory {\em reggeization} of fermions had been understood\cite{gglmz}
by this time, but a bound-state Pomeron remained elusive. Meanwhile the
abstract formalism of multi-regge theory\cite{bdw} based on analyticity and
unitarity, and the related Reggeon Field Theory formalism\cite{gr}, had begun
its development and the theoretical virtues of a Pomeron Regge pole began to
be appreciated. When string theories containing something\cite{lov} very close
to the desired Pomeron came into vogue, concern about the absence of such
behavior in $QED$, or field theory in general, essentially dissipated. As we
emphasised in Part I, the culmination of the general formalism is the Critical
Pomeron solution\cite{mpt} of the Reggeon Field Theory referred to
above - which still provides the only known (``non-trivial'') solution of
all analyticity and unitarity constraints at asymptotic energies.

Since the acceptance of $QCD$ as a field theory of the strong interaction the
question of whether a Pomeron Regge pole, and more importantly the Critical
Pomeron, can occur in field theory has re-emerged as of prime significance.
Particularly since essentially the same leading-log calculations and
{\em summations} performed for $QED$ were repeated\cite{klf,cl} for
(spontaneously-broken) non-abelian theories and showed the same phenomenon, in
the vacuum channel, of the dominance of large transverse momenta leading (after
summation) to non-Regge behavior. (This is essentially the ``Lipatov
Pomeron''\cite{lnl} which we shall return to below). Of course, quark
confinement is believed to be a major part of the dynamical solution of
$QCD$ and it could be that the Pomeron describing hadron scattering
originates entirely from a confining ``string-like'' solution of the theory
which is completely divorced from perturbation theory. However, if this were
the case, we would lose all connection between the vector nature of the
gluon and the approximate constancy of hadronic total cross-sections (and as
a consequence any understanding of how or why Critical Pomeron behavior
should occur). It would probably also imply much less validity for the
parton model, and perturbative calculations in general, than is
conventionally assumed.

As we emphasised in Part I, and touched on above, our viewpoint is that the
Pomeron and its consistency with analyticity and unitarity provides a true
testing-ground for the compatibility of conventional assumptions about the
short-distance and long-distance behavior of $QCD$. Consequently a complete
understanding of the Pomeron is not to be found in either a purely
``perturbative'' (multi-)gluon exchange picture or in some form of
``non-perturbative'' flux-tube (string-)picture of confining $QCD$. Rather we
have to understand how and why it is that the Pomeron achieves a marriage of
these two contrasting pictures. The Sub-Critical Pomeron described by
conventional Reggeon Field Theory has a very natural association with the
non-perturbative flux-tube picture, whereas we argued in Part I that the
Super-Critical Pomeron should describe a spontaneously-broken gauge theory - in
which perturbation theory (at low transverse momentum) has at least partial
validity. This clearly suggests that the Critical Pomeron should be related to
some form of transition between the perturbative and non-perturbative pictures.
It also suggests that if we wish to start from perturbative calculations and
reach the Pomeron of non-perturbative confining $QCD$ we can anticipate that,
as we discussed in Part I, it will be necessary to go through the
Super-Critical
Phase.

While the general arguments and analysis presented in Part I clearly underly
our
whole approach in this Part, we do believe that our procedure for studying
$QCD$ can be directly motivated without reference to abstract Pomeron theory.
Indeed, with the hope of attracting the interest of young theorists raised
entirely on $QCD$, we shall try to emphasise the direct motivation as much as
possible and minimise the reference to abstract Pomeron theory until the later
stages of our analysis. Our starting point is Regge limit {\em low transverse
momentum} perturbation theory calculations. In effect this is unavoidable
because perturbation theory is all we have to concretely define $QCD$ for
our purposes. Also, since the Higgs mechanism is the only available {\em
unitary} infra-red cut-off, we inevitably begin with calculations in
spontaneously-broken gauge theories. Therefore our starting point does not
require any reference to Super-Critical Pomeron theory or to the virtues of the
multi-Regge S-Matrix which we discuss below, to justify it. We should
emphasise, however, that the summation and limiting procedures involving
reggeon diagrams that we employ to reach confining $QCD$ from our
``perturbative'' starting point are only directly valid under the very
limited circumstances that the gauge coupling does not grow in the infra-red
region. That the results we obtain are then also consistent with a flux-tube
picture can be viewed as welcome independent justification for our procedure.

We should also note the current variety of interpretations given by different
authors to the words ``the Pomeron in $QCD$''. Recently it has become popular
to
distinguish phenomenologically the ``soft'' (or ``non-perturbative'') Pomeron
from the ``hard'' (or ``perturbative'') Pomeron. While it is certainly not
clear that any absolute distinction can be made, either experimentally or
theoretically, we can distinguish correspondingly distinct theoretical starting
points. The ``hard Pomeron'' appears in the study\cite{lnl,glr,bl} of problems
which are direct extensions of those which can be consistently formulated
within
conventional perturbative $QCD$. The presence of the large transverse momentum
of a hard scattering process provides the initial justification\cite{bl} for
use of the perturbative (leading log) calculations of Lipatov and
others\cite{klf,cl} and it is argued\cite{ahm} that the resulting ``Lipatov
Equation'' controls, for example, the small-x behavior of the parton
distributions appearing in deep-inelastic scattering and even\cite{ahmc} that
the corresponding ``quark scattering amplitude'' can be detected in very high
energy production processes involving multiple jets. The study of the
``Lipatov Pomeron'', and corrections to it, at large transverse momentum can
be viewed, therefore, as a problem of summing more and more complicated
diagrams within short distance, perturbative, $QCD$. As such, the problem
may well be insurmountable (if not undefined) in its full generality.

Our objective, on the other hand, is to find the so-called ``soft'' Pomeron
which describes, within $QCD$, the {\em low momentum transfer} high-energy
scattering of {\em hadrons,} and satisfies all the S-Matrix constraints of {\em
analyticity, crossing and unitarity} elaborated in Part I. Though this may
appear to be an even more difficult problem it is surely well-defined. Indeed
we start from essentially the same leading log perturbative calculations as in
the large transverse momentum problem. However, it is important that we
explicitly start in {\em spontaneously-broken} $QCD$ and that we impose an
{\em upper} transverse momentum cut-off. We keep, therefore, the momentum space
region where the general Analytic Multi-Regge Theory of Part I, and in
particular the closely related program of Bartels\cite{jb} (which calculates
non-leading multi-regge corrections in spontaneously-broken theories directly
from analyticity and unitarity) is directly applicable. In effect, we begin
with the perturbative description of the (low transverse momentum) region which
in the deep-inelastic problem, for example, a conventional perturbative $QCD$
analysis would ``subtract'' into the non-perturbative part of parton
distribution functions\cite{jce}. However, we actually consider (multi-regge
regions of) very complex multiparticle quark and gluon scattering amplitudes
which lie well outside of any existing calculations within perturbative
$QCD$. This is essential to see the emergence of confinement in our formalism.

Returning to the historical development of the subject we note that the absence
of a perturbative Pomeron Regge pole in spontaneously-broken non-abelian gauge
theories was seen as particularly disappointing by many theorists in that the
reggeization of {\em all} the elementary vector bosons and fermions of the
theory\cite{gs} seemed to be such a {\em striking verification} of the
fundamental concepts of Regge theory - {\em in the context of field theory}.
Indeed this feature apparently culminates in the complete multi-regge
behavior of the {\em low transverse momentum} perturbative S-Matrix - with
{\em only} reggeised quarks and gluons. The remarkable extent to which this
behavior exactly reproduces the general Multi-Regge Theory results of Part I
provides, from our perspective, (more than) sufficient justification for
making it the starting-point of our study of the high-energy behavior of
$QCD$.

In starting from spontaneously-broken $QCD$, therefore, we begin with a theory
whose low momentum transfer high-energy S-Matrix is, in principle, well
understood. In practise this means that we can build up the general multi-regge
behavior of arbitrary amplitudes by using quark and gluon reggeon diagrams to
straightforwardly extrapolate existing perturbative results. There will
be {\em no Pomeron} in this theory. Indeed, the challenge, or effectively
{\em our formulation of the problem of the ``Pomeron in $QCD$''}, is then to
obtain the high-energy S-Matrix of $QCD$ through some form of (infra-red)
limiting procedure which can be legimately applied to this known formalism and
which also clearly preserves all the desirable properties. In particular if
Reggeon Unitarity is to be explicitly maintained, we must find a route (via
Reggeon Field Theory phase transition analysis etc.) whereby the {\em
reggeisation of gluons and quarks} in the spontaneously-broken theory leads
to the appearance of a Regge pole Pomeron (and consequent reggeon diagrams)
in the {\em hadron} S-Matrix of pure $QCD$.

The part of our program which simply extends the existing perturbative
calculations (of spontaneously-broken theories) by use of the reggeon diagram
technology described in Part I has some significance in its own right. Although
we should emphasise that the work of Bartels\cite{jb}, which is conceptually
very close, actually goes much further in providing detailed calculations of
non-leading logarithms. However, we believe that we make fundamental progress
when, as anticipated above, we introduce the abstract Super-Critical Pomeron
formalism\cite{pt1} of Reggeon Field Theory to describe the infra-red limit of
(perturbative) Reggeon diagrams as the gauge symmetry of $QCD$ is {\em
partially restored}. This provides the framework within which we can interpret
the divergences that occur. It is also the key to an understanding of both the
relationship between broken and unbroken $QCD$, and how it is possible that the
abstract Pomeron of Part I, and in particular the Critical Pomeron, can emerge
from the $QCD$ starting point of a field theory of quarks and gluons. The
crucial feature, which we enlarge on below, is the appearance (from the
infra-red divergence structure) of a reggeon condensate which {\em directly
converts an  odd-signature reggeised gluon into an even signature Pomeron}.

The infra-red analysis we employ is conceptually elaborate in that we find
it necessary to study (essentially) the complete quark/gluon multi-regge
S-Matrix embedded in general multiparticle amplitudes. To understand just how
the combination of massless quark interactions with the infra-red massless
gluon limit does indeed produce a {\em confining} hadronic multi-regge
S-Matrix, it is essential to {\em simultaneously} study the formation of
hadrons, as reggeon singularities, together with the development of a
Pomeron Regge pole. This requires the study of rather complicated
multi-regge amplitudes and is closely related to the emphasis placed in Part
I on the necessity to simultaneously study hadrons {\em and} the Pomeron in
order to consistently describe the Super-Critical Pomeron.

The infra-red limits we study involve, of course, the decoupling of the
mass-generating Higgs sector and as we have implied, this is carried out in
two stages for $QCD$. In the first stage, $SU(2)$ gauge symmetry is restored
by decoupling just one of the two color triplet Higgs scalars employed to
break the full $SU(3)$ symmetry. It is at this stage that the presence of
{\em massless quarks} is vital for obtaining a result consistent with
confinement. The physical reason for this is actually quite simple. A physical
(i.e. on-shell) massless gluon can decay into a pair of physical quarks {\em
carrying zero transverse momentum}, whereas it can not decay into a pair of
gluons because of helicity conservation. Consequently, infra-red regions of
massless quark loops can lead to (reggeized) gluon interactions which do not
vanish at zero transverse momentum.  These interactions produce further
exponentiation of those (color zero) massless gluon singularities which
survive the exponentiation of divergences due to gluon reggeization. Such
singularities would be inconsistent with confinement and their presence is
what determines that the massless gluon theory without quarks is
inconsistent.

Essential use has to made of the structure of interactions implied by
multi-regge theory to see that this last class of interactions do not destroy
the gauge invariance of the theory manifested in the reggeisation of gluons.
Even though the interactions do violate properties that follow naively from
the Ward Identities of the theory. We should also note that the analysis of
these interactions highlights a conflict between the use of a transverse
momentum cut-off, which we argue is essential as a matter of principle, and
Pauli-Villars regularisation of quark transverse momentum diagrams.

The most important contribution of massless quarks is, however, in what we call
the ``anomalous interactions'' produced by transverse momentum {\em quark
triangle diagrams}, within certain triple-regge contributions to {\em
multi-regge} amplitudes. In effect the Regge limit enhances the triangle
diagram singularity and in doing so brings the full structure of the axial
anomaly into the infra-red behavior of reggeon diagrams. We believe this {\em
feature is what allows a spectrum consistent with confinement and chiral
symmetry breaking to emerge from our analysis}. The ``triangle anomaly'' does
not occur within triple-regge vertices that appear in the reggeon diagrams for
elastic scattering or inclusive cross-section amplitudes but rather occurs
in an obscure component of a general vertex that contributes only in very
special kinematic limits. As a consequence the additional divergences produced
do {\em not exponentiate} but instead are associated with the development of a
zero transverse momentum configuration, or ``{\em reggeon condensate}'',
accompanying the creation of hadrons and the Pomeron. (As we elaborate there is
a sense in which this can be interpreted as a generalised ``winding-number
condensate''\cite{man}.) The reggeon condensate is, not surprisingly,
ultimately
identified as the Pomeron condensate of the Super-Critical Phase\cite{pt1}.

It is the ``triangle-anomaly'' phenomenon which has caused us the greatest
difficulty as we have attempted to develop our formalism over the years. We
have understood for a long time that it lies at the heart of the mechanism for
the development of a reggeon condensate and so is vital for the relationship
between spontaneously-broken $QCD$ and the Super-Critical Pomeron theory. We
have described this qualitatively in numerous publications\cite{arw}. However,
it is only recently that we have properly understood the relationship
between the anomalous interactions and the structure of the triple-regge
vertex. We believe this is the essential ingredient that enables us to both
properly expose the phenomenon and to begin to analyse its full
significance.

The reggeon condensate can be seen as directly responsible for confinement and
chiral symmetry breaking at the first stage of symmetry restoration (that is
the
restoration of $SU(2)$ gauge invariance). The condensate can be viewed as
originating from the regularisation of the massless quark sea in the presence
of the anomaly. In effect, {\em only} the states consistent with confinement
and chiral symmetry breaking can be consistently regularised. This closely
parallels the role of the regularisation of the fermion sea in the solution of
the two-dimensional massless Schwinger model\cite{man}. Indeed we find that
our multi-regge infra-red analysis is both self-consistent and
self-contained only when applied to $QCD$ with the maximum number of
massless fermions allowed by asymptotic freedom. In this case both the
infra-red and the ultra-violet behavior of the theory are as close to
perturbation theory as is possible in that there is no infra-red (or
ultra-violet) growth of the gauge coupling. In field-theory language the
only ``non-perturbative'' aspect of this theory is the contribution of
topological gauge fields and the related problem of the regularisation of
the fermion sea. We might anticipate therefore that, in these circumstances,
any ``condensate'' related to gauge-fields can necessarily be thought of as
topological in origin.

The restoration of $SU(3)$ gauge symmetry is achieved by the decoupling of
the second Higgs scalar field. This limit involves the {\em disappearance of
the
condensate} and is described by the Critical Pomeron. Nevertheless, it is the
development of the intermediate $SU(2)$ condensate which ultimately leads to a
Pomeron with all the right physical properties to give a unitary high-energy
limit for $QCD$. The condensate directly converts an {\em $SU(3)$ colored
reggeized gluon} to an {\em even signature Pomeron Regge pole} which becomes
an {\em $SU(3)$ color-singlet} in the symmetry limit. In this manner the
Regge behavior of (massive) gluons is {\em directly related} to the Regge
behavior of the Pomeron. Clearly the quark sea plays an absolutely essential
role in the emergence of a Regge pole Pomeron and ultimately in the
asymptotic consistency of $QCD$. It is perhaps not surprising that the
``simplest'' starting point for understanding this phenomenon is when the
fermion sea is most significant dynamically - that is when the maximum
number of massless fermions is present. In contrast, the pure gauge theory
has no fermion sea and no physically sensible Pomeron. Consequently it is
not a good starting point for understanding the vacuum properties of the
theory related to the Pomeron.

It is also interesting that the method of analysis can be extended
to a general $SU(N)$ gauge group. The existence of $N - 2$ Pomeron trajectories
comes perhaps as a surprise given the simple relationship expected between the
1/N expansion and Reggeon Field Theory. As we describe, the spectrum of
trajectories can be directly understood in terms of the topological properties
of transverse flux-tubes determined by the center of the gauge group\cite{gth}.
However, it should be emphasised that this structure only emerges as
significant when there are a large number of fermions in the theory. The
special relationship between the Critical Pomeron and $SU(3)$ gauge invariance
may have profound significance in that it could imply that $QCD$ really is
uniquely selected as describing the {\em maximal strong interaction consistent
with unitarity}. The flavor dependence could also be significant for
understanding how $QCD$ should be defined as a matter of principle and also
whether a further very massive quark sector, perhaps associated with
electroweak
symmetry breaking, is actually required for asymptotic consistency.

\newpage

\mainhead{2.  SUMMARY}

We have three presentational goals. We want to present the results of our own
research, to provide some review of the material which defines the context of
our work, and also to provide enough development that a newcomer to the field
can use the article as a learning vehicle and reference guide. There is clearly
potential for conflict in these goals, which we shall try as far as possible to
resolve. It will, however, mean that there is an uneven oscillation from
Section to Section in terms of the level of presentation, the proportion of new
to review material etc.. We should apologise for the extreme length of the
article. Unfortunately we have developed most of the material simultaneously
and feel that it would be very hard to break it up without losing the cohesion
and inter-relation of the various components. There are also a number of places
where we feel the overall strengths of the argument we are making may be better
than the details we formulate. To understand this it is essential to keep in
mind the global perspective which the complete article provides. We hope that
the lengthy summary which we give in the following goes some way in mitigating
the negative effects of the overall length.

We begin in Section 3 with a brief description of some general reasons why we
would expect that if a Regge pole Pomeron appears in $QCD$ at all, then the
Critical behavior will occur when the theory contains the maximal number of
hadronic states. This corresponds to adding the maximal number of quark flavors
without losing asymptotic freedom. As we then elaborate there are a number of
field-theoretic properties of this ``saturated'' theory which make it
compatible
with Critical Pomeron behavior and with the property that a Super-Critical
Pomeron condensate will develop if a Higgs sector is added. That the Pomeron
condensate might be ``topological'' in origin and related to the regularisation
of the fermion sea is even suggested by the general field-theoretic properties
we discuss. This Section makes no use of reggeon diagrams and is intended to
prepare the reader for the results that emerge from the more technical analysis
of later Sections.

To begin our review of the Regge properties of spontaneously-broken gauge
theories, the t-channel unitarity analysis\cite{gs} of the general reggeization
of all elementary fermions and vector bosons in a non-abelian theory is briefly
described in Section 4. This analysis is complimentary to almost all of the
diagrammatic analysis and development of the succeeding Sections. We include it
for completeness and also because of its simple elegance and generality. We
also emphasise that the Regge poles produced by reggeization in field
theory are isolated in the angular-momentum plane. As a result they are better
candidates to link up with the isolated Regge poles of an effective string
theory (describing the confining flux-tube solution of a gauge theory) than the
typical spectrum of bound state Regge poles. The isolation property also allows
reggeized gluons (or quarks) and their associated Regge cuts to be responsible
for the {\em complete} leading power behavior (that is including {\em all}
logarithms) of a theory at high energy.

          The detailed program of constructing perturbative multi-regge
behavior
is begun in Section 5. We discuss the general reduction of Feynman diagrams to
transverse momentum diagrams in the Regge limit. We begin with the simplest
scalar box diagram. We then move on to low-order diagrams in an abelian gauge
theory (massive $QED$), where the elaborate large transverse momentum
cancellations that characterise a gauge theory can be most simply illustrated.
We then present the results of the 10th order leading and non-leading log
calculations of Cheng and Lo\cite{cl} for $SU(2)$ gauge theory. We emphasise
the
origin of the leading-log reggeization of the gluon in the ``close-to'' on
mass-shell regions of ladder diagrams. The unitarity plus dispersion relation
derivation of these results by Lipatov and collaborators\cite{klf,lnl} is then
reviewed to prepare for the general insight that is derived from this
approach in the following Section. We also mention, but do not review, the
all-orders (of leading log and non-leading logs) formalism of Sen\cite{as}. His
results guarantee that transverse momentum diagrams fully describe the
leading power (plus all logs) Regge limit of a gauge theory.

          The organisation of transverse momentum diagrams into the reggeon
diagram language of Part I is the subject of Section 6. The 10th order SU(2)
results (originating from several hundred Feynman diagrams) can be rewritten as
just five reggeon diagrams. We discuss why unitarity requires that this must be
the case, and how the Lipatov et al. method\cite{klf,lnl} (and its extension by
Bartels\cite{jb}) leads to the complete multi-regge behavior of all leading and
non-leading log multiparticle amplitudes. We then show that all of the reggeon
diagrams that occur conform with the general formalism of Part I in the
simplest possible manner. That is the diagrams are reproduced by a construction
procedure based on reggeon unitarity which employs only the simplest form for
reggeon vertices. We take this to mean that we can straightforwardly
extrapolate
the existing results to the most general amplitudes and multi-regge limits.

          In Section 7 we begin to move outside the perturbative multi-regge
framework of the previous Sections by discussing one of the basic questions of
principle posed in the Introduction. Namely are there conditions under which we
can reach the high-energy Regge limit of unbroken (confining) QCD, starting
from
(one of) the spontaneously-broken theories that we apparently have under such
good control. The lattice gauge theory principle of complimentarity\cite{fs}
provides the basic answer. This tells us that, with a {\em transverse momentum
cut-off}, and with {\em fundamental representation} Higgs scalars used to first
give gluons a mass, and then decoupled to remove this mass, there is {\em no
confinement phase-transition}. That is the ``Higgs'' and ``confinement''
regions of parameter space are {\em smoothly} related. The second question we
pose is whether the smooth transition from a perturbative massive gluon theory
to a confining massless gluon theory can be followed (even in principle) in the
high-energy reggeon diagrams. We first discuss the dynamical significance of
transverse momentum divergences in $QED$ and then argue that {\em if the
gauge-coupling of a non-abelian theory does not grow in the infra-red region}
then such divergences can be the origin of the transition to a confining
theory. Eliminating the growth of the gauge coupling provides our first
motivation for the introduction of massless fermions.

Our infra-red analysis of reggeon diagrams starts in Section 8 by analysing the
$SU(2)$ theory with reggeized gluons only. In this analysis the significance of
reggeization is that it produces a ``$t$-channel'' exponentiation (in rapidity
space) of infra-red divergences. By relating this exponentiation to the
inversion of reggeon singularities (in the angular momentum plane) due to
reggeon unitarity, we show that it simply sends to zero, all reggeon diagrams
which carry {\em non-zero (total) color} in the $t$-channel. In color zero
channels the divergences cancel and instead there is a set of infra-red finite,
scaling, reggeon interactions and off-shell amplitudes. However, because
the reggeon interactions vanish whenever any sub-set of the interacting
reggeons carries zero transverse momentum,(this is the Ward-Identity property
referred to in the Introduction), there are infra-red
divergences of color-zero on mass-shell (gluon) scattering amplitudes which are
not inverted. As a result the ``pure-glue'' reggeon S-Matrix does not have a
sensible infra-red limit. We note that the scale-invariance property of the
color-zero kernels can be used to generate the large transverse momentum
``Lipatov Pomeron'' singularity referred to in the Introduction by first
removing the transverse momentum cut-off and then taking the infra-red limit.
We argue that, for our purposes, this is the {\em wrong-order} in which to
{\em take limits} and that this is why a finite (with suitable scattering
states) but {\em non-unitary} result is obtained.

We finally introduce reggeized quarks into our discussion in Section 9. We note
first that in the leading-log approximation quarks have analagous reggeization
and multiperipheral bootstrap properties\cite{fs1} to gluons. We also note that
the zero-mass limit for quarks is smooth and does not destroy their
reggeization
properties. We then go on to discuss the contribution of quark-loops to
reggeized gluon vertices. In $QED$, electron loops provide the leading-log
interactions for the iteration of photon pair exchange. The analagous quark
loop interactions are buried at non-leading log level in a non-abelian theory.
However, we show that there is a vital new reggeized gluon interaction due to
the exchange of a colored quark/antiquark pair which is a well-defined
leading-log Regge region contribution. With massless quarks and a transverse
momentum cut-off this interaction necessarily eliminates the ``Ward-Identity''
zeroes discussed in the previous paragraph. Almost all gluon infra-red
divergences are now inverted (or exponentiated in rapidity space). We then note
that the {\em infra-red} scaling properties of the color zero kernels could
produce further divergences which {\em are not inverted} if the gluons involved
form an odd-signature ``Odderon'' configuration which has ``{\em anomalous
color-charge parity}'' - that is the gluons form an ``anomalous Odderon''.
However, we show that such a configuration can not appear at all in the
theory unless there are {\em divergent} massless quark transverse momentum
loops to which it can couple.

Section 10 contains, perhaps, the most fundamental new result that we derive.
We
begin with what is essentially an omission from Part I, that is a discussion of
triple-Regge limits and the structure of the triple-Regge vertex. We recall
that
this vertex is a sum\cite{dw} of three familiar terms (which appear in
inclusive
cross-sections and in elastic scattering reggeon diagrams) and a fourth
somewhat mysterious term which appears only in {\em strictly multiparticle}
helicity-pole and triple-Regge limits. This term has previously seemed to be of
academic significance only. However, when we ask whether an infra-red reggeon
diagram divergence involving (simultaneously) three quark propagators can
occur, we find that it requires a very particular discontinuity (and helicity)
structure. This implies that such a divergence can occur in (and only in) the
``mysterious'' fourth term of the triple-Regge vertex. This result is (not
surprisingly) closely connected to the helicity and momentum structure of the
triangle anomaly. The important point for our analysis is that it immediately
follows that the ``anomalous Odderon'' gluon configuration {\em will couple}
via this anomaly and the resulting divergence {\em will not be inverted} via
reggeon interactions. This implies that this divergence can be interpreted as
associated with a {\em reggeon condensate}. Furthermore this condensate is
clearly linked to the anomaly and the massless fermion sea. We reserve further
explanation of the significance of this relationship until after we discuss the
Schwinger model in Section 11. We finish the Section by describing further
anomalous interactions involving the production (or absorption) of a massive
gluon reggeon pair or the emission (or absorption) of a zero transverse
momentum
anomalous odderon configuration from a quark/antiquark reggeon state.

         The construction of the high-energy S-Matrix for $SU(2)$ gauge theory
(with many fermions) is the subject of Section 11. We first summarise essential
results established in the previous Sections and then discuss, qualitatively at
least, how the full set of multi-regge amplitudes generates the physical
spectrum of reggeon (and therefore particle) states together with the physical
S-Matrix. Indeed with just $SU(2)$ gauge symmetry {\em there is no Pomeron} and
the only surviving reggeon states are {\em pseudoscalar mesons} formed by
quark/antiquark reggeons in the background of the reggeon condensate. In effect
we have a {\em confining (and chiral symmetry breaking) spectrum and a
non-interacting Regge limit S-Matrix}. In order to emphasise the close
parallels that can be made we then go on to present a brief review of the
relationship between the fermion sea, confinement, and a winding number
condensate, in the two-dimensional Schwinger model\cite{man}. In particular we
note the parallel between confinement at zero and non-zero electron mass in
the Schwinger model, and at infinite and finite momentum in $QCD$.

In Section 12 we finally begin explicit discussion of $QCD$, beginning with the
construction of the reggeon diagrams when the gauge symmetry is completely
broken with the aid of two color triplets of Higgs scalars. An $SU(2)$ global
symmetry again remains and is the ``color'' that is ultimately confined by our
infra-red analysis. We give an outline of how, as $SU(2)$ gauge symmetry is
restored, the anomaly divergences both {\em convert an $SU(2)$ singlet
reggeized
gluon to an even signature Pomeron} and build up the full structure of the
Super-Critical Pomeron. In this theory we have a very interesting
``Parton-Model'' picture in which there are $SU(2)$ color singlet reggeized
gluons and quarks which have elementary interactions but their Reggeon
signature properties are transformed by the condensate. In particular we have
an
even signature Pomeron which couples\cite{dl} ``just like a photon''.

Section 13 begins with a description of how the Super-Critical Pomeron gives
the Critical Pomeron as $SU(3)$ gauge symmetry is restored. We emphasize the
role played by color charge parity in determining, in particular, that there
are no ``glueballs'' on the Pomeron trajectory. No transverse cut-off is
required if the maximum number of quark flavors is present. If there are fewer
flavors present we argue that the Critical Pomeron can still be obtained
with a specific {\em critical} value of the cut-off. (We postpone discussion
of the significance of this result until Section 14.) We then go on to
extend the analysis to $SU(N)$ gauge theory and explain how the special
properties of the $SU(3)$ gauge group lead to the presence of a single Regge
pole Pomeron. We do this by relating the number of Pomeron trajectories
built up, by our symmetry restoration procedure, to the role of the center
of the gauge group in determining the transverse flux-tube structure that is
possible with $SU(N)$ gauge fields and in determining the Reggeon Field
Theory cutting-rules that we expect. The conclusion is that there are $N-2$
trajectories, with alternating signatures, in $SU(N)$ gauge theory and the
number that are critical depends on how close the fermion flavor number is
to "saturation".

Finally we discuss in Section 14 how we relate all of our work to the physical
world as we currently understand it. We begin with a brief description of how
we think our results relate to the current phenomenological understanding of
the Pomeron. We then consider implications for the construction of the theory
with a small number of flavors. We also discuss how a flavor-doublet of
color sextet quarks could both complete the ``flavor-saturation'' of $QCD$
(producing the Critical Pomeron) and be responsible for dynamical electroweak
symmetry breaking\cite{wjm}. We conclude by briefly elaborating how generalised
Super-Critical Pomeron theory may be the appropriate formalism to analyse
the dynamical solution of (particular) unified gauge theories.

\newpage

\mainhead{3.  QCD AND THE CRITICAL POMERON - GENERAL ARGUMENTS}

We present in this Section some general arguments which support our
expectations
as to how and why the Critical Pomeron occurs in $QCD$ but which do not require
the full technology of gauge theory reggeon diagrams which we develop in
succeeding Sections. We shall find that we are rather straightforwardly led
towards the physics that the Critical and Super-Critical Pomeron are associated
with.

First let us suppose only that $QCD$ is indeed a theory of hadrons and that
{\em
low transverse momentum} hadron production is described by Pomeron diagrams as
outlined in I-6.8. To support this assumption we can appeal to a ``1/N'' or
``topological'' expansion of the theory\cite{cmv}, or to the closely related
idea that an effective closed string theory of flux tubes describes the
Pomeron in $QCD$. Within the resulting ``multiperipheral'' model, it is
straightforward to show\cite{gt} that increasing the number of hadron states
increases the ``bare'' Pomeron intercept. Therefore we anticipate that within
$QCD$, ``flavoring the Pomeron'' by increasing the number of quark flavors
directly increases the bare intercept. Consequently, if we can establish that
the Pomeron is Sub-Critical for some specific number of flavors then we can
move towards the Critical theory simply by increasing this number.

We also know from the (Reggeon Field Theory) weak-coupling formula\cite{sw}
given in Part I that increasing the transverse momentum cut-off increases the
bare {\em critical} intercept. This implies that by raising this cut-off, while
keeping the other parameters of the theory fixed, we can move towards,
or further into,the Sub-Critical Phase. It is therefore very reasonable that
the theory with a small number of flavors will be Sub-Critical, if the
transverse cut-off is taken {\em sufficiently large}. If this is the case then
we can indeed expect to approach the Critical Pomeron by increasing the number
of quark flavors. However, $QCD$ is not expected to significantly depend on the
flavor number until it radically changes its character as asymptotic freedom is
lost. Therefore a natural point for the Critical behaviour to occur would be
just at the maximum number of flavors allowed by asymptotic freedom (referred
to as ``flavor saturation'' in the following). Indeed there are further general
arguments that it is ``flavor-saturated $QCD$'' that gives the Critical
Pomeron,
as we now describe.

A special property of flavor-saturated $QCD$ is\cite{gw,cel} that a (complex)
color-triplet Higgs scalar sector can be added - with both the gauge-coupling
{\em and} the Higgs self-coupling asymptotically free. Let $g(t)$ and
$h(t)$ be the respective scale-dependent couplings, then
$$\eqalignno {
{{dg} \over {dt}} =& \beta(g,h)&\num\label{3.1}\cr
=& -{1\over 2}b_0t^3 + \cdots&\num\label{3.2} \cr }
$$
where
$$
b_0 = {1\over {8\pi^2}} \left[11 -{4\over 3}S_3(f)-{1\over 6}\right]
\auto
$$
The 1/6 is due to the triplet scalar and $S_3(f)$ depends on the number (and
representation) of the quarks present. Similarly
$$\eqalignno{
{{dh}\over {dt}} =& \tilde{\beta} (g,h)&\num\label{3.4}\cr
=& Ah^2 + Bg^2 + Cg^4 + \cdots &\num\label{3.5}\cr }
$$
where
$$
A = {7\over {8\pi^2}},\ B = -{1\over {\pi^2}}~~~ and~~
C = {{13}\over {48\pi^2}}.
\auto
$$
We can have $h \rightarrow 0$ consistently in (3.5) if
$h = xg^2 + 0(g^3)$.
 This gives a stability equation for $x$, that is
$$
{{dx}\over {dt}} = g^2 \left( Ax^2 + B^\prime x + C \right)
\auto
$$
where $B^\prime = B + b_0$.  When the stability condition
$(B^\prime)^2 > 4AC$ is
satisfied there are two fixed-points of (3.7) and the smaller is stable for
$t \rightarrow \infty$.  The stability condition gives
$$
\left( 1-\pi^2b_0\right) ^2 > {{91}\over {96}}
\auto
$$
which for $b_0$ small gives
$$
{5\over {24}} > 8\pi^2 b_0
\auto
$$
If all quarks are color triplets then  $S_3(f) = N_f/2$, where $N_f$ is the
number of flavors. We now observe that if $N_f = 16$ then
$$
8\pi^2 b_0 = {1\over6} < {5\over{24}}
\auto
$$
whereas if $N_f = 15$ then
$$
8\pi^2 b_0 = {5\over6} > {5\over{24}}
\auto
$$
We conclude that when the number of flavors saturates at exactly 16 we can use
the Higgs mechanism to break the $SU(3)$ gauge symmetry to $SU(2)$ and so
smoothly introduce a (single) massive vector into the theory while
{\em maintaining} the short-distance asymptotic freedom of the theory.

We now recall from Part I that the Super-Critical Pomeron phase is
characterized
by the smooth entry into the theory of a massive reggeized vector particle.
Equivalently, as the critical behavior is approached from the Super-Critical
phase a (reggeized) vector particle becomes massless and smoothly decouples
from the theory just at the critical point. The above analysis shows that
{\em just} at the flavor saturation point, a smooth parameter variation can
indeed introduce a massive vector into the theory via the Higgs mechanism,
without destroying the ultra-violet properties of the theory. This argument -
based on properties of the Super-Critical Phase - provides a remarkable
confirmation of the previous argument (which is totally independent in that it
is based on properties of the Sub-Critical Phase) that the critical theory
should be flavor-saturated $QCD$. Also, we already see some motivation for the
suggestion made in Part 1, that partially-broken $QCD$ can be identified with
the Super-Critical Pomeron phase containing a vector reggeon degenerate with
the
Pomeron. Indeed, we emphasize that flavor-saturated $QCD$ is the {\em simplest}
gauge theory\cite{gw,cel} to which an asymptotically free Higgs sector can
be added, and a massive vector introduced via the partial spontaneous
breaking of the gauge symmetry.

Consider now what the role of quark masses might be with respect to the nature
of the Pomeron. To the extent that the quark mass scale is directly
related to that of the hadron spectrum, we would expect this scale to be
relevant only in determining the energy and momentum transfer scales at which
the asymptotic behavior due to the Pomeron sets in. We would not expect the
functional form of the  asymptotic behavior to be directly affected by smooth
variation of the quark mass scale. From this point of view lowering the quark
mass scale would simply be expected to bring down the scale of ``asymptopia".
Of course, the asymptotic behavior can be singular in the massless limit and
indeed we would expect that asymptotic properties are most likely to be
directly
related to finite energy properties of the theory when all quarks are massless.
It is interesting therefore to consider specific properties of flavor-saturated
$QCD$ which hold only if the quarks are massless.

A vital property can be seen by studying the three-loop $\beta$-function. In
$QCD$ with sixteen flavors of quarks the $\beta$-function (in the momentum
subtraction scheme) is\cite{tvz}
$$
\beta(g) = {g^3\over 16\pi^2} \left[\beta_0 + \beta_1 {g^2\over 16\pi^2}
+ \beta_2 {g^4\over (16\pi^2)^2} + \cdots\right]
\auto
$$
where
$$
\beta_0 = {1\over 3}, \qquad \beta_1 = -137^{1/3} \qquad {\rm and} \qquad
\beta_
2
\left({\beta_0\over \beta_1}\right)^2 \sim 0.01.
\auto
$$
Keeping only the first two terms in (3.12), there is a zero of $\beta(g)$
at

$$
\alpha_s = {g^2\over 4\pi} = {-4\pi\beta_0\over \beta_1} = {4\pi\over 412}
\sim {1\over 33}.
\auto
$$

The smallness of the third-term in (3.12) when (3.14) is satisfied, is
emphasized by both (the last part) of (3.13) and by noting that the expansion
parameter of (3.12) is actually
$$
{g^2\over 16\pi^2} \sim {1\over 412},
\auto
$$
Consequently it is clear that the zero should survive to very high order in the
perturbation expansion and therefore should be present in any reasonable
non-perturbative definition of $\beta(g)$.  A more formal version of this
argument can be given\cite{bz} by making an ``$\epsilon$-expansion'' with
$$
\epsilon = [16{1\over2} - N_F],
\auto
$$
where $N_F$ is the number of flavors and $16{1\over2}$ is the value of $N_F$
at which asymptotic freedom is lost.  It can then be argued that a zero
of $\beta(g,\epsilon)$ is present for sufficiently small $\epsilon$ and
persists
to at least $\epsilon = {1\over 2}$.

If all the quarks are massless, then the zero of $\beta(g)$ is a genuine
infra-red fixed-point and so implies that g {\em does not evolve} in the
infra-red region.  Although we have specifically
argued for the zero in the case $N_F = 16$, the same phenomena can also
be produced\cite{cel} by adding a much smaller number of higher color
fermions. In particular, if we consider that five conventional quark flavors
have been discovered, then a $\beta$-function almost identical to (3.12) is
produced by
$$
\eqalign{
\left.\matrix{
{\rm i)}&\hbox{~~ 6 color triplet quarks${}+{}$2 color sextet quarks\hfill}
\cr
{}~~~\cr
{\rm ii)}&\hbox{~~ 5 color triplet quarks${}+{}$1 color sextet quark${}+{}$1
color octet quark\hfill}.\cr} \right\}}
\auto
$$

As we shall review in Section 14, possibility i) is actually physically
realistic if color sextet chiral symmetry breaking is the mechanism for
breaking the electroweak gauge symmetry\cite{wjm}. There may also be, as we
briefly discuss in Section 14, a deeper reason why a higher color sector is
necessary - involving the ``Strong $CP$'' problem. Therefore it will be
particularly important that ``flavor-saturated $QCD$'' could refer to option
i) above. However, for much of this article we shall, for simplicity, refer
only to the ``unrealistic'' sixteen triplet quark flavor theory as the
``saturated'' theory. If the sextet quark option is not utilized by nature,
the sixteen triplet theory may still be a good theoretical starting point from
which to understand the Pomeron in $QCD$ with only a small number of flavors.

The infra-red fixed point of massless, saturated, $QCD$ also has a potential
direct relationship to Critical Pomeron behavior as we now discuss. To see this
we first consider the infrequently discussed ``strong-coupling" limit of the
Critical Diffraction peak\cite{sw}. The scaling property I-(6.48)
of the elastic differential cross-section can be written in two closely related
forms, that is
$$
{{d\sigma}\over {dt}} \centerunder{$\large\sim$}{\raisebox{-2mm}
{$\scriptstyle s\to \infty$}}\left[ (lns)^\eta f(t(lns)^\nu /K)\right]^2
\auto
$$
$$
\eqalign{= \left[ t^\delta g (t(lns)^\nu /K)\right]^2}
\auto
$$
where $\nu \delta + \eta = 0$.  (3.19) is much less familiar than (3.18).
(3.18) is most commonly used because the finiteness of $f(0)$ gives directly
the prediction of an increasing total cross-section. However, $g(\infty)$ is
also finite, while $f(\infty)$ and $g(0)$ are not. Therefore if we wish to
study the limit $K \rightarrow 0$, for example, then (3.19) is what we should
use. This gives
$$
\eqalign{
{{d\sigma}\over {dt}} &\centerunder{$\rightarrow$}{\raisebox{-6mm}
{$K \rightarrow 0$}} t^{2\delta}\left[ g(\infty)\right]^2}
\auto
$$
Since
$$
K = (\alpha ^\prime_0)^\nu /r_0^{2\nu-2}
\auto
$$
it follows that (3.20) will, in particular, be the result of the
``strong-coupling" limit $r_0 \rightarrow \infty$.  That is in this limit the
diffraction peak shrinks to zero width and  gives a {\em vector-like}
singularity, with an anomalous dimension, at t = 0.

In general, massless (quark) $N_f$-flavor $QCD$ is expected to correspond to
a chiral limit of the massive theory in which all pseudoscalar (Goldstone
boson) masses go to zero. Since we expect the scale of the triple Pomeron
coupling $r_0$ (which has the dimension [mass]$^{-1}$) to be set by the
lightest particle in the theory we anticipate that this chiral limit will give
$r_0 \rightarrow \infty$. If the slope parameter $\alpha^\prime_0$ remains
finite (or has a much milder divergence) then we have the strong-coupling limit
discussed above. If the theory is at the Critical Point then the vector-like
singularity of the diffraction peak discussed above must be a property of the
massless chiral limit we are considering. Given the infra-red fixed point
discussed above, we anticipate that the $\beta$-function for massless $QCD$
evolves as a function of the number of flavors as illustrated in Fig.~3.1. The
vector-like singularity (with anomalous dimensions) that we are looking for
should then appear in amplitudes as {\em a natural consequence of the infra-red
scaling associated with the fixed-point}. This again would happen just at the
flavor saturation point. Once again, therefore, we are led to the conclusion
that properties of the Critical Pomeron match closely with those of
``flavor-saturated $QCD$''.

It is important to note that properties of the infra-red fixed point could be
related to properties of chiral limits in a variety of indirect ways. We noted
in Part I that Adler zeroes could be directly related to Critical Pomeron
modifications of Regge residue functions. It is clearly plausible that such
zeroes could be related to infra-red anomalous dimensions associated with the
fixed point. Note also that we expect the strong-coupling limit $r_0
\rightarrow \infty$ to result from the vanishing of any pseudoscalar mass and
so to be a property of any chiral limit in flavor-saturated $QCD$. That is to
say that if all quark masses are initially non-zero (so that the complete~
$SU(16)\otimes SU(16)$ chiral symmetry is broken) and then a chiral
$SU(2)\otimes SU(2)$ symmetry limit is taken, the strong coupling limit of
the Critical Pomeron will necessarily be involved and so can produce the
same infra-red effects as in the completely massless (saturated) theory. If
the Critical Pomeron does play the crucial role in chiral limits that we are
suggesting, then this implies that massive (as yet undiscovered) quarks are
actually important (via the mixture of infra-red and ultra-violet effects in
the form of the Critical Pomeron) in the way the chiral limit is realised
for massless pions.

As we now discuss, there is an additional theoretical significance to
saturating
$QCD$ with massless quarks which will eventually enable us to understand why,
in
this case, we may be able to construct the Pomeron from gluon reggeon diagrams.
To see this it will be useful to first recall why, in general, we can not
define $QCD$ from perturbation theory. That is we discuss the ``wild
divergence"
of the perturbation series due to vacuum fluctuations.

The existence of series of vacuum self-energy diagrams, such as that
illustrated
in Fig.~3.2, within the perturbation expansion for general amplitudes gives
rise to series having the qualitative form

$$
f(g^2)  = \sum_n a_n(g^2)^n \int_o {{dp} \over {p}} (ln\ p)^n p^m
\auto
$$

$$
\sim \sum_n a_n(g^2/m)^n n!
\auto
$$
where, as we have illustrated, the n! in (3.23) arises from integration over
the infrared region of momenta. The nature of the divergence (3.21) is
characterized by transforming to the Borel plane. That is we write
$$
f(g^2) = \int^\infty_o db\ e^{-b/g^2} \tilde{f} (b)
\auto
$$
where $\tilde{f}(b)$ will have the same perturbation expansion (in powers of b)
as (3.23) but without the n! factors. It can then be shown\cite{gth1,fd} that
in general amplitudes, series of sub-diagrams, such as those of Fig.~3.2, lead
to a sequence of ``renormalon" branch-points in the b-plane (as illustrated in
Fig.~3.3) at b = b$_1$, b = b$_2$, $\cdots$, whose discontinuities are
proportional to gluon vacuum condensates i.e.
$$\rm {residue\ at}\ b = b_1 \sim \langle 0|F^2_{\mu\nu}|0\rangle
\auto
$$

$$\rm {residue\ at}\ b = b_2 \sim \langle 0|F^4_{\mu\nu}|0\rangle
\auto
$$
etc.

Because of the presence of the renormalon singularities the integration in the
Borel representation (3.24) is not uniquely determined (the theory is not Borel
summable). Indeed condensates such as (3.25), (3.26) etc. are also not uniquely
determined\cite{fd} (since they do not characterize the breaking of any
symmetry) but rather in any particular renormalization scheme they
characterize the non-perturbative ambiguity of the theory. In particular if
we wish to study the low $p_\bot$ Pomeron in terms of conventional
perturbative gluon exchange we can expect that gluon condensates will have
to be invoked in a vital manner (as, for example, in the work of Landshoff
and Nachtmann\cite{ln}).

Consider now the effect of increasing the number of flavors of massless quarks.
As is now (we hope) very clear, we anticipate that this will be related to the
Pomeron approaching criticality. As $N_f$ increases, the renormalon
singularities move to the right in the Borel plane and expose multi-instanton
singularities as illustrated in Fig.~3.4. (The first renormalon singularity
occurs at $b = b_1 = 24\pi/(33-2N_f)$, whereas the first
instanton/anti-instanto
n
singularity occurs at $b = 4\pi$). When $N_f$ reaches the flavor-saturation
point, the first renormalon discontinuity is known to vanish\cite{fd} and the
expectation is that the renormalons disappear completely leaving only the
``topological" multi-instanton singularities. This occurs simultaneously with
the development of the infra-red fixed-point in the $\beta$-function discussed
above. For the present discussion the most important property of the
fixed-point is the implication that the infrared contributions of the massless
quarks exactly cancel those of gluons to prevent the growth of the
gauge-coupling in the infrared region.

           We conclude therefore that when $N_f$ is a maximum the $QCD$
perturbation expansion is much less divergent. Also the ``non-perturbative
vacuum problem'', characterized by the array of gluon condensates discussed
above, reduces to the simpler problem of the effects of topological fields and
the related problem of the regularization of the massless quark sea in the
presence of the anomaly. We can hope therefore that, in principle at least,
this simplification could allow us to calculate the Pomeron by summing
multi-gluon exchanges (together with the careful removal of cut-offs and
exploitation of unitarity via reggeon diagrams that we will be discussing in
the following). If this is the case, the remaining ``non-perturbative''
effects, namely instanton interactions, must (as we discuss again in Section
11)
be introduced by the process of infinite summation and removal of cut-offs.
However, we can anticipate that if we utilize a large number of
massless quarks to ``simplify'' our problem, it must be that all the dynamical
issues of confinement etc. are shifted to those problems which are enhanced by
the presence of the extra fermions. This can only be the problem of the effects
of instantons and their relation to the regularization of the quark sea.
Therefore this problem must enter our analysis at some point, at least
indirectly, if we are indeed confronting the basic vacuum properties of the
theory.

This last discussion gives us our first glimpse of what the physics of the
Critical Pomeron could be associated with. We see that the ``saturation'' of
$QCD$ with quarks leads naturally to the dominance of ``topological'' problems
connected to the anomaly and the regularization of the quark sea. In the
example of the Schwinger model, which we discuss at length in a Section 11,
it is well known\cite{man} that this problem leads to the development of a
"winding-number" vacuum condensate. Perhaps then the Super-Critical Pomeron
condensate that is produced as we go through the critical point (by adding
more and more quarks) can be seen as a ``topological condensate'' resulting
from
the dominance of the divergence of the quark sea. After much technical
development and analysis of reggeon diagrams in the following Sections we shall
eventually be able to realise this idea explicitly.

\newpage

\mainhead{4.  ELASTIC UNITARITY AND THE REGGEIZATION OF NON-ABELIAN THEORIES}

        In Section I-5.1 we discussed the general relationship between
two-particle unitarity (in the t-channel) and Regge pole trajectories. In this
Section we begin our description of the Regge properties of non-abelian gauge
theories by briefly reviewing the argument\cite{gs} that in a general
{\it renormalizable}, spontaneously-broken, {\it non-abelian} gauge theory (in
which {\it all} gauge bosons acquire a mass), the two-particle unitarity
approximation shows that {\it all} vector bosons and fermions (corresponding to
the elementary fields of the theory) lie on Regge trajectories. However, this
analysis is not sufficient to demonstrate that there are {\it no additional
trajectories} in the neighborhood of J = 1 or J = 1/2. This requires the
high-order leading-log calculations which we describe in later Sections.
Indeed, the weakness of the two-particle unitarity analysis is that it is
vulnerable, in higher orders, to an {\it overestimate} of the number of Regge
poles that are present - in particular by suggesting the misidentification of
Regge cut terms as Regge poles.

        As we noted in the Summary, it will be an important part of our
argument in subsequent sections that the {\em only} Regge poles in the theory
are those of the elementary fields, and that multiparticle t-channel unitarity
is {\it completely} accounted for by the Regge cuts generated by such Regge
poles. We emphasize that the reggeization of an elementary particle of the
theory is the only known mechanism within field theory for generating a Regge
pole trajectory which is isolated in the manner that we would like the Pomeron
to be isolated in $QCD$. Bound state Regge pole trajectories are typically
generated in infinite sets with accumulations points of trajectories etc..

        The essential ingredients of reggeization are the existence of
``nonsense'' states - as defined in I - due to the presence of vector particles
and the dispersive N/D solution of elastic unitarity. If we consider
vector-vector scattering in the t-channel then at J = 1 we can (by choosing the
external helicities to sum to the sense values of n = -1, 0, 1, or the nonsense
values of n = -2, 2) identify physical sense/sense amplitudes -$A_{ss}$,
and by Froissart-Gribov (FG) analytic continuation, the unphysical
sense/nonsense amplitudes -$A_{sn}$, and nonsense/nonsense amplitudes
-$A_{nn}$.  Because of the nonsense square-root branch-points discussed in
I-4.5, these amplitudes have the following form near J = 1
$$\eqalignno{
A_{ss} &= V_{ss} (t) \delta_{J,1} + \cdots &\num\label {4.1}\cr
A_{sn} &= V_{sn} (t)/(J-1)^{1\over 2} + \cdots &\num\label {4/2}\cr
A_{nn} &= V_{nn} (t)/(J-1) + \cdots &\num\label {4.3}\cr}
$$
In lowest-order perturbation theory
$$
V_{ss} (t) = g^2/(t-M^2)
\auto
$$
where $M$ is the vector-meson mass (which for simplicity we temporarily take
to be common for all vectors).

        The J = 1 singularity of $A_{nn}$ acts as a ``driving" potential in the
analytically-continued elastic unitarity equation I-(5.19) and taking the
simplest N/D solution gives for the analytically continued $A_{ss}$
$$
A_{ss} (t,J) = -V_{sn} K\left[ J-1-V_{nn}K \right]^{-1}V_{ns}
\auto
$$
where
$$
K(t) = {1\over \pi} \int {{dt^\prime \rho(t^\prime)} \over {(t^\prime-t)}}
\auto
$$
and $\rho(t)$ is a phase-space factor. (4.5) generates a Regge pole
trajectory
$$
\alpha(t) = 1 + V_{nn}(t)K(t)
\auto
$$
and is consistent with (4.1) if
$$
V_{ss} = V_{sn} \left[ V_{nn}\right]^{-1}V_{ns}
\auto
$$
which is the well-known ``factorization'' condition for reggeization. In
particular (4.4) requires that $V_{nn}$ has a zero at the vector particle
pole, or (if there are distinct masses $M_v$)
$$
V^{-1}_{nn} = \sum_v R_v/(t-M^2_v)
\auto
$$
$V_{nn}$ is calculated from the lowest order Feynman diagrams,
and after some lengthy calculation it can be shown\cite{gs} that
for vectors a,b,c,d in a theory with (broken) gauge symmetry group G
$$
(V_{nn})_{ab,cd} = \sum_e \left[ C_{abe}C_{cde}\right] (t-M^2_e)
\auto
$$
where the $C_{abe}$ are structure constants of the group G. This is sufficient
to give the factorization condition (4.8) and (4.9) in the vector adjoint
representation of G {\it provided} that G is semi-simple. That is G should have
no abelian subgroups. Therefore the (massive) gauge bosons of a general
non-abelian group do indeed reggeize. A very similar argument can be given to
show that the elementary fermions of the theory also reggeize.

      The strength of the above argument is that it is straightforward to carry
through for a general gauge group and also for fermions in general
representations. It's weakness is that it anticipates particular N/D Regge
pole contributions to amplitudes and if the process is extended to
non-leading order or to lower values of J then ambiguities arise (CDD
ambiguities in particular) which can lead to conflicting results or, (as we
discuss in Section~6) the misidentification of Regge cut contributions as
additional Regge poles. However, all existing explicit calculations of
high-energy logarithms are consistent with the general reggeization of
elementary gauge bosons and fermions described above.

\newpage

\mainhead{5.  TRANSVERSE MOMENTUM DIAGRAMS IN GAUGE THEORIES}

In this Section we begin our summary of what we shall need to know about the
perturbative high-energy behavior of gauge theories. Our basic purpose, in this
and the next Section, will be to demonstrate that the great complexity of the
Feynman perturbation expansion in (spontaneously broken) gauge theories
reduces,
in the Regge limit, to the simplest structures expected from the general Regge
theory we have described in Part I. The contributions of internal fermion loops
will be of special importance in our analysis but we will postpone their
discussion until later Sections.

We begin by giving a brief overview of the reduction of Feynman diagrams to
transverse momentum diagrams in the Regge limit. In the next Section we shall
describe the organisation of transverse momentum diagrams into reggeon
diagrams.

\subhead{5.1 Transverse Momentum Diagrams}

Consider first the box-diagram\cite{mw} illustrated in Fig.~5.1. We shall
initially neglect numerators so that we have, in the notation illustrated,
$$
\eqalign{
I(s,t,m^2) = &\int d^4k \left[k^2-m^2+i\epsilon\right]^{-1}
\left[\left(p_1-{q\over 2}-k\right)^2-m^2+i\epsilon\right]^{-1}\cr
&\times \left[(q+k)^2-m^2+i\epsilon\right]^{-1}\left[\left(
p_2+{q\over 2}+k\right)^2-m^2+i\epsilon\right]^{-1}.}
\auto
$$
If we write
$$
\eqalignno{
p_1&=\left(\sqrt{s/2}+O\left({1\over \sqrt s}\right),\sqrt {s\over 2},
\,\til{0}\right)&\num\label{5.2}\cr
p_2&=\left(\sqrt{s/2}+O\left({1\over \sqrt s}\right),-\sqrt {s\over 2},
\,\til{0}\right)&\num\label{5.3}\cr
q&=\left(O\left({1\over \sqrt s}\right),O\left({1\over \sqrt s}
\right),\underline{q}\right)\qquad q^2=t&\num\label{5.4}\cr }
$$
we obtain
$$
\eqalign{
I\centerunder{$\large\sim$} {\raisebox{-2mm} {$\scriptstyle s\to \infty$}}\;
{1\over 2}\int
&d^2\underline{k}_{\,\perp} dk_+dk_- \left[ k_+k_--k^2_{\,\perp} + m^2+
i\epsilon\right]^{-1} \bigg[ \left( k_+-\sqrt{2s}\,\right) k_-\cr
&-\left( \underline{q}\,_{\,\perp}/2+\underline{k}\,_{\,\perp}\right)^2
-m^2+i\epsilon\bigg]^{-1} \left[ k_+k_-- \left(
\underline{q}\,_{\,\perp}+\underline{k}\,_{\,\perp}\right)^2-m^2+i\epsilon\right
]^{-1}\cr
&\times \left[k_+\left(\sqrt {2s} +
k_-\right)-\left(\underline{q}\,_{\,\perp}/2
+\underline{k}\,_{\,\perp}\right)^2-m^2+i\epsilon\right]^{-1}.}|
\auto
$$
To obtain a non-zero answer by closing the $k_+$ contour in (5.5) we must
have
$$ -\sqrt{2s}<k_-<0.
\auto
$$
In this case closing in the lower-half $k_+$-plane, picking up the pole
in the last square-bracket, and making the approximation
$|k_-|\lsim\lambda\sqrt s,\ \lambda\ll 1$ gives
$$
\eqalign{
-\pi i\int d^2\underline{k}\,_{\,\perp} &\int^0_{-\lambda\sqrt s} \,dk_-\left(
\sqrt{2s}\,\right)^{-1}\left[\underline{k}^2_{\,\perp} + m^2\right]^{-1}
\left[\left(\underline{q}\,_{\,\perp} +
\underline{k}\,_{\,\perp}\right)^2+m^2\right]^{-1}\cr
&\times
\left[-\sqrt{2s}k_--\left(\underline{q}\,_{\,\perp}+\underline{k}\,_{\,\perp}
\right)^2
-m^2+i\epsilon\right]^{-1}.}
\auto
$$
This approximation is valid for obtaining both the {\em leading real logarithm}
which now comes from the $k_-$ integration close to the pole(in the last
bracket) and the {\em leading imaginary part} which comes directly from the
same pole. That is
$$
\eqalignno{
{1\over \sqrt {2s}}\int^0_{-\lambda\sqrt s}&\,dk_-\left[-\sqrt{2s}k_--
\left(\underline{q}\,_{\,\perp}/2+\underline{k}\,_{\,\perp}\right)^2-m^2+i
\epsilon\right]
^{-1}
&\num\label{4.8}\cr
&\centerunder{$\large\sim$}{\raisebox{-2mm} {$\scriptstyle s\to \infty$}}\;
{1\over 2s}\left(\ln s-i\pi\right),
&\num\label{4.9}\cr}
$$
where the non-leading contribution to the real part is $O\left(1/s\right)$ and
to the imaginary part is $O\left(1/sln~s\right)$.
We therefore obtain
$$
I\parens{s,t,m^2}\centerunder{$\large\sim$}{\raisebox{-2mm} {$\scriptstyle s\to
\infty$}}\;-{\pi i\over 2s}\left(\ln s-\pi i\right)K(t),
\auto
$$
where $K(t)$ is the {\em transverse momentum integral}
$$
K(t)=\int d^2k_{\,\perp} \left[\underline{k}^2_{\,\perp}
+m^2\right]^{-1}
\left[\left(\underline{q}\,_{\,\perp} +
\underline{k}\,_{\,\perp}\right)+m^2\right]^{-1}.
\auto
$$

Note that this integral can be obtained directly from (5.5) by taking
$$
k_+ \sim k_- \sim {1\over \sqrt s}\Rightarrow k_+k_-\to 0,
\auto
$$
and performing the $k_+$ and $k_-$ integrations by putting two propagators
on-shell. Indeed this is how the transverse  momentum diagram emerges if
we use a dispersion relation to calculate the high-energy behavior of the
box-diagram. That is we first calculate the discontinuity associated with
the cut shown in Fig.~5.2 using elastic unitarity. This will give the
``$\pi^2$'' term in (5.10) with $K(t)$ obtained directly due to the cut
propagators being on-shell. The $\ln s$ term in (5.10) is then
reconstructed from the dispersion relation. In fact the use of $s$-channel
unitarity to compute high-energy behavior is a powerful technique exploited
extensively by Lipatov \etal\cite{klf,lnl} and Bartels\cite{jb} which we shall
briefly elaborate on later in this Section and in the next Section. For the
moment we note only that it provides an underlying explanation for why the
transverse momentum integrals obtained in the Regge limit can always be
obtained from the underlying Feynman diagram(s) by putting appropriate lines
on-shell.

In general we might expect the justification for using $k_+$ and $k_-$
integrations to produce logarithms---with the coefficients as transverse
momentum integrals---to be that the integrals involved are uniformly convergent
in the high-energy limit. For scalar field theories there are no numerators
to add to (5.1) and indeed for such theories the transverse momentum
integrals obtained are always sufficiently convergent. For the vector (gauge)
theories that we shall be interested in this is not the case for individual
Feynman diagrams. In fact in such theories gauge invariance produces a vast
complexity of cancellations of divergent transverse momentum integrals.
As we elaborate below the result is nevertheless that, after such
cancellations, the only logarithms that survive are those produced by
longitudinal $k_+$ and $k_-$ integrations ``close to" on mass-shell regions.
In addition, the coefficients can always be written as the corresponding
transverse momentum integrals.

The reduction to transverse integrals appears to be a deep consequence of
renormalizability or equivalently gauge-invariance. Alternatively, if it did
not
hold it is very doubtful that the theory would be describable in terms of the
Regge theory we have developed in Part I. As we described in I we believe Regge
behavior to be intrinsic to obtaining a unitary high-energy $S$-Matrix and so
if a theory did not reduce (at high-energy) to the transverse momentum
integrals
which naturally generate Regge poles we would be very doubtful of its
self-consistency. However, as is well known from the case of massive
$QED$\cite{fgl} the emergence of transverse momentum diagrams is not in
itself a guarantee of Regge behavior and high-energy unitarity.

To illustrate both the general spin structure and the general form of large
transverse momentum cancellations - and also to make an additional point, we
consider some low-order diagrams in an abelian gauge theory.

\subhead{5.2 Low-Order Diagrams in QED}

We consider specifically massive $QED$, that is both the photon and the
electron
are massive. Consider first the coupling $G_1$ for a single fast electron to
couple to a single photon as illustrated in Fig.~5.3. A fast electron
propagator
gives
$$
\eqalignno{
{{\gamma\cdot p+m}\over{p^2-m^2}}
\quad&\centerunder{$\large\sim$}{$\raisebox{-4mm}
p_+\rightarrow\infty$}\quad
{{\gamma_-p_+ + \st{p}_\perp +\cdots} \over {p^2-m^2}}&\num\label{5.13}\cr
&\equiv {{\gamma_-+\gamma_\perp\cdot(p_\perp/p_+)+ 0(1/p^2_+)}
\over {\left[ p_- - {{p^2_\perp - m^2} \over {p_+}}\right]}}
&\num\label{5.14}\cr
}
$$
For an electron initially and finally on-shell, we simply remove the
$(p^{2} - m^{2})^{-1}$ factor from (5.13) and so in lowest-order perturbation
theory we obtain, before inserting external spinor factors,
$$
\eqalignno{
G_{1\mu}&\sim ep^2_+\gamma_-\gamma_\mu\gamma_- \sim \gamma_-p^2_+
\hspace{0.5in} {\rm if}\
\gamma_\mu = \gamma_+&\num\label{5.15} \cr
&= 0 (|p_\perp|~p_+) \hspace{0.5in} {\rm otherwise}&\num\label{5.16}\cr }
$$
Therefore the leading power behavior (as $p_+\to\infty$) is obtained if the
spin
of the scattering electron is conserved, that is there is helicity
conservation.
Inserting external spinors allows the replacement of one $\gamma_-p_+$ factor
in (5.15) by m, giving finally
$$
G_{1\mu}\sim emp_+\delta_{\mu -}
\auto
$$

To understand the spin structure of transverse-momentum diagrams that we shall
exploit it will be sufficient to consider photon exchange via a simple Feynman
propagator. Note first that
$$
{{g_{\mu \nu}} \over {q^2+M^2}} \sim {{g_{\mu \nu}} \over {q^2_\perp + M^2}}
\hspace{0.5in} q_+q_- \sim 0
\auto
$$
In the limit given by $s\to\infty$ in (5.2)-(5.4), that is
$$
p_{1+},\ p_{2-} \ \longrightarrow\ \infty\hspace{0.5in}
 p_{1-},\ p_{2+} \longrightarrow 0
\auto
$$
single photon exchange gives, as illustrated in Fig.~5.3, the
(helicity-conserving) amplitude for electron scattering
$$
\eqalignno{
e^2m^2 &p_{1+}\delta_{+\mu}
\left[ {{g_{}\mu\nu} \over {q^2_\perp + M^2}}\right] \delta_{\nu -}p_{2-}
&\num\label{5.20}\cr
&\sim {{e^2m^2S} \over {q^2_\perp + M^2}}&\num\label{5.21}\cr}
$$

The above discussion extends straightforwardly to the exchange of $N$ photons
between fast electrons. We can suppose that each intermediate state propagator
is placed on-shell by $k_-$ and $k_+$ integrations. The denominator is thus
removed from (5.14), giving in analogy with (5.15)
$$
\eqalignno{
G_{N\mu_1\cdots \mu_N} &\sim e^N\gamma_-\gamma_{\mu_1}\gamma_-
\cdots \gamma_-\gamma_{\mu_N}\gamma_- &\num\label{5.22}\cr
&\sim e^N\gamma_-p_+\ \hspace{0.5in}{\rm if}\ \mu_1 = \mu_2 = \cdots = \mu_N =
+
&\num\label{5.23}\cr
&= 0\hspace{0.5in} {\rm otherwise}&\num\label{5.24}\cr}
$$
So again the electron spin structure is preserved, and using the propagator
(5.18) the Feynman diagram of Fig.~5.4 will generate the transverse-momentum
diagram of Fig.~5.5, that is the helicity conserving amplitude is
$$
\eqalign{
\Gamma_N = e^{2N}S &\int d^2\underline{k}_1\cdots d^2\underline{k}_N\delta^2
\left( \underline{q}-\underline{k}_1-\underline{k}_2
\cdots -\underline{k}_N\right)\cr
&\times {{1} \over {\underline{k}^2_1 + M^2}}\cdots
{{1} \over {\underline{k}^2_N + M^2}}\cr
&\times [{\rm possible\ logarithms\ from}\ k_+,k_- \
{\rm integrations}]\cr}
\auto
$$

Clearly the same amplitude is given by diagrams in which the exchanged photons
are emitted and absorbed by the electron lines in all possible orders. In $QED$
there are no group factors involved and the symmetry (or antisymmetry) of the
diagrams with respect to rotation of one electron line to be a positron line
determines their signature. That is transverse momentum diagrams with an even
number of photon lines appear in the even signature amplitude, while those with
an odd number appear in the odd-signature amplitude. As a result, {\em all}
the logarithms generated by $k_+$ or $k_-$ integrations cancel among the
diagrams and (5.25) holds without any logarithms in the bracket. As we shall
discuss further shortly, this is particular to $QED$ and is effectively the
reason why the photon does not reggeize.

Moving on to the subject of large transverse-momentum cancellations, we
consider
now the simplest sixth-order ladder diagram illustrated in Fig.~5.6. The
photons
will couple to the external electrons as in (5.20) and so as the diagram is
reduced to a transverse momentum integral the product of exchanged electron
numerators gives\cite{mw}
$$
[-\underline{\st{k}}_2 -\underline{q} + m] \gamma_-[\st{k}_1 + \st{k}_2 +
\st{q} + m] \gamma_+ [-\underline{\st{k}}_1 - \underline{\st{q}} +
m]
\auto
$$
where the central bracket will reduce to a transverse-momentum factor after the
$\gamma_-$ is commuted through to the $\gamma_+$. Consequently the resulting
high-energy behavior is
$$
\eqalign{
&-g^6{1\over 2}\left[ \ln^2 s-2\pi i\ln s\right]\int {d^2\underline{k}\,_1\over
(2\pi)^3} {d^2\underline{k}\,_2\over
(2\pi)^3}\left[\underline{k}^2_1+M^2\right]^{-1}\left[\underline{q}+
\underline{k}\,_1+m^2\right]^{-1}\cr
&\times \left[\parens{\underline{q}+\underline{k}\,_2}^2+m^2\right]^{-1}
\left[\underline{k}^2_2+m^2\right]^{-1}\left[-\underline{\st k}_2-\underline
{\st q}+m\right] \left[\underline{\st k}_1+\underline{\st k}_2+\st q+m\right]
\left[-\underline{\st k}_1-\underline{\st q}+m\right]\cr}
\auto
$$
where $M$ is the photon mass, $m$ the electron mass and, as usual, / indicates
multiplication by (two-dimensional) Dirac matrices.

The presence of the electron numerators in (5.27) gives a divergent
integral---both the $\underline{k}\,_1$ and $\underline{k}\,_2$ integrations
diverge. This implies that a more careful analysis of the original Feynman
diagram would show that there are additional factors of $\ln s$ that should
be associated with the diagram. However, the divergences are {\em cancelled
by other diagrams} that we will describe shortly. First we want to show
that the (ladder diagram) transverse momentum diagram (5.27) actually
has a {\em well-defined finite part} (which is ultimately the correct answer).

The numerator in (5.27) can be expanded as follows
$$
\eqalign{
\left[-\kbarsl\,_2-\qbarsl+m\right]&\left[\kbarsl\,_1+\kbarsl\,_2+m\right]
\left[-\kbarsl\,_1-\qbarsl+m\right]\cr
={}&-\left[\kbarsl\,_2+\qbarsl-m\right]\left[\qbarsl+m\right]\left[\kbarsl\,_1
+ \qbarsl-m\right]\cr
&+\left[\left(\kbar\,_2+\qbar\right)^2+m^2\right]
\left[-\kbarsl\,_1-\qbarsl+m\right]+
\left[(\kbar\,_1+\qbar)^2+m^2\right]\left[-\kbarsl\,_2-\qbarsl+m\right].\cr}
\auto
$$
Each of the latter two terms vanishes when one or other electron line goes
on-shell. Hence at the electron-photon threshold the last two terms can
be dropped. This gives a convergent transverse momentum integral which in
the even signature amplitude contributes
$$
g^2\left[\ln^2s+(-\ln s)^2\right]{\left[\alpha(\st q)-1\right]^2\over \st q
+m},
\auto
$$
where $\alpha(\st q)$ is the electron trajectory function
$$
\alpha(\st q)= 1 + {g^2\over 8\pi^3}(\st q+m)\int {d^2\kbar~~(\qbarsl -m)
\over \left[(\qbar+\kbar)^2+m^2\right]\left[\kbar^2+M^2\right]}.
\auto
$$
That is (5.29) gives just what is required for the sixth-order
contribution to the reggeization of the electron. That the t-channel threshold
behavior of ladder diagrams gives the full sixth-order Regge behavior is a
consequence of a general feature of renormalizable field theories containing
vectors which we shall enlarge on in the discussion of dispersion relation
methods in sub-Section 5.4. [It is also directly related to the
success of the two-particle t-channel unitarity formalism discussed in the last
Section in demonstrating the reggeization of the electron.]

Note that the transverse momentum integral (5.27) is (like that of (5.11))
obtained from the original ladder diagram of Fig.~5.6 by placing
the three vertical lines on-shell. (The $\ln s$ factors are generated ``close
to'' this on-mass-shell configuration.) Since the relevant part of this
transverse momentum integral is the $t$-channel threshold region where
{\em all} of the horizontal lines of the ladder graph are then close to
on-shell, this provides our first explicit illustration of a general feature
we have already emphasized. That is the Regge pole behavior of a vector
gauge theory originates, in first approximation, from the region of planar
Feynman diagrams in which all internal lines are close to the mass-shell.

The additional diagrams needed to cancel the divergences of (5.27)
are those shown in Fig.~5.7. Remarkably, perhaps, the first diagram gives
a transverse momentum integral which exactly cancels the contribution to
(5.27) of the second term in (5.28) while the second diagram
cancels the contribution of the third term. We can see that this is an
elementary example of a (generalized) Ward identity cancellation by noting
that the divergence of the $\kbar\,_2$ integration is cancelled by adding
the sub-diagrams illustrated in Fig.~5.8. This amounts to attaching the
photon line involved at all possible points around the electron (plus photon)
loop. The resulting softened large transverse momentum behavior is a
consequence of the Ward identity involved. The corresponding Ward identities
for a general spontaneously broken non-abelian gauge theory are clearly much
more complicated. Nevertheless the full set of needed identities has been
derived by Sen\cite{as} in his all logarithms formalism which we referred to in
the Introduction.

For several years the large transverse momentum cancellation was effectively
assumed to take place by first calculating with a transverse momentum cut-off.
It was then proved (or in high-order cases assumed) that within the leading
ln~s approximation all logarithms of the cut-off cancel in the sum over all
diagrams. In this way McCoy and Wu were able\cite{mw} to calculate, in massive
QED, leading and next-to-leading logs explicitly up to twelfth order, and to
generalize the results to arbitrary order. The leading logs simply describe
the reggeization of the electron, and to all orders continue to be attributable
to the ``on mass-shell" regions of ladder diagrams. The intricacy of the
``off-shell" large transverse momentum cancellations can be appreciated by
noting that at twelfth-order 142 diagrams are involved\cite{mw}.

Using the same ``transverse-momentum cut-off" technique, Cheng and Lo\cite{cl}
were able to calculate up to tenth order in SU(2) Yang-Mills theory. The
leading log result is now the reggeization of the gauge bosons via
``on-shell" ladder diagram contributions, confirming directly the elastic
unitarity analysis described in the last Section. Since we shall explicitly
utilize the Yang-Mills results in the discussion of Reggeon diagrams in the
next Section we reproduce these results in some detail.

\subhead{5.3 Transverse Momentum Diagrams for SU(2) Gauge Theory}

The most immediate consequence of a non-abelian symmetry is that, because of
group factors, the diagrams involving the exchange of gluons no longer
symmetrise and antisymmetrise according to the number of gluons as in $QED$.
This implies that the factors of $\ln s$ allowed for in (5.25) do not
cancel. This, combined with the existence of gluon interaction vertices,
results in many different diagrams contributing to leading-order
calculations, with (at first sight) no distinctive topology involved.

For an $SU(2)$ Yang-Mills theory with the gluons given a mass by the Higgs
mechanism (involving one {\em fundamental representation} $SU(2)$ doublet of
scalar fields), Cheng and Lo have derived leading and non-leading log results
up to tenth order\cite{cl}. There is an $SU(2)$ global symmetry of the
theory and the results depend on the $t$-channel ``isospin'' of amplitudes
under this symmetry. We can summarize the Cheng and Lo results for (massive)
gluon-gluon scattering as follows
$$
\eqalign{
T_0\sim &{}isg^4\bigg\{J_1(q)+\left[2J_2(q)-2\left(q^2+{5\over 4}M^2\right)
J_1(q^2)\right]g^2\ln s\cr
&+\bigg[J_3(q) + \skew6\tilde J(q) - 4\left(q^2+{5\over
4}M^2\right)J_1(q)J_2(q)\cr
&+2\left(q^2+{5\over 4}M^2\right)^2 J^3_1(q)\bigg]\left(g^2\ln s\right)^2
 +\bigg[{2\over 3}J_4(q)-{1\over 3}J_A(q)\cr
&-J_B(q)+{4\over 3}J_C(q) + {2\over 3}J_D(q) - {4\over 3}\left(q^2+{5\over
4}M^2\right)\Big(J^2_2(q)+\cr
&J_1(q)J_3(q) + J_1(q)\skew6\tilde J(q)\Big) + 4\left(q^2 + {5\over
4}M^2\right)^2J^2_1(q)J_2(q)\cr
&-{4\over 3}\left(q^2+{5\over 4}M^2\right)^3J^4_1(q)\bigg](g^2\ln
s)^3+\cdots \bigg\}}
\auto
$$
$$
\eqalign{
T_1\sim{} &{sg^2\over q^2+M^2} \left\{1-\left(q^2+M^2\right)
J_1(q)g^2\ln s + {1\over 2}\left[\left(q^2+M^2\right)J_1(q)g^2
\ln s\right]^2\right.\cr
&\left.-{1\over 6}\left[ \left( g^2+M^2\right) J_1(q)g^2\ln s\right]^3+
{1\over 24}\left[ \left( q^2+M^2\right) J_1(q)g^2\ln s\right]^4+\cdots\right\}}
\auto
$$
and
$$
\eqalign{
T_2\sim {}&isg^4\left\{ J_1(q) + \left[ -4J_2(q) + \left( q^2+2M^2\right)
J^2_1(q)\right]g^2\ln s\right.\cr
&+\bigg[ 4J_3(q) + 4\skew6\tilde J(q) - 4\left(
q^2+2M^2\right)J_1(q)J_2(q)\cr
&+{1\over 2}\left( q^2+2M^2\right)^2J^3_1(q)\bigg]\left(g^2\ln s\right)^2+
\left[ -{4\over 3}J_4(q)-{4\over 3}J_4(q)-{4\over 3}J_A(q)\right.\cr
&-4J_B(q)-{8\over 3}J_C(q)-{4\over 3}J_D(q) - {8\over 3}\left( q^2+2M^2\right)
\left(J^2_2(q)\right.\cr
&\left.+J_1(q)J_3(q)+J_1(q)\skew6\tilde J(q)\right)-2
\left( q^2+2M^2\right)^2J^2_1(q)J_2(q)\cr
&\left.\left.+{1\over 6}\left( q^2+2M^2\right)^3J^4_1(q)\right]\left( g^2\ln
s\right)^3+\cdots\right\}.}
\auto
$$

$J_1(q)$, $J_2(q)$, $J_3(q)$, $\skew6\tilde J(q)$, $J_A(q)$, $J_B(q)$, $J_C(q)$
and
$J_D(q)$, are all transverse momentum integrals with the diagrammatic
representation shown in Fig.~4.9 and having the following explicit form
$$
\eqalignno{
J_1(q)&=\int {d^2k\over (2\pi)^3}\left[k^2+M^2\right]^{-1}
\left[(q-k)^2+M^2\right]^{-1}\equiv {1\over (2\pi)^3}K(q^2)&\num\label{5.34}\cr
J_n(q)&=\int {d^2k\over (2\pi)^3}{J_{n-1}(k)\over
{\left[(q-k)^2+M^2\right]}}
\qquad {n=2,3}&\num\label{5.35}\cr
\skew6\tilde J(q)&=\int {d^2k\over (2\pi)^3}{(k^2+q^2)J^2_1(k)\over
\left[(q-k)^2+M^2\right]}&\num\label{5.36}\cr
J_A(q)&=\int {d^2k\over (2\pi)^3}{(k^2+M^2)^2J^3_1(k)\over (q-k)^2+M^2}
&\num\label{5.37}\cr
J_B(q)&=\int {d^2k\over
(2\pi)^3}(k^2+M^2)^2J^2_1(k)J_1(q-k)&\num\label{5.38}\cr
J_C(q)&=\int {d^2k\over (2\pi)^3}{(k^2+M^2)^2J_1(k)J_2(k)\over (q-k)^2+M^2}
&\num\label{5.39}\cr
J_D(q)&=\int {d^2k_1d^2k_2\left(k^2_1+M^2\right)J_1(k_1)
\left(k^2_2+M^2\right)J_1(k_2)\over
(2\pi)^6\left[\left(q-k_1\right)^2+M^2\right]\left[\left(q-k_2\right)^2+M^2
\right]\left[\left(q-k_1-k_2\right)^2+M^2\right]}.&\num\label{5.40}\cr}
$$

The complexity of these results is a testimony to the enormous effort
involved in performing the underlying Feynman diagram calculations. (Several
hundred diagrams contribute at tenth order.) The most striking feature is
the simple form of the $T_1$ amplitude which, as we noted above, contains only
the reggeization of the gluon. That is (5.32) is clearly an expansion of the
Regge pole amplitude
$$
T_1= {{g^2s^{\alpha(t)}} \over {q^2+M^2}},\hspace{0.5in}
 \alpha(t) = 1 + g^2 J_1(q)
\auto
$$
The first $\ln s$ term in the expansion comes directly from box diagrams as in
(5.10). The crossed and uncrossed diagrams do not cancel in the odd-signature
amplitude, as in $QED$, but instead give just the right factor for (5.41). This
is what the t-channel unitarity analysis of the last Section anticipates. That
analysis also anticipates that {\em only two-particle t-channel states are
important in the leading-log result to all orders}. This is indeed the case,
and
can be demonstrated by showing, in analogy with our discussion of the
reggeization of the electron in massive $QED$, that (5.32) is correctly
reproduced by keeping only the leading {\em two-particle threshold} ``finite
part'' of ladder diagram transverse momentum integrals. That is {\em the
leading-log reggeization of the gluon comes directly from the (close to) on
mass-shell regions of ladder diagrams, with the remaining hundreds of diagrams
involved simply producing (at the leading-log level) an extremely elaborate set
of cancellations}. We should note that the absence at leading-log level of
higher-order ladder diagram contributions involving more than two states in the
t-channel (where some lines are necessarily off-shell) is, in part, due to a
cancellation that does not involve divergences. This is an important feature
which is not adequately illustrated by our discussion of $QED$ diagrams. It
ultimately includes contributions from general planar diagrams and will be part
of the ``Multi-Regge Bootstrap'' described in the next Section. We shall give
a more detailed discussion of this cancellation in the next sub-Section.

We might have expected general planar diagrams with all lines close to
mass-shell to also contribute to the leading-log result. This would presumably
be the case if we were considering the generation of bound-state Regge poles.
However, as we emphasized in the last Section, a complicated (infinite)
spectrum
of trajectories would then also be anticipated. In the special case of the
reggeization of an elementary gluon that we are discussing the simplification
that only two-particle states in the t-channel are involved at the leading-log
level is very important and is what allows the complete set of all
(leading-power) logarithms to be described by the exchange of an isolated Regge
pole and its associated Regge cuts (as we describe in the following). We have
emphasized the importance of this for ultimately arriving at an isolated
Pomeron
pole describing hadron interactions.

The non-leading log results for $T_0$ and $T_2$ are clearly more complicated,
even though they also involve very complex cancellations.  In the next Section
we shall address the problem of reducing the results for $T_0$ and $T_2$ to a
simplicity corresponding to that of $T_1$. This is, in fact, what will be
achieved by the introduction of {\em reggeon diagrams}. In the rest of this
Section we shall describe an alternative formalism and further results which
both strengthen and extend the above results and also provide insight into
their origin.

We consider now the dispersion relation, s-channel unitarity, calculational
method utilized by Lipatov and collaborators\cite{klf,lnl} and extensively
developed by Bartels\cite{jb}. This method is particularly powerful when
applied to vector theories because vector exchange is, as is very well
known, very close to violating unitarity bounds. As a result unitarity (and
renormalizability) strongly constrain the general form of amplitudes and can
be exploited very effectively, at high energy, in reducing the complications
of the perturbation expansion for non-abelian gauge theories.

\subhead{5.4 Renormalizability and Dispersive Calculations of Vector Gauge
Theories}

Consider first the calculation of the high-energy Regge behavior of the
{\em set of tree diagrams} for gluon-gluon scattering in a non-abelian theory.
It is well-known, of course, that $t$-channel exchange of a vector gives
an amplitude proportional to $s$. However, there are actually several diagrams
giving comparable behavior. They are illustrated in Fig.~5.10. Each diagram
gives a result of the form
$$
A(s,t)\centerunder{$\large\sim$}{\raisebox{-2mm} {$\scriptstyle s\to
\infty$}}\;s\,f(t).
\auto
$$
Now renormalizability (or equivalently the unitarity bound) implies that
the {\em sum of all diagrams} grows no faster than $|t|^{1/2}$ in the limit
$s\sim t\to \infty$. Therefore since we are considering only a finite number
of diagrams, there can be no non-uniformities in the limits involved, and
for the sum of all diagrams in Fig.~5.10 we must have
$$
f(t)\centerunder{$\large\sim$}{\raisebox{-2mm} {$\scriptstyle t\to \infty$}}\;
|t|^{-1/2}.
\auto
$$
This implies that $f(t)$ satisfies an unsubtracted dispersion relation
$$
f(t)={1\over 2\pi}\int {dt'\over (t'-t)}\,{\rm Im}\,f(t').
\auto
$$
But the only diagram in Fig.~5.10 with a singularity in the $t$-channel is
the pole term. Therefore we can construct $f(t)$ from the dispersion relation
and use the single diagram only. We can then infer that the
{\em sum of diagrams} gives a non-zero result {\em only if} the t-channel
quantum numbers allow exchange of a vector meson, and that in this case
$$
M_{ab\rightarrow cd} (s,t) \sim \gamma_{ac}\gamma_{bd}\ {{s} \over {(t-M^2)}}
\auto
$$
where the residues $\gamma_{ac}$ and $\gamma_{bd}$ are the ``on-shell"
couplings evaluated at $t = M^{2}$. The complicated problem of evaluating and
summing the diagrams in Fig.~5.10 is therefore completely by-passed by a simple
exploitation of unitarity and analyticity!

The real power of the above argument is apparent when it is applied in the
multi-Regge limit, as we briefly illustrate in the following. (A final
description of the results has to await the multi-Regge Bootstrap described in
the next Section.) Using the kinematics illustrated in Fig.~5.11, the
multi-Regge limit can be specified by writing
$$
q_i = \beta_i P_b - \alpha_i P _a + q_{i\perp}
\auto
$$
fixing $q_i$, i = 1,...,n+1, and requiring
$$
\eqalign{
1 >> \alpha_1 >> \alpha_2 \cdots >> \alpha_{n+1} \sim M^2/s\cr
1 >> \beta_{n+1} >> \beta_n >> \cdots \beta_1 \sim M^2/s}
\auto
$$
The diagram cancellation argument above can be generalized to show that
$$
M_{ab\rightarrow cd+n} \sim s\gamma_{ab} {{1} \over {(q_1^2+ M^2)}}
\gamma_1 {{1} \over {(q^2_2+M^2)}} \gamma_2\cdots \gamma_n
{{1} \over {(q^2_{n+1}+M^2)}}\ \gamma_{bc}
\auto
$$
where the (complicated) residue functions $\gamma_i = \gamma_i(q_i, q_{i+1})$
are determined uniquely by the spin properties of the produced particles
together with unsubtracted dispersion relations in the momentum transfer
variables (exploiting the renormalizability/unitarity boundedness properties of
on-shell Born amplitudes).

The technique of Lipatov et al.\cite{klf,lnl} is to combine (5.48) with an
approximation to n+2 - body phase-space in the region (5.47) $i.e.$
$$
\eqalign{
d\rho_{n+2} \sim {{1} \over {(2\pi)^{3n+2}2^{n+1}s}}\quad
\prod_{i=1}^n
{{d\alpha_i} \over {\alpha_i}}\quad
\prod_{j=1}^{n+1}
d^2\underline{q}_{j\perp} }
\auto
$$
Integration over the $\alpha_i$ in the region (5.47) then gives
$$\eqalign{
d\rho_{n+2}\longrightarrow ln^n(s/M^2)\
{{1} \over {(2\pi)^{3n+2}2^{n+1}s\ n!}}
\prod_{j=1}^{n+1}
d^2\underline{q}_{j\perp} }
\auto
$$
and so the unitarity equation gives
$$
 M_{ab\rightarrow cd} (s,q^2) = {\sum^\infty_{n=0}}
\ \int d\rho_{n+2} M_{ab\rightarrow n+2}M^*_{cd\rightarrow n+2}
\auto
$$
with $M_{ab\to cd}$ satisfying a twice subtracted dispersion relation
$$\eqalign{
M = {{s^2} \over {2\pi i}}\int {{ds^\prime} \over
{(s^\prime-s)(s^\prime)^2}}\quad
\centerunder{$disc$}{\raisebox{-1mm} {$\scriptstyle s^\prime$}}
M &+ {{u^2} \over {2\pi i}} \int
{{du^\prime} \over {(u^\prime-u)(u^\prime)^2}} \quad
\centerunder{$disc$}{\raisebox{-1mm} {$\scriptstyle u^\prime$}}M \cr
&+ C_1s + C_2\cr}
\auto
$$
where u  is the usual invariant variable and $C_1$ and $C_2$ are
subtraction constants. The logarithms produced by (5.50) dominate the
subtraction terms in (5.52) and hence inserting (5.50) into (5.51)
and performing the s  and u  integrations in (5.52) amounts to the
replacement
$$
s~ln^n \left(s/M^2\right) \rightarrow -\bigg[s~ln^{n+1}\left(-s/M^2 \right)
\pm (-s)~ln^{n+1}
\left(s/M^2\right)\bigg]\bigg[2\pi (n+1) \bigg]^{-1}
\auto
$$
with the plus sign for positive signature (t-channel I=0,2) and the minus sign
for negative signature (I=1).

Clearly integrating (5.48) over the phase-space (5.50) will give a product
of transverse momentum integrals as the coefficient multiplying (5.53). That
the integrals are convergent follows directly from the fall-off properties of
the $\gamma_i$ discussed above. Indeed from this perspective, the large
transverse momentum cancellations found in the Feynman gauge calculations of
Cheng and Lo are simply a manifestation of renormalizability, or (presumably)
equivalent unitarity, bounds for on-shell production amplitudes.

Conversely, once we know that all transverse momentum integrals are convergent,
then the {\em leading-log results can only come from vector-exchange in the
multi-regge regions of phase-space} and it immediately follows from (5.53) that
the leading real logarithms will be in the odd-signature channel. However,
within the unitarity construction we are outlining there is a vital additional
cancellation which goes beyond the finiteness of transverse momentum integrals
and is responsible for the extreme simplicity of the odd-signature $T_1$
amplitude. This is what we referred to earlier as the cancellation of the
finite regions of ladder diagrams with more than two states in the t-channel.
We can illustrate this cancellation at the two-loop level, where it first
appears, and where it can be described without reference to the multi- Regge
bootstrap.

The multi-Regge region for three-particle (s-channel) intermediate states
contributes two transverse momentum integrals. The first is the product of
bubbles illustrated in Fig.~5.12(a)  while the second originates from
central-region vector production and is illustrated in Fig.~5.12(b). In $QED$
the ladder diagram of Fig.~5.6 generates the second diagram only at the
next-to-leading log level, whereas in a non-abelian theory it appears at the
leading log-level. However, as illustrated in Fig.~5.13, s-channel iteration of
the two-particle (s-channel) state produces the same diagram (in a non-abelian
theory) and at the leading log-level, that is in the $T_1$ amplitude, these
distinct unitarity contributions cancel. As a result the $T_1$ amplitude
contains only Fig.~5.12(a), at the two-loop-level, and so contains {\em only
two-particle states in the t-channel}. As we elaborate further in the next
Section, a generalization of this cancellation result persists to all orders as
the Multi-Regge Bootstrap, and explains why the two-particle unitarity analysis
of the last Section correctly demonstrates the reggeization of non-abelian
gauge bosons. It also confirms that (in the t-channel) the reggeization can be
understood as originating from that region of ladder diagrams where {\em all}
internal lines are close to on mass-shell.

Note also that the close relationship between Regge singularities and the
corresponding $t$-channel thresholds, which the general analysis of Part I is
based on, is clearly beginning to emerge. {\em Isolated} Regge pole behavior is
associated with two-particle unitarity while, as we shall shortly demonstrate,
the higher thresholds produce Regge cuts (and modify the trajectory functions
of Regge poles) but {\em do not} produce any {\em new} Regge Poles . The
importance of this is that we can then expect {\em multiparticle t-channel
unitarity to be completely represented by the reggeon unitarity of the reggeon
diagrams we introduce}.

In general terms it is clear that the dominance and simplicity of the
mass-shell regions in a non-abelian gauge theory is just what we
expect the dispersion-relation based Regge theory of Part I will
be able to exploit. The ``short-distance'' off mass-shell region, which
would surely require field theoretic techniques to analyse, cancels out
entirely. That the off-shell large transverse momentum cancellations do
indeed take place {\em at all orders of logarithms} is a major consequence
of the work of Sen\cite{as}. The techniques used by Sen involve a powerful
exploitation of the relationship between gauge invariance and Lorentz
invariance and straightforwardly give, for example, the fourth order term in
the electron trajectory function. However, although we believe these
techniques may be exploited further in the future we will not describe them
here since their major significance for our program (at this point) is
simply to guarantee that we can safely neglect that part of the theory which
is not controlled by ``Analytic Multi-Regge Theory''.

\newpage

\mainhead{6.  REGGEON DIAGRAMS}

We begin this Section by rewriting (5.31)--(5.40) in terms of reggeon diagrams
following the work of Bronzan and Sugar\cite{bs}. This will produce what is,
at first sight, a remarkable simplification of the results for $T_0$ and
$T_2$. We shall then describe the results of Fadin, Kuraev and
Lipatov\cite{klf} which extend (5.46)--(5.53) to a ``multi-regge bootstrap''
and in doing so, explain why this simplification has to occur.

It is interesting, and perhaps not too surprising, that initially several
authors were tempted by the low-order results for $T_0$ to introduce a vacuum
Regge pole with the trajectory
$$
\alpha^\circ
 (\qbar^2) = 1 -2 g^2 \left(\qbar^2 + {{5}\over {4}} M^2 \right)J_1(\qbar^2)
\auto
$$
Fortunately this is not only an unnecessary complication but is also
demonstrably wrong, as we shall shortly elucidate. Both $T_0$ and\ $T_2$ are
completely described by reggeon diagrams which give a two-reggeon cut as the
{\em only} Regge singularity in these amplitudes.

\subhead{6.1 SU(2) Leading and Next-to-Leading Log Diagrams}

We write odd-signature reggeon diagrams in the same notation that we used for
Pomeron reggeon diagrams in Section~6 of I, that is we use $E$ and  $\kbar$ as
variables. For an odd-signature reggeized vector we must, as in Section~7 of I,
include the particle pole in the reggeon propagator. Therefore we write
$$
\Gamma_{1,1}\left(E_1,\kbar^2\right)={1\over \left[E-\Delta\left(\kbar^2\right)
\right]}{1\over \left[\kbar^2+M^2\right]},
\auto
$$
where now
$$
\eqalignno{
\Delta(\kbar^2)&= 1-\alpha(\kbar^2)&\num\label{6.3}\cr
&=g^2(k^2+M^2)J_1(k^2).&\num\label{6.4}\cr}
$$

The two-reggeon diagram illustrated in Fig.~6.1 is written in $E$--space as
$$
F^I_2(E,\qbar^2) = \int
{{dE_1d^2\kbar} \over {(2\pi)^3}}
{{\left[ \beta^I_2 \right]^2}
\over {\left[ E_1-\Delta(\kbar^2) \right] \left[ \kbar^2 + M^2\right]
\left[ E-E_1-\Delta((\qbar -
\kbar)^2)\right] \left[ (\kbar - \qbar)^2 + M^2\right]}}
\auto
$$
where $\beta^{I}_2$ is the coupling of two-reggeons, with t-channel isospin I,
to the external particles. (In general this coupling can be a function of both
$\qbar$ and $\kbar$ although in the leading-log order to which we work this
dependence will be absent). Performing the $E_1$--integration and writing the
(even-signature) S-W transform gives
$$
F^I_2(s,\qbar^2) =  is \int {{dE} \over {2\pi i}} s^{-E} \int
{{d^2\kbar} \over {(2\pi)^3}}
{{\left[ \beta^I_2 \right]^2} \over {\left[ E-\Delta(\kbar^2)-\Delta
((\qbar-\kbar)^2) \right] \left[ \kbar^2 + M^2\right] \left[ (\kbar - \qbar)^2
+ M^2\right]}}
\auto
$$
Now using the expansion
$$
\int {{dE} \over {2\pi i}} {{s^{-E}} \over {[E-A]}} = 1-A\ lns \ + {{1} \over
{2!}}\ (Alns)^2 + \cdots
\auto
$$
we can expand the denominator of (6.6) and, if we simply
take $\beta^{I}_2 = g^{2}$, we obtain
$$\eqalignno{
F^I_2 (s,q^2) =\ {\rm is}\ g^4\  \bigg\{ J_1(q^2) &- 2J_2 (q^2) g^2 \ lns\ +
\left[ J_3(\qbar^2) + \tilde{J} (\qbar^2)\right] (g^2\ lns)^2\cr
&- \left[ {{1}\over {3}} J_A(\qbar^2) + J_B (\qbar^2)\right] (g^2\ lns)^3
+ \cdots \bigg\} &\num\label{6.8}\cr }
$$
This gives the first term in the expansions for $T_0$ and $T_2$ exactly, with
higher-order terms also partially reproduced.

Proceeding in the same fashion, Bronzan and Sugar\cite{bs} were able to show
that the {\em complete} 10th order results for $T_0$ and $T_2$ are exactly
produced by adding the four reggeon diagrams shown in Fig.~6.2 if we
introduce the four-reggeon interaction
$$
\eqalign{
R^I_{2,2}(\qbar,\kbar\,_1,\kbar\,_2)= a_Iq^2+b_IM^2+c_IV\left(\kbar\,_1,
\kbar\,_2,\qbar\right),}
\auto
$$
where the notation is illustrated in Fig.~6.3. $V$ is universal and has the
complicated structure
$$
\eqalignno{
V(\kbar\,_1,\kbar\,_2,\qbar)= &{{\left(\kbar^2_1+M^2\right)\left(\left(
q-\kbar\,_2\right)^2+M^2\right)+\left(\kbar^2_2+M^2\right)\left(\left(
\qbar-\kbar\,_1\right)^2+M^2\right)}\over
{\left(\kbar\,_1-\kbar\,_2\right)^2+M^2}}\cr
+ &{{\left( \kbar^2_1+M^2\right)\left( \kbar^2_2+M^2\right)+
\left( (\qbar-\kbar\,_1)+M^2\right)\left( (\qbar-\kbar\,_2)^2+M^2\right)}
\over {\left( \qbar - \kbar\,_1 -
\kbar\,_2\right)^2+M^2}}.&\num\label{6.10}\cr
}
$$
For both I=0 and I=2 we have
$$
a_I = C_I~g^2,~~\ b_I = {{3C_I-1} \over {2}}~g^2,~~\ c_I = -{{C_I}
\over {2}}~g^2
{}~~\ {\rm with}\ C_I = 2-{{I(I+1)} \over {2}}
\auto
$$

Note that all of the parameters of (6.10) are completely determined by the
expansions up to sixth order only. Therefore the third and fourth reggeon
diagrams shown in Fig.~6.2 are {\em explicit predictions} of the eighth and
tenth order contributions respectively. Consequently it is clear that in
addition to extrapolating the reggeization of $T_1$ to be the leading log
result to all orders, we can also extrapolate the reggeon diagram series
illustrated in Fig.~6.2 to give $T_0$ and $T_2$ to all orders as the
next-to-leading log result. This is the desired simplification of the results
for $T_0$ and $T_2$.

Bronzan and Sugar also allowed for the addition of further Regge poles such as
(6.1) together with the corresponding reggeon diagrams. They found that
{\em all} such contributions {\em completely decoupled} in their final
expressions.

\subhead{6.2 The Multi-Regge Bootstrap}

A fundamental explanation of the simplicity of the above results is found in
the following extension of the s-channel unitarity calculations outlined in
Subsection 5.4. If it is assumed that the basic argument for the convergence of
transverse momentum integrals extends appropriately then the leading log result
for {\em all production amplitudes in the multi-regge limit} will be given by
the exchange of {\em reggeized} vector bosons. Fadin, Kuraev and
Lipatov\cite{klf} checked that this was the case for some low-order
production amplitudes and then assumed that in the general multi-regge
leading log approximation, (5.47) becomes
$$
M_{ab\rightarrow cd+n} \sim \gamma_{ab}
\left( {{s_1} \over {M^2}}\right)^{\Delta\left(\qbar_I^2\right)}
{{\gamma_1} \over {\left( \qbar_1^2+M^2\right)}}
\left( {{s_2} \over {M^2}}\right)^{\Delta\left(\qbar^2_2\right)}
{{\gamma_2} \over
{\left(\qbar_2^2 + M^2\right)}} \cdots
\vspace{.1in}
\cdots {{\gamma_n} \over {\left(\qbar^2_{n+1}+ M^2 \right)}}
\left({{s_{n+1}} \over {M^2}} \right)^{\Delta\left( \qbar^2_{n+1}\right)}
\gamma_{bc}
\auto
$$
where $\Delta(\qbar^{2})$is again defined by (6.3) and (6.4). The $\gamma_i$
actually have a complicated form which is important for the consistency of
(6.12) with the general form of multi-regge amplitudes given in Part I.
However, we shall not dwell on this point here. The vital result we want to
emphasize is that when (6.12) is inserted into the unitarity equation
(5.51), as illustrated in Fig.~6.4, and (5.52) is used {\em the complete set
of leading and non-leading results} (5.31)--(5.33) is reproduced.

There are several comments to be made. First we note that the cancellation of
Fig.~5.12(b) by Fig.~5.13 is completely accounted for by the reggeization of
the
exchanged vector mesons. That is once reggeization is included then, at the
leading-log level, we only need consider the ``multiperipheral'' production
processes illustrated in Fig.~6.4 . Secondly we note that such production
processes can only possibly produce the reggeon diagrams found by Bronzan and
Sugar. Consequently the ``remarkable'' simplification that we described above
is simply {\em a consequence of the leading log reggeization of production
processes}.

The complete set of leading and next-to-leading log results extrapolated to all
orders can be compactly summarized by one integral equation. The on-shell
gluon-gluon scattering amplitudes $T_0$, $T_1$, $T_2$ are given by
$$
T_I \left( E,\qbar^2\right) = T_2^I\left( E, \qbar, \kbar\right)
_{\kbar^2 = (\qbar-\kbar)^2 = M^2}
\auto
$$
where the $T^{I}_2$are ``off-shell'' gluon-reggeon scattering amplitudes (see
Fig.~6.5 for notation) which satisfy
$$\eqalignno{
\left[ E-\Delta(\kbar^2)-\Delta \left( (\kbar-\qbar)^2 \right)\right]
&T_2^I\left( E,\kbar, \qbar\right)= {{E} \over {a_I\qbar^2 + b_I M^2}} + \cr
& {{g^2} \over {2\pi}} \int
{{d^2k'\ R^I_{2,2}\left( \qbar,\kbar,\kbar'\right) T_2\ ^I (E,\kbar',\qbar)}
\over {\left[ \kbar^{\prime 2} + M^2\right] \left[ \left( \qbar-\kbar^\prime
\right)^2 + M^2\right]}}&\num\label{6.14}\cr}
$$
where $R^{I}_{2,2}$ is defined by (6.9)--(6.11), {\em also for} $I = 1$.
(Note that since $T_0$ and $T_2$ are even-signature amplitudes the zero-order
contribution of the inhomogeneous term in (6.14) is essentially irrelevant).
For
$I = 0$, (6.14) is the massive version of the ``Lipatov equation'' which has
been the focus of attention in general discussions of ``small-x physics'' and
the ``perturbative Pomeron''.

The simple reggeization of $T_1$ is now restated by noting that,
for I = 1, (6.14) has the {\em exact} solution
$$
T^1_2(E, \kbar, \qbar) = {{E} \over {(\qbar^2+M^2)(E-\Delta(\qbar^2))}}
\auto
$$
Given that it is multiperipheral production which leads to (6.14) we can
aptly describe this solution as a ``multiperipheral bootstrap'' of the kind
that was suggested historically for the Pomeron. It is clearly very tempting to
believe that an ultimate understanding of the Pomeron in QCD should in some way
or other carry over the simple bootstrap properties of the gluon to the
Pomeron.
For the present we shall (later in this Section) simply exploit the bootstrap
to show how the reggeized gluon diagrams can be directly generated from reggeon
unitarity in a manner very similar to the derivation of Pomeron RFT in Section
6 of Part I.

\subhead{6.3 Multi-Regge Behavior and the Transverse Momentum Cut-Off}

We emphasized in the previous Section that the transverse momentum diagrams
derived in leading and next-to-leading log approximations are actually obtained
by first imposing a transverse momentum cut-off within individual Feynman
integrals. In general it is found that a finite limit is obtained if this
cut-off is removed after the high-energy limit is taken and after the sum over
all diagrams (of the appropriate order) is performed. Of course, just because a
limit exists it is not necessarily correct to take it. Indeed we shall discuss
in the next Section some general issues of principle which determine that if we
want to reach the high-energy behavior of an unbroken gauge theory, it is
imperative that the cut-off not be removed until after the infra-red limit of
massless gluons is taken.

At this point we want to note a simple, practical, reason why a transverse
momentum cut-off should be employed in the multiperipheral calculation which
inserts (6.12) into (5.50) (even though this may lead to violations of the
bootstrap equation, for example, that vanish only as the cut-off is removed).
We are inserting multi-regge behavior with the trajectory given, in first
approximation, by (6.4). This trajectory function has the qualitative shape
illustrated in Fig. 6.6, where the units on the horizontal axis are
multiples of $M^2$ and on the vertical axis are multiples of $(g^2/16\pi^2)$.
For finite $M^2$ and small $g^2$ it is clearly necessary to go to large
$q^2$ before the trajectory decreases by a single unit in the angular
momentum plane. Therefore the production amplitudes we begin with are good
leading-power approximations (apart from the Regge cut behavior that is
generated by the higher order reggeon diagrams that we construct) out to
reasonably large, {\em but not infinitely large}, transverse momentum.
However, as $M^2 \to 0$ the trajectory function goes to - $\infty$, {\em
except at} $q^2 = 0$. Consequently, {\em in this limit}, our initial
multi-Regge amplitude is not even a leading power approximation at finite
transverse momentum. The trajectory of the two-reggeon cut is also shown in
Fig. 6.6 and when this is incorporated it clearly extends the leading power
approximation out to larger $q^2$. Nevertheless, in the massless limit, the
complete set of reggeon diagrams that we construct can at best describe the
full amplitude only at transverse momenta very close to zero - {\em
certainly not at infinite transverse momenta}. Hence, in our opinion, the
transverse cut-off is mandatory.

\subhead{6.4 General Structure of the Reggeon Diagrams}

{}From the point of view of the general multi-regge theory of Part I it is
clear
that the existing perturbative calculations have exposed only an extremely
small subset of the full reggeon structure which from the general arguments
must be present. We would, of course, like to use the general formalism to
reliably extrapolate the perturbative results to the complete structure,
particularly the multiparticle amplitudes which we shall need in the following
Sections. In the first instance at least, we shall not need as much detail
as appears in the weak-coupling perturbative calculations. Instead we would
be satisfied with a formalism where the general features are sufficiently under
control for us to discuss reliably the infra-red and critical limits that will
concern us, in a manner comparable with the ``critical phenomenon" treatment of
the Critical Pomeron discussed in Part I. As we elaborated, the Critical
Pomeron
formalism has the very attractive feature that we need to know only the
existence of an even signature Regge pole with intercept $\alpha(0)$ near one
and a finite non-zero triple Pomeron interaction. We can then justify the
neglect of {\em non-singular} higher-order interactions and (in effect) use the
reggeon unitarity relations alone to predict the dominant scaling behavior of
the diffraction peak in the limit $\alpha(0)\to 1$.

       In analogy with the Pomeron formalism, therefore, we are immediately
led to ask how much of the structure found in the perturbative Yang-Mills
calculations can we reproduce from the general formalism if we suppose we know
only the existence of odd-signature reggeons, with a particular group
structure, together with low-order {\em non-singular} interactions. At first
sight the answer is very little indeed, since we apparently would not
anticipate
the four-reggeon vertex (6.10). This vertex looks very complicated and also is
singular as a function of its transverse momenta. Clearly we can expect further
non-leading logarithms to generate arbitrarily high-order reggeon interactions
in all possible isospin channels with comparably singular transverse momentum
structure. This has immediate consequences.

If we were to try to apply the renormalization group, as we did for the Pomeron
RFT in Section~6 of I, to look for possible critical behavior of the full
reggeized gluon RFT, we would find a very important difference. For the Pomeron
RFT we found that all but the lowest-order vertices had negative scaling
dimensions and so were irrelevant at the fixed-point we found. Assuming that
the transverse momentum singularity structure described above generalizes, then
all reggeized gluon vertices will have the {\it same scaling dimension.}
Consequently a fixed-point, if it existed, would involve all vertices, making
the finding of such a feature, given only the structure of the lowest order
vertices, almost impossible. The transverse momentum singularities are due to
the vector particle on the gluon trajectory and it seems that to successfully
apply the renormalization group we need some form of RFT without such
singularities. This clearly suggests that we must understand confinement and
uncover the Pomeron before we can apply the RFT renormalization group.

On a much less ambitious level the complicated form of the four reggeon vertex
apparently makes it difficult to extrapolate to even the general structure of
the high-order vertices without performing the corresponding calculations. If
this were the case then the initial success of the general formalism - that
at 10th order hundreds of Feynman diagrams are completely summarized by just
five reggeon diagrams - would be essentially worthless. That is, the ultimate
goal of completely summarizing the high-energy behavior of spontaneously broken
gauge theories in terms of reggeon diagrams would clearly be hopeless.

Fortunately there is a further vital simplification to be observed. We shall be
able to argue in the following that the singularity of the interaction, and
even the singularity of the trajectory function (6.4), is an effect of
formulating reggeon diagrams with only those interactions that
satisfy {\em signature conservation} rules. In particular, the signature of the
two-reggeon cut generated by two reggeized gluons is necessarily even, while
the gluon itself, of course, has negative signature. Therefore if signature is
conserved (as it is in the reggeon diagrams of four-particle amplitudes) we
should not have diagrams in which a three-reggeon vertex directly couples a
reggeized gluon to a two-reggeon cut. In terms of general multi-regge theory, a
three reggeized gluon vertex is certainly possible and we would expect such a
vertex to make major contributions to multi-regge amplitudes.

If a three-reggeon vertex exists then signature conservation should be
implemented in the reggeon diagrams of four-particle amplitudes by the presence
of a ``nonsense-zero'' of the form (in the notation of Fig.~6.7)
$$\eqalignno{
\Gamma_{12}(E,\kbar,\kbar\,_1,\kbar\,_2)&\sim
\left[E-\Delta(k_1)-\Delta(k_2)\right]r_{12}&\num\label{6.17}\cr
&\centerunder{$\large\sim$}{\raisebox{-2mm} {$\scriptstyle E\sim\Delta(k)$}}\;
\left[\Delta(k)-\Delta(k_1)-\Delta(k_2)\right]r_{12}.&\num\label{6.18}\cr}
$$
Indeed it follows from (6.14) and (6.15) that $T^{I}_{2,2}$ is defined so that
in lowest order it contains a {\em nonsense-pole} factor of the form of (6.16).
Consequently if the full three-reggeon vertex is defined from the residue of
the Regge pole in (6.14), with the nonsense-pole factor extracted, it will
indeed have the form (6.16).

Given the existence of the nonsense zero, it is natural to consider whether the
complicated four-reggeon vertex might actually originate from a product of
three-reggeon vertices (containing such zeroes) and a reggeon propagator.
Indeed the purpose of the next sub-Section will be to show that given only the
existence of the nonsense zero in the triple reggeon vertex and the
`multiperipheral' origin of the reggeized gluon, we can essentially reconstruct
all of the general structure of the reggeon diagrams from t-channel unitarity
and the hexagraph formalism. We shall, however, have to develop a rather
elaborate construction procedure.

\subhead{6.5 Reggeon Diagrams from Hexagraph Loops}

The essential difference between the reggeon unitarity equation for
odd-signature reggeons and that for even signature Pomerons is that the
signature factor behavior I.(6.3) will be replaced by
$$\eqalignno{
\sin \ {{\pi}\over{2}} (\alpha_1 - \tau_1') \sim {{\pi}\over{2}}
\Delta (t_1), ~~~\cdots,~~~
&\sin {{\pi}\over{2}} (\alpha_N - \tau_N') \sim
{{\pi}\over{2}}\Delta(t_N)\cr
&t_1 \sim \cdots \sim t_N \sim M^2 \sim 0
&\num\label{6.18}\cr}
$$
We noted above that the propagator for an odd-signature reggeon contains a
signature factor (particle pole) - that is unless a discontinuity is taken
through the reggeon as we discuss below. In general the effective propagator
for an N-reggeon intermediate state in a hexagraph loop diagram contains
signature factors for each `new' reggeon which appears (supposing for the
moment
that we construct the diagram by proceeding from left to right across the
rapidity axis) in addition to the nonsense pole ``energy denominator". That is
we write for such a propagator
$$
\left[ E-\Delta \left( \kbar_1^2\right)- \cdots \Delta \left( \kbar_N^2\right)
 \right]^{-1} \left[ \Delta\left( \kbar_r^2\right)\cdots \Delta
\left( \kbar^2_{r+s}\right)\right]^{-1}
\auto
$$
where $r,\cdots,\ r+s$ are the new reggeon lines and, as yet, the trajectory
function $\Delta(k^{2})$ can have a general functional form. The signature
factors are an immediate source of transverse momentum singularities and the
issue we have to resolve is whether they can actually produce all the
singularities of gluon reggeon diagrams via a systematic construction
procedure.

Consider now the transverse momentum loop contribution (6.4) to the trajectory
function. We would like to understand this as originating from the simplest
hexagraph loop, that is the diagram shown in Fig.~6.8. At first sight there are
two nonsense-zeroes of the form (6.16) from the two triple reggeon vertices,
just one of which is sufficient to cancel the reggeon propagator from the
intermediate two-reggeon state. Therefore we might easily conclude that this
diagram has essentially no meaning. However, if we consider the
`multiperipheral' cut through the diagram, as illustrated, then for any
finite upper limit on the mass of the states contributing to the cut
reggeons the diagram should make sense as a two-reggeon loop amplitude. This is
because the nonsense zero of (6.15) appears only after the infinite sum over
states is performed - and the integral equation (6.14) is satisfied.
Consequently the diagram should make sense if it is first evaluated as a
reggeon loop with general vertices as illustrated in Fig.~6.9 which are then
replaced by a regge pole plus nonsense zero vertex as illustrated - to
represent
the limit of an infinite sum over states.

Note that we can also represent the (I = 1) sum of multiperipheral diagrams
generating the gluon reggeon via the integral equation (6.14) as in Fig.~6.10,
where the bubbles are first defined to include only a finite number of
intermediate states. Taking the limit of an infinite number of states then
implies that we can replace the bubbles by a Regge pole with nonsense zero
vertices. This give the series of diagrams shown in Fig.~6.11, with the
single loop diagram being that of Fig.~6.8.

Cutting a reggeon simply means that a discontinuity is taken - which has the
effect of removing the particle-pole signature factors for the cut reggeons.
Therefore if we evaluate the diagram of Fig.~6.8 with one nonsense-zero
cancelling the two-reggeon propagator and write the other in the form (6.17) we
obtain
$$
\left[{1\over E-\Delta (\qbar)}\right]^2 r^2_{12}\int {d^2\kbar\left[
\Delta (\qbar)  -\Delta(\qbar-\kbar)-\Delta(\kbar)\right]\over
\Delta(\kbar)\Delta(\qbar-\kbar)}.
\auto
$$
If we also introduce a transverse momentum cut-off $\lambda$ in (6.20) then,
conceptually at least, we can suppose that, in the renormalization group sense,
some range of transverse momenta greater than $\lambda$ has already been
integrated out to give a reggeon trajectory function that is analytic at small
transverse momenta. As a first approximation to this trajectory function we can
then take
$$
\Delta (\qbar)=\alpha'(q^2-M^2),\hspace{0.5in} \alpha' \sim M^2/\lambda
\auto
$$
(Note that $\alpha'$ could also be `non-perturbative' in origin.) Inserting
(6.21) into (6.20) then {\em near the two-particle threshold} we obtain
$$
\left[{r_{12}\over E-\Delta(\qbar)}\right]^2{1\over \alpha'}(q^2-M^2)K(q^2)+
\cdots.
\auto
$$
If we now sum the infinite series of diagrams shown in Fig.~6.11 (in all of
which the multiperipheral discontinuity is taken) we obtain a
renormalization of the trajectory (6.21), that is
$$
\left[E-\alpha'(q^2-M^2)\right]^{-1}\to \left[E-\alpha'(q^2-M^2)-
{r^2_{12}\over \alpha'}(q^2-M^2)K(q^2)\right]^{-1}.
\auto
$$

Thus the reggeon diagrams of Fig.~6.11 consistently reproduce the perturbative
threshold contribution to the gluon trajectory function if we identify
$$
{r^2_{12}\over \alpha'}~~\sim~~g^2 c_{ijk}^{2}
\auto
$$
where $c_{ijk}$ is a group-dependent factor that we have ignored until this
point. Actually if we considered in more detail the symmetry properties of
diagrams which follow from their signature properties we would find that the
$c_{ijk}$ must have the antisymmetry properties which identify them as the
structure constants of a non-abelian group. However, we will not focus on this
aspect and will essentially ignore the $c_{ijk}$ in the following discussion.

The `multiperipheral' cut of Fig.~6.8 gives the leading threshold
behavior (compared to a cut which goes through one or both of the loop
reggeons) since it leaves both loop propagators uncut and so leaves the maximum
number of particle poles in the loop integral. Note also that if the
transverse momentum cut-off is inserted in (6.20) it would not appear in the
leading singular behavior at the two-particle threshold.

Clearly the nonsense zero is responsible for ensuring that the diagrams of
Fig.~6.11 simply produce reggeization. Indeed if we {\em take $\alpha'\to 0$
with $g$ finite} (which is achieved by taking $\lambda\to \infty$) then
(6.23) {\em gives exactly the perturbative result}. However, we can also say
that we have a straightforward ``non-perturbative'' formulation of how the
{\em odd-signature} reggeon diagrams of Fig.~6.11 contribute to the
odd-signature amplitude and the reggeon trajectory function in particular.
$\alpha'$ can be regarded either as non-perturbative in origin, or as
originating from the transverse momentum region above the cut-off. In either
case the gauge theory perturbative analysis is exactly as we expect - with a
simple {\em non-singular three reggeon vertex}.

Before discussing the diagrams involving the singular four reggeon vertex,
we must first elevate to a major point of principle the need to take a
discontinuity in evaluating the diagrams of Figs.~6.8 and 6.11. The
derivation of the reggeon unitarity equations in Part I was critically
dependent on the breakdown of multiparticle amplitudes into component
discontinuities. When odd-signature reggeons are involved, distinct multiple
discontinuities have quite distinct transverse momentum singularities -
because of the removal of different products of signature factors. Therefore
it is not surprising that the hexagraph loop expansion is a well-defined
iteration of the reggeon unitarity equations for an individual hexagraph
amplitude only when an appropriate (multiple) discontinuity of the hexagraph
is specified.

Although we will not give a detailed construction procedure for a general
amplitude we note that a single loop contribution to a multiple
discontinuity of a hexagraph amplitude can always be constructed unambiguously
via a particular reggeon intermediate state. The corresponding full
amplitude can then be obtained (effectively from an asymptotic dispersion
relation) by adding the signature factors of the appropriate
Sommerfeld-Watson representation. This (loop) amplitude can then be used for
the construction of further loop contributions to other hexagraph
discontinuities and so on - giving a complete set of multiple loop amplitudes.
The full iteration procedure is particularly important for the construction
of multiparticle amplitudes but will also be what we implicitly use below to
construct the diagrams for elastic scattering which apparently involve the
very complicated four-reggeon vertex. As we emphasized above, in the first
instance at least, we shall be satisfied with constructing only the leading
transverse momentum singularities of a particular hexagraph loop amplitude.
This requires that the maximum number of propagators remain uncut - or
equivalently that the maximal {\em multiperipheral} discontinuity, or
discontinuities, be taken.

Consider now the one-loop hexagraph contribution to the three-reggeon vertex
illustrated in Fig.~6.12. The discontinuities shown leave the maximum number
(two) of uncut propagators. As we described in detail in Part I, there is a
direct relationship between taking a discontinuity of a hexagraph amplitude and
the removal of a particular (often complicated) signature factor. For our
present purpose (although not for the more elaborate discussion of effects
associated with fermion loops discussed in Section 9) it will be a good enough
approximation to take all signature factors to be simple particle poles, as in
(6.19), and to assume that a cut through any {\em adjacent} set of lines in
a hexagraph removes the signature factor associated with one and only one of
the lines cut. In Fig.~6.12 this is sufficient to determine that the pole
factors associated with lines a, b, c and f are removed by the discontinuities
taken. If this diagram appears as a sub-diagram in an `energy-conserving'
reggeon diagram, there will be two energy denominators to insert and nonsense
zeros for each of the vertices. If we assume that the energy integration for
the loop has been performed by using the Regge pole for line b we obtain
$$\eqalignno{
r^3_{12} \int d^2\kbar &[E-\Delta_b-\Delta_e][\Delta_e(E-\Delta_b-
\Delta_e)]^{-1}\cr
& \times[(E-\Delta_e)-\Delta_c-\Delta_d][\Delta_d(E-\Delta_c-\Delta_d-
\Delta_f)]^{-1}\cr
& \times [\Delta_f-\Delta_d-\Delta_e]&\num\label{6.25}\cr}
$$
Clearly both internal reggeon propagators are simply cancelled by the
nonsense zeroes in the triple-regge vertices. Consequently the diagram
generates no Regge cuts, but instead gives only a renormalization of the
triple-regge vertex corresponding to the transverse bubble diagram shown in
Fig.~6.13, i.e.
$$
\delta \Gamma_{12} = {{r^3_{12}} \over {\pi^2}} \left( q^2_f - M^2\right)
K^{(2)}_\lambda \left( q^2_f\right)
\auto
$$
If we add this contribution, together with the analogous diagram in which
c and f are interchanged, to the original triple regge vertex we obtain
$$
\Gamma^R_{12} = r_{12} \left[ E-\Delta_R \left( \kbar^2_1\right)- \Delta_R
\left( \kbar^2_2 \right) \right]
\auto
$$
and $\Delta_R$ is given by (6.23). Consequently the diagram of Fig.~6.12 simply
renormalizes the vertex by shifting the nonsense zero to match the shift of the
trajectory produced by the diagrams of Fig.~6.11. That is the vertex acquires
the two-particle threshold in each of its transverse momenta. Indeed we
emphasize that the calculation keeping only the multiperipheral cut in
Fig.~6.12 is only accurate near this threshold. It is possible to show in a
similar manner that diagrams of the form illustrated in Fig.~6.14 build up the
higher thresholds in the trajectory and vertex function in a similar manner.

Consider next diagrams of the form illustrated in Fig.~6.15 which are produced
by the {\em even signature} coupling of two reggeons to external particles. To
build up these diagrams we begin with the single loop diagram shown in
Fig.~6.16. The evaluation of Fig.~6.16 is almost identical to that of
Fig.~6.12, the only difference being that the energy denominator $\Gamma_2$
is not removed by a nonsense zero. If we insert this diagram into Fig.~6.17
and use (6.24) and $\frac{\beta_2}{\alpha'} = g^{2}$, we obtain
$$\eqalignno{
g^6 \int d^2k_1d^2k_2 &{{1} \over {\Delta(\kbar^2_1)}}
\left[ E-\Delta(\kbar^2_1) -\Delta \left( \kbar_1-\qbar)^2\right)\right]^{-1}
{{1} \over {\Delta\left( (\kbar_1-\kbar_2)^2\right)}} \cr
& \times \left[ E- \Delta(\kbar_2^2)- \Delta\left( (\kbar_2-\qbar)^2\right)
\right]^{-1}
{{1} \over {\Delta\left( (\kbar_2-\qbar)^2\right)}}&\num\label{6.25}\cr}
$$
Clearly this is exactly the contribution to the second diagram of the expansion
in Fig.~6.2 from one term of what we have been regarding as the singular part
of the four-reggeon vertex ($V(\kbar\,_1,\kbar\,_2,\qbar)$). Indeed the
sum of all diagrams of the form of Fig.~6.17 clearly produces the complete
contribution of $V(\kbar\,_1,\kbar\,_2,\qbar)$. If we examine the nature of the
approximations made, by taking multiperipheral cuts etc., we find that our
evaluation of these diagrams is accurate {\em close to the three particle
threshold}.

The above analysis can be extended to the complete series of diagrams shown in
Fig.~6.2 (and any other diagram apparently involving the singular four-reggeon
interaction) to show that only triple-Regge vertices with nonsense zeroes are
involved. In fact the `four-reggeon' interaction has the simple  hexagraph
representation illustrated in Fig.~6.18. As a result we can conclude that in
reality there are no singular reggeon interactions in the reggeized gluon RFT.
All transverse momentum singularities are due to the signature factor, or
particle pole, in the reggeized gluon propagator {\em if we construct the
diagrams via their discontinuities and utilize a three-reggeon vertex with a
nonsense zero}.

This last conclusion is clearly very comforting since it implies that there is
no significant mystery or complexity in the reggeon diagrams occurring in a
spontaneously broken gauge theory. Apart from the group-structure involved,
which we have ignored in our general discussion since it is straightforward,
the
structure of all interactions is just the simplest expected (for odd-signature
reggeons) from the general Regge theory of Part I of this article. Therefore we
can, in principle at least, construct the reggeized gluon loop contributions to
a general hexagraph amplitude directly from the general formalism that we used
to discuss Pomeron amplitudes.

In the most recent work of Bartels\cite{jb} extensive non-leading log
results are obtained by directly iterating the asymptotic dispersion relations
of Part I. Conceptually this should be equivalent to the iteration of hexagraph
amplitudes through reggeon unitarity that we have described. However, the
approach of Bartels is clearly capable of keeping more detail than we have
attempted. In particular Bartels is able to show the existence (to the
non-leading order at which he works) of the general class of bootstrap
equations
which are required to maintain signature rules in higher order diagrams. As we
shall discuss further in Section 8 these ``self-consistency'' equations for
the full reggeization of the theory can be regarded as the low transverse
momentum consequence of the underlying gauge invariance of the theory,
{\em which must be maintained} by our treatment of massless fermions, for
example.

\newpage

\mainhead{7. COMPLIMENTARITY AND THE INFRA-RED ANALYSIS OF TRANSVERSE MOMENTUM
DIAGRAMS}

{}From the last Section, it is clear that reggeon diagrams provide a complete
description of spontaneously broken gauge theories in the Regge limit.  We
specifically considered SU(2) gauge theory but as we partially described in
Section 4, there are a large number of results in the literature demonstrating
that essentially the same structure emerges for any gauge group and with any
fermion and scalar content.  Provided the gauge group is non-abelian (or a
product of non-abelian groups) and the gauge symmetry is completely broken so
that all vector bosons acquire mass, then {\it all} vectors and fermions
reggeize. (In general the scalars do not reggeize). The full Multi-Regge Theory
of Part I of this article is therefore applicable to the leading high-energy
behavior which originates from the vector bosons and fermions. Given the simple
structure of the resulting reggeon diagrams, as described in the last Section,
we are clearly encouraged to push the formalism as far as is possible within
its own limitations.

Our ultimate purpose is to study unbroken gauge theories and $QCD$ in
particular. Consequently we must consider if, even in principle, we can hope to
obtain information about the Regge behavior of an unbroken gauge theory from
that of the broken theory. The major problem we have to face, of course, is
that in general there are expected to be confinement and chiral-symmetry
phase-transitions separating the region of parameter space where our starting
perturbative calculations are valid, from the `physical' region of unbroken
gauge invariance. This suggests that the S-Matrix for (massive) gluons and
quarks is totally unrelated to the hadron S-Matrix that we would like to study.
The purpose of this Section is to establish that under certain circumstances
the problem of confinement (and chiral symmetry breaking) can be confronted and
we can argue that an infra-red limit of the reggeon diagrams of a
spontaneously-broken gauge theory can give results relevant to the high-energy
behavior of the unbroken theory. This will lead us to consider the potential
physical significance of the transverse momentum infra-red divergences which
occur in this limit. We begin by discussing the lattice gauge theory
result\cite{fs} that there can be a smooth relation between the
``confining'' and ``Higgs'' regimes of a theory.

\subhead{7.1 Complimentarity for SU(2) Gauge Symmetry}

Consider SU(2) gauge theory with a fundamental (that is doublet) Higgs
field as discussed in the last Section.  The lagrangian is
$$
{\cal L} = -{1\over 4} F_{\mu\nu}^2(A) + |(\partial_\mu - igA_\mu)\phi|^2
- {1\over 2} h(\phi^*\phi)^2 + \mu^2\phi^*\phi .
\auto
$$
Since $\phi(x)$ is in the fundamental representation, we can write
$$
\phi(x) = \Omega(x) {0\choose\rho(x)}  ,
\auto
$$
where $\Omega(x)$ is an SU(2) matrix.  If we write
$$
B_\mu = \Omega^+ \left({1\over g}\partial_\mu - A_\mu\right)\Omega  ,
\auto
$$
we obtain
$$
{\cal L} = -{1\over 4} F_\mu\nu^2(B) + (\partial_\mu\rho)^2 +
\rho^2g^2[B_\mu^+B\mu]_{22} + V(\rho) .
\auto
$$

Because gauge transformations on A and $\phi$ can be absorbed by
$\Omega$, we can define B and $\rho$ to be gauge independent.  Also since
$\Omega$ does not appear in (7.4), we can regard this as an expression
for the gauge-invariant lagrangian in terms of gauge-invariant variables.
(In fact $B_\mu$ and $\rho$ are equal to A and $\phi$ in the unitary gauge.)
For a fermion field $\psi(x)$ in the fundamental representation, we can also
write
$$
\chi(x) = \Omega^+(x)\psi(x)  ,
\auto
$$
so that $\chi(x)$ is a gauge-invariant fermion field.

Consider now the path-ordered line-integral
$$
\phi^*(x)\left[\exp - \int_x^y dx_\mu gA_\mu\right]\phi(y)  ,
\auto
$$
which is, of course, gauge-invariant.  This integral is defined as a limit
of a product of the form
$$
\phi^*(x)\left[\prod_i\left({\delta\over \delta x_i} -
gA(x_i)\right)\delta x_i\right]\phi(y)  ,
\auto
$$
$$
\sim\rho(x)\left[\prod_ig B(x_i) \delta x_i\right]\rho(y)  ,
\auto
$$
if we use the approximation
$$
\Omega(x_i)\Omega(x_{i+1}) \simeq 1   .
\auto
$$
Consequently if the Higgs field $\phi$ develops a continuous classical
component (that is a vacuum expectation value) it is possible for a
gauge-invariant non-local operator such as (7.6) to factorize into a product of
local operators of the form (7.8). This is clearly straightforward if only a
finite product is involved in (7.7), as it would be in a lattice theory, so
that there is no subtlety involved in defining the ``continuum limit'' (7.6).

In the confining region of the parameter space for (7.1), which we anticipate
to include $\mu^2$ large and positive, we expect the physical states to
be created by operators of the form (7.7) (as well as loop operators involving
$A_\mu$ only and expressions of the form (7.7) involving fermion fields). In
the Higgs region, which we anticipate to include $\mu^2$ large and
negative, we expect the physical states to be created instead by local
gauge-invariant operators of the form of $\rho(x)$ and $B(x)$.  The above
discussion suggests that the two kinds of states are not really distinct
if (7.8) holds, and so in this case it should be possible to go smoothly
from one region of parameter space to the other {\it without encountering a
phase-transition}. This smoothness property, known as
{\it complimentarity}\cite{fs}, has indeed been verified in a number of
lattice studies of the theory with the phase-diagram of Fig.~7.1 shown to be
the appropriate description. It has also been shown that if $\phi(x)$ is in
a higher representation than the fundamental (so that (7.2), (7.3), (7.5),
{\it and} (7.8) can no longer be written), then the line of
phase-transitions in Fig.~7.1 extends across the whole phase-diagram and so
it is impossible to go from the Higgs regime to the confining regime without
encountering a phase-transition.

{}From our point of view, the significance of complimentarity is that
``string-like'' states can form smoothly from perturbative gluon states. This
is clearly vital if we expect to build a Pomeron related to a flux-tube picture
starting from perturbative calculations. Note that it is only for
strong-coupling that there is no phase-transition encountered in Fig.~7.1.
Since the continuum limits in the confining and Higgs regime would be expected
to be taken as shown, we would not expect complimentarity to hold, in general,
in the continuum. (The continuum limit of the Higgs theory may very well not
exist since it is not asymptotically free.)  We conclude therefore that to
apply complimentarity straightforwardly in an infra-red analysis of reggeon
diagrams, we {\em must keep an ultra-violet cut-off}. In general there is no
gauge-invariant alternative to the lattice cut-off. Fortunately, the transverse
momentum and reggeon diagrams we have described in the last Section are
gauge-invariant and so imposing a {\em transverse momentum cut-off} in such
diagrams {\em is a gauge-invariant ultra-violet cut-off in the Regge limit}.
As we have discussed, to derive the diagrams from the underlying Feynman
diagrams, it is technically convenient (if not necessary) to impose a
$\underline{k}_\perp$ cut-off from the outset and only remove it when a
gauge-invariant result is obtained.  We are now arguing that this cut-off
should not be removed until {\it after} the infra-red limit we consider has
been taken. Indeed we reached exactly the same conclusion in the last Section
by considering the validity of the multi-Regge approximation for production
amplitudes.

Our first conclusion is therefore that if we consider the limit in which the
vector-boson mass M given by (6.4) goes to zero, we should take limits in the
order
$$
i)~~ M \equiv g\langle\rho\rangle \rightarrow 0, \qquad ii)~~ \lambda
\rightarrow \infty ,
\auto
$$
where $\lambda$ is the transverse momentum cut-off.  This will correspond
to going around the phase-transition line in Fig.~7.1 by first going from the
Higgs to the confinement regime (at a finite value of the coupling) and then
taking the continuum limit in the confinement regime. As we have discussed this
procedure allows the appropriate string-like states to form before the
ultra-violet cut-off is removed.

An additional subtlety we should discuss is that there are two parameters
in the Higgs sector of (7.1), whereas in the lattice phase-diagram of
Fig.~7.1, there is only one parameter, that is $\langle\rho\rangle$.  This is
because lattice theories effectively fix the magnitude of the Higgs field and
so fix $h$ in the continuum theory.  We shall assume that we can exploit
complimentary and also decouple the Higgs field entirely in the limit {\it i)}
of (7.10) if we take $\langle\rho\rangle \rightarrow 0$ by taking

$$
-\mu^2 \rightarrow \infty \quad {\rm with} \quad {\mu^2\over h}
\rightarrow 0 \quad {\rm and} \quad {\mu^4\over h} \rightarrow \infty.
\auto
$$
Note that adding $h$ to any diagram in perturbation theory always involves
the addition of two accompanying propagators. Therefore, {\em if there is an
ultra-violet momentum cut-off}, such an addition multiplies the diagram by a
factor of
$$
\int{h\over (p_1^2-\mu^2)(p_2^2-\mu^2) }~~~~~
\centerunder{$\longrightarrow$}{$\scriptstyle\mu^2 \rightarrow \infty$}~~~
{h\over \mu^4}  ,
\auto
$$
and so decouples such a diagram in the limit (6.11). Consequently the Higgs
sector decouples entirely in this limit and we are left with only the massless
gauge field sector. (Note that that the large momentum region would, in
general, lead to a violation of (7.12) if there was no momentum cut-off).

\subhead{7.2 SU(3) Gauge Symmetry}

If we add to an SU(N) gauge theory R scalar fields which are N-tuplets under
the gauge group and then give expectation values to all such fields, the gauge
symmetry is, in general, broken\cite{cel} from SU(N) to SU(N-R).  To break
SU(3)
gauge symmetry completely, and also exploit complimentarity, we must therefore
add two triplets of Higgs scalars.  Giving the triplets distinct expectation
values produces {\em two} vector-boson mass-scales $M_1^2$ and $M_2^2$.  We
shall extract maximal information from our analysis if we separate the limits
in which these mass-scales go to zero.  That is we decouple the two triplet
scalars separately so that

$$
\eqalign{
\left.\matrix{\hbox{no gauge symmetry} & \cr
\hbox{+ SU(2) global symmetry}} \right\}  & \centerunder{$\longrightarrow$}
{$\scriptstyle M_1^2 \rightarrow 0$} SU(2) {\rm gauge\ symmetry} }
\auto
$$
$$
\centerunder{$\longrightarrow$}{$\scriptstyle M_2^2 \rightarrow 0$} SU(3) {\rm
gauge\ symmetry}.
\auto
$$

For reasons that we shall describe, we shall eventually prefer to remove the
$\lambda$ cut-off immediately after taking the first infra-red limit.
However, we can expect the additional infra-red limit(s) to be insensitive
to the cut-off only if the {\em complete spontaneously-broken theory} is
asymptotically free - so that a cut-off is effectively generated
dynamically.  For this and other reasons already discussed in Section 3, we
will eventually add the maximum number of quarks consistent
with the asymptotic freedom of QCD.  In this special case the theory with one
Higgs triplet that we obtain after the first limit (7.13) is taken, is indeed
asymptotically free in both the gauge coupling and the Higgs coupling.  We can
therefore take the limit $\lambda \rightarrow \infty$ at this stage and
hope that (an extended version of) complimentarity will then allow us to take a
limit of the form (7.11) in which $M_2^2 \rightarrow 0$, and smoothly obtain
unbroken QCD - with a very large number of quarks! (Of course, adding the
quarks necessarily involves us in the additional problem of chiral symmetry
breaking which we shall discuss shortly). In this special case, therefore,
we may be able to reach unbroken QCD by taking limits in the order
$$
i)~~ M_1^2 \rightarrow 0 \qquad ii)~~\lambda \rightarrow
\infty \qquad iii)~~M_2^2 \rightarrow 0 .
\auto
$$

When $M_1^2 \rightarrow 0$ and we have SU(2) gauge symmetry, the massive
and massless vector bosons have the group structure discussed in detail in
Section 11. There are two massive SU(2) doublets, which will not produce
physical vector particles, and one SU(2) singlet.  This singlet {\it will give
a physical particle} and if we have removed the ultra-violet cut-off, as in
{\it ii)} of (7.15), we can identify $M_2$ in {\it iii)} as the actual
{\it physical mass} of this vector particle.  As we shall discuss in later
sections, this vector remains reggeized in the infra-red limit $M_1^2
\rightarrow 0$.  Consequently we have a {\em reggeized, massive vector which
becomes massless}, and (presumably) {\it decouples from the physical spectrum},
as the full SU(3) gauge symmetry is restored. This already suggests, as we
noted in Section 3, a potential relationship between the restoration of the
full gauge symmetry of QCD and the limit giving the Critical Pomeron from the
Super-Critical Pomeron discussed in Section 7 of Part I.  However, it will
require considerable further analysis in the following sections to
illuminate all aspects of this relationship!

We now go on to discuss how it is that the infra-red divergences of reggeon
diagrams can have the physical significance required to give a confining
gauge theory. That is {\em the appropriate physical states} can be obtained
simply by considering the infra-red limits discussed above.  To this end we
briefly consider the infra-red divergences of transverse momentum diagrams in
QED.

\subhead{7.3 Divergences of QED Transverse Momentum Diagrams}

Because of the photon numerators, the box-diagram for electron scattering
gives (from (5.10))
$$
\sim \alpha^2s(\ln s -{\rm i}\pi) K(t,M^2),
\auto
$$
where $K(t,M^2)$ is given by (5.11) and M is now the photon mass.  Adding
the crossed diagram as in Fig.~7.2 gives

$$
\sim \alpha^2 i\pi K(t,M^2),
\auto
$$
$$
\centerunder{$\longrightarrow$}{\raisebox{-1mm} {$\scriptstyle M^2
\rightarrow 0$}} ~~2\alpha^2i\pi \ln[M^2/t],
\auto
$$
with each pole in $K$ contributing additively to the logarithmic divergence.

Higher-order infra-red divergences come only from the sum $\sum$ of photon
exchange diagrams of the form of Fig.~5.4, which also includes single photon
exchange.  In the Regge limit, this sum gives the sum of transverse momentum
diagrams shown in Fig.~5.5.  The sum $\sum$ is well-known to eikonalize
and as a consequence the infra-red divergences of the diagrams of Fig.~5.5
exponentiate to give

$$
\sum ~~~\centerunder{$\sim$}{\raisebox{-1mm} {$\scriptstyle M^2
\rightarrow \infty$}}~~~{\alpha\over t}
\exp \left[ i\alpha~ln(M^2/t) \right] ~~\equiv~~ {\alpha\over t}~
\hbox{``~$e^{i\theta}$~''}
\auto
$$
which demonstrates that the transverse momentum diagrams correctly
give\cite{cm} ``$\theta$'' the {\it infinite Coulomb phase} of QED.

In the $t$-channel (where $s$ is negative) the sum $\sum$ contains only real
radiation divergences which are well known to exponentiate and
give\cite{stw,yfs}

$$
\eqalign{
\sum~~~ \centerunder{$\sim$}{\raisebox{-1mm} {$\scriptstyle M^2
\rightarrow 0$}}~~~ &
\exp\left[\sum_{i,j=1,2} e_ie'_jp_ip'_j \int {d^4q\over (q^2 +
M^2)(p_i\cdot q)(p'_j\cdot q)}\right] \times {\rm finite\ part}}
\auto
$$
$$
\equiv \exp[R] \times {\rm finite\ part}.
\auto
$$
R is real and can, of course, be factorized and absorbed into
the definition of external electron states. Also\cite{stw} if R
is continued from the $t$-channel to the $s$-channel, the divergent Coulomb
phase is generated as the imaginary part of $R$, that is

$$
[R]~~~ \centerunder{$\longrightarrow$}{\raisebox{-2mm} {s
{}~ above~ threshold}}~~~[R] + i[\theta]
\auto
$$

It is well known that a photon mass can be added smoothly to QED without
producing a phase transition. It is also well-known that the factorization of
divergences (as the mass is removed) into eikonal factors absorbed into the
definition of external states as in (7.21) is equivalent to the modification of
the field operator for a charged fermion by the incorporation of a
line-integral of the form appearing in (7.6). If this line-integral extends to
infinity, then a gauge-invariant operator is obtained which simultaneously
creates a charged particle and a ``cloud'' of soft photons. Therefore we can
summarize the foregoing discussion by saying that in QED, transverse momentum
diagram divergences determine the divergent {\it $s$-channel imaginary-part} of
the ``non-local'' string-like factors needed to define gauge-invariant
$t$-channel states.

\subhead{7.4 Infra-red Divergences in Non-Abelian Gauge Theories}

{}From the point of view of Regge theory, which links $t$-channel bound-states
to $s$-channel high-energy behavior, it is natural that transverse momentum
divergences of $s$-channel high-energy diagrams determine $t$-channel states.
In
QED the result is a relatively trivial $s$-channel exponentiation which
reflects
the well-known necessity to define gauge-invariant states as including a cloud
of soft-photons. In a non-abelian theory, the reggeization of the gluons leads
to an entirely distinct ``$t$-channel exponentiation'' of divergences, that is

$$
s^{\alpha (t)}  ~\equiv~ s^{1 + (t - M^2) g^2/16\pi^2 K(M,t)}
\auto
$$
$$
\centerunder{$\sim$}{\raisebox{-2mm} {$\scriptstyle M^2 \rightarrow 0$}}
{}~~~s\exp[g^2/16\pi^2 t\ln s \ln (M^2/t)]
\auto
$$

The physical significance of this new exponentiation has, until this point, not
been well-understood. It is the purpose of the following Sections to analyze
it's ramifications in detail, using the full technology of the multi-Regge
theory developed in Part I and in previous Sections.  The major question will
be whether we can, under appropriate conditions, {\em link the divergences to
confinement}.

There is yet a further complication in attempting to extract the infra-red
divergence structure of a non-abelian theory from transverse momentum diagrams.
The infra-red growth of the gauge coupling (which is absent in QED) provides a
source of infra-red problems which we {\em can not analyse or control} in terms
of transverse momentum diagrams.  Of course, it is precisely this growth which
provides the basis of the conventional understanding of confinement in
non-abelian theories.  Not surprisingly, perhaps, we have to conclude that the
conventional confinement mechanism can not be studied within our formalism.

In fact the conventional confinement possibility is eliminated and
{\it transverse momentum divergences are the only divergence problem}
if we not only add the maximum number of quarks allowed by asymptotic freedom
(motivated by our desire to obtain an asymptotically-free spontaneously-broken
theory) but, as we discussed in some detail in Section 3, we also make
all quarks massless. This produces an {\it infra-red fixed-point} for the gauge
coupling. Consequently this coupling no longer grows in the infra-red region
and the conventional form of confinement definitely can not be realised.
However, we also anticipated in Section 3 that a related form of confinement
(involving the quark sea and the anomaly in a crucial manner) can take place.
In effect, the removal of the large distance growth of the gauge coupling
implies that the problem of confinement reduces to (the non-abelian version of)
the S-Matrix infra-red divergence problem. In this case we might anticipate
that, in parallel with our discussion of QED, complimentarity will allow some
approximation to the appropriate "string-like" states to emerge from the
infra-red limit - that is {\em from the infra-red divergences of reggeon
diagrams}. Clearly the question of chiral symmetry breaking is also reduced to
a
study of the infra-red limit. That is the divergences either do, or do not,
produce a particle spectrum consistent with chiral symmetry breaking.

We emphasize that if we demonstrate some form of confinement through
our analysis, then it will be in circumstances in which {\em the conventional
understanding of confinement does not hold}. Indeed we shall compare our
results with the various manifestations of confinement in the Schwinger model
in Section 11. It is also interesting at this stage to note Gribov's
argument\cite{gr1} that the infra-red growth of the gauge-coupling is
eliminated
in $QCD$ if quark loops are not Pauli-Villars regulated. Gribov also argues for
a close relationship between confinement and the quark sea anomaly in this
case.
In fact we shall discover in Section 9 that we are forced to not Pauli-Villars
regulate quark reggeon diagrams by our transverse momentum cut-off requirement.
It could be therefore that it is a natural extension of our procedure to assume
that the gauge-coupling simply does not grow in the infra-red region and that
although {\em it is essential that some massless quarks be present during our
analysis} we only need introduce a large number of flavors into our discussion
when we require asymptotic-freedom for the Higgs sector we have added.

For the moment we shall assume that the large number of massless flavors is
necessary to halt the infra-red growth of the gauge coupling. Therefore we
propose to study ``flavor-saturated QCD'' and to study the infra-red
divergences
of the reggeon diagrams obtained by adding two triplets of Higgs scalars and
using the Higgs mechanism.  We shall restore the gauge symmetry through the
sequence (7.13)--(7.14) with the goal of taking limits in the order (7.15). As
we shall see, it will be essential that (some) quarks remain massless during
the infra-red limit {\it i)} of (7.15).  If confinement is produced by the
infra-red divergences of limit {\it i)} as our analysis will suggest, then
quark-masses can be smoothly added after limits {\it ii)} and {\it iii)}
are taken.

\newpage

\mainhead{8. INFRA-RED ANALYSIS OF SU(2) REGGEON DIAGRAMS}

       In this Section we begin our infra-red analysis of reggeon diagrams.
Our ultimate aim is to discuss the sequence of limits (7.15) for QCD. However,
we shall initially study the simple limit $M^2 \to 0$ of the pure SU(2)
gauge theory diagrams. Much of our discussion will also apply to
the restoration of any gauge symmetry by a single limit in which all gluons
become massless and the Higgs sector decouples completely.

\subhead{8.1 Trajectory Function and Reggeon Interaction Divergences}

        It is straightforward to pick out the infra-red divergences of the
transverse momentum integrals (5.31)-(5.40) as the various propagators go
on-shell. To serve our general purpose we shall immediately describe these
divergences in the triple-reggeon diagram formalism outlined at the end of
Section 6.

        Consider first the trajectory function integral $\Delta(q^2)$ generated
by the reggeon diagram of Fig. 6.8 as discussed in Section 6. This diverges as
follows
$$\eqalignno{
\Delta &\left( \qbar^2\right)\quad\centerunder{$\sim$}
{$M^2\rightarrow 0$}\quad
{{-g^2\qbar^2} \over {(2\pi)^3}}\
\int {{d^2\kbar} \over {\left( \kbar^2 + M^2\right)
\left( (\qbar - \kbar)^2 + M^2\right)}}&\num\label{8.1}\cr
\sim & ln \ \left[ {{M^2} \over {\qbar^2}}\right]\
{{g^2\pi} \over {\left( 2\pi\right)^3}}
\left[ \qbar^2 \int {{d^2\kbar \delta^2(\kbar)} \over {(\qbar -\kbar)^2}}
+ \qbar^2  \int \
{{d^2\kbar \delta^2 (\qbar - \kbar)}
\over {\kbar^2}}\right]&\num\label{8.2}\cr
\sim & \ {{g^2} \over {\left( 2\pi\right)^2}} ln \left[ {{M^2} \over
{\qbar^2}}\right]&\num\label{8.3}\cr}
$$
Therefore the full divergence is the sum of the divergences resulting from
one propagator or the other going on-shell. We also see a feature that we
shall exploit extensively. The coefficient of the logarithmic divergence from
single transverse propagator is simply given by replacing the propagator by
a $\delta$-function. If we denote the divergence due to an on mass-shell
massless gluon by a dashed line, then the trajectory function divergence
is associated with the two diagrams of Fig.~8.1.

As we have discussed, the singular four-reggeon interaction is derived from
triple reggeon diagrams in the infra-red region and the infra-red divergences
when either of the propagators in (6.10) goes on shell are represented as in
Fig.~8.2. The resulting $\delta$-functions simply imply that transverse
momentum is conserved in the through-going reggeon lines, giving
$$
\eqalign{
V&\left( \kbar_1,\kbar_2,\qbar\right)\quad\centerunder{$\longrightarrow$}
{$M^2\rightarrow 0$}\quad
\left\{ \pi \delta^2(\kbar_1 - \kbar_2)\right.
\left[ \kbar_1^2 (\kbar_2 - \qbar)^2 + \kbar_2^2
(\kbar_1 - \qbar)^2\right]\cr
+&\pi \delta^2 (\qbar - \kbar_1 - \kbar_2)\left[ \kbar_1^2 \kbar_2^2 +
\left.(\qbar - \kbar_1)^2(\qbar-\kbar_2)^2 \right]\right\} ln M^2\cr
}
\auto
$$
Therefore the exchange diagrams of Fig.~8.2 generate logarithmic divergences
with identical transverse momentum dependence to those generated by the
self-interaction diagrams of Fig.~8.1. This motivates the incorporation of the
self-interaction diagrams in a full reggeon kernel defined as follows (we
include all momentum conservation $\delta$-functions in the definition and
normalize all two-dimensional integrals by $(2\pi)^{-3})$. As illustrated in
Fig.~8.3 we define
$$
\eqalign{
K_2^I (\qbar, \qbar^\prime, \kbar_1,\kbar_2)
&= (2\pi)^3 \delta^2(\qbar - \qbar^\prime) R_{2,2}^I
(\qbar, \kbar_1, \kbar_2) \cr
&+ (2\pi)^6 \delta^2 (\qbar - \qbar^\prime)(\kbar_1^2 + M^2)
((\qbar -\kbar_1)^2 + M^2)
\left[ \Delta(\kbar_1^2) \right. \cr
&+ \left.\Delta ((\kbar_1 - \qbar)^2) \right]
\left[ {{1} \over {2}} \delta^2\,
(\kbar_1 - \kbar_2) + {{1} \over {2}} \delta^2\, (\qbar - \kbar_1 - \kbar_2)
\right] \cr
}
\auto
$$
where $R^I_{2,2}$ is given by (6.9) and the remaining terms are given by the
self-interaction diagrams. Utilizing (6.10) and (6.11) together with (8.3)
we see that the $\log M^2$ divergences in $K^0_2$ directly cancel. This is a
central result which we write as
$$\eqalign{
K_2^0\quad&\centerunder{$\longrightarrow$}
{$M^2\rightarrow 0$}\quad
\tilde{K}^0_2 (\qbar, \qbar^\prime, \kbar_1, \kbar_2)\cr }
\auto
$$
where $\tilde{K}^0_2$ is a complicated scale-invariant distribution having the
f
orm
$$
\eqalign{
\tilde {K_2^0} ~~~\sim~ & \delta^2
(\qbar -\qbar^\prime)\delta^2(\kbar_1-\kbar_2)
\kbar_1^2 (\qbar-\kbar_1)^2 ln \left[ \kbar_1^2/(\kbar_1-\qbar)^2\right]\cr
& + \cdots \cr}
\auto
$$

{}From (6.11) we see that $C_I$ decreases as I increases and so for any
non-zero
I there is no cancellation of the $\log M^2$ divergences in $K^I_2$, that is
$$\eqalign{
K_2^I\quad&\centerunder{$\longrightarrow$}
{$M^2\rightarrow 0$}\quad \infty \hspace{1in} I \neq 0 \cr }
\auto
$$
Consequently for non-zero I, the reggeization divergences of the
self-interaction diagrams are not removed and for I$\geq$2 are actually
enhanced by the exchange diagrams.

There is also an important zero of $R^I_{2,2}$ when $M^2 \to 0$ which can be
thought of as {\em a consequence of gauge-invariance} and is vital for the
general infra-red divergence structure. First we note that
$$
\eqalign{
V(0,\kbar_2,\qbar)\quad&\centerunder{$\longrightarrow$}
{$M^2\rightarrow 0$}\quad
2\qbar^2\cr}
\auto
$$
Therefore since $a_I = - \frac{C_I}{2}$ it follows from (6.9) that
$$
R_{2,2}^I \left( \kbar_1=0, ~\qbar,~ M^2 = 0\right) ~~= 0
\auto
$$
and in fact
$$
\eqalign{
R^I_{2,2} \left( \kbar_1,\kbar_2,\qbar,M^2 = 0\right)\quad
\centerunder{$\sim$}
{$\kbar_1\rightarrow 0$}\quad
\left[ -4\qbar + 2\, {{\qbar^2} \over {\kbar_2^2}}\, \kbar_2 +
2 {{\qbar^2(\qbar-\kbar_2)} \over {(\qbar-\kbar_2)^2}}\right]\cdot \kbar_1\cr}
\auto
$$
In general this linear zero is sufficient to remove a $(\kbar^2_1)^{-1}$
divergence from a propagator - leaving only an integrable
$(\kbar^2_1)^{\frac{1}{2}}$ singularity.

The zero (8.10) can be understood as following from a Ward Identity for the
$\kbar_1$ gluon and (8.11) is, of course, the form of zero that we would
expect as a consequence of gauge invariance. There is, however, another way of
understanding the origin of the zero (8.10) in terms of the S-Matrix on
mass-shell formalism we are developing. This will be fundamental for our
discusion of the contribution of massless quarks, which begins in the next
Section. We consider a general reggeized gluon amplitude, as illustrated in
Fig.~8.4 and consider the behavior of the amplitude as $\kbar^2 \to 0$, where
$\kbar$ is the transverse momentum of one gluon. We consider reconstructing
this amplitude via the unitarity equation in the corresponding $t$-channel,
or equivalently use the $s$-channel unitarity plus dispersion relation
construction (5.49)-(5.53). In either case (as we have been emphasizing) we
reconstruct the reggeon amplitude via the contribution of on-shell
intermediate states, as illustrated in Fig.~8.4. It now follows that if the
reggeized gluon is massless and goes on shell it must decouple (that is the
zero (8.10) must appear) because a physical transition into the $t$-channel
gluon intermediate states is not allowed. We shall see that {\em this
argument does not go through if the $t$-channel intermediate states are
massless quarks}.

\subhead{8.2 The Infra-Red Finiteness of I = 0 Reggeon Diagrams}

Consider now the full four-reggeon amplitude $T^I_{2,2}$ defined by the
sequence of diagrams illustrated in Fig.~8.5 and which we can represent
formally
as
$$\eqalignno{
T^I_{2,2} (E,\qbar,\kbar_1,\kbar_2) &= K^I_2 + \int d\Omega_2 K^I_2
\Gamma^I_{2,2} K^I_2 + \cdots&\num\label{8.12}\cr
&\equiv K + K\Gamma K + K\Gamma K\Gamma K + \cdots&\num\label{8.13}\cr
}
$$
where
$$
\int d\Omega_2 = \int d^2\kbar d^2 \kbar^\prime \delta
(\qbar-\kbar-\kbar^\prime)
\auto
$$
and $\Gamma^I_{2,2}$ is the two-reggeon propagator
$$
\Gamma^I_{2,2} (E,\kbar, \kbar^\prime) = \left[ \kbar^2 + M^2\right]^{-1}
\left[ (\kbar^\prime)^2 + M^2\right]^{-1}
\left[ E-\alpha^\prime(\kbar^2 + M^2)-
\alpha^\prime((\kbar^\prime)^2 + M^2)\right]
\auto
$$
where, as discussed in Section 6, $\alpha'$ can be viewed as non-perturbative
in
origin or as originating from interactions above the transverse momentum
cut-off. ($\alpha'$ can equally well be set to zero in the following
discussion.)

For $\qbar$, $\kbar_1$, $\kbar_2$  non-zero, the zeroes of the form (8.11) are
sufficient to remove those divergences in $T^I_{2,2}$ which would result from
the propagator poles in $\Gamma^I_{2,2}$. In effect {\em a consequence of gauge
invariance is that, in off-shell reggeon amplitudes, there are no divergences
directly associated with reggeon intermediate states}. Divergences can come
only from the kernels $K^I_2$. Therefore, for I =0, (8.6) implies that
each term in the series (8.12) is finite as $M^2 \to 0$, while (8.8) implies
that for I non-zero each term in the series is infinite. From the definition
(8.12) it follows that
$$\eqalign{
T^I_{2,2}\quad\centerunder{$\longrightarrow$}{\centerunder
{$\kbar^2_1\rightarrow M^2$}{$(\kbar_1-\qbar)^2\rightarrow M^2$}}\quad
T^I_2\quad\centerunder{$\longrightarrow$}{\centerunder
{$\kbar^2_2\rightarrow M^2$}{$(\kbar_2-\qbar)^2\rightarrow M^2$}}\quad
T^I\cr
}
\auto
$$
where $T^I_2$ is the gluon-reggeon scattering amplitude appearing in (6.13)
and (6.14). The combination of (8.8) and (8.12) is sufficient to prove that
$T^0_2$ is finite (for non-zero $\qbar,~\kbar$) as $M^2 \to 0$.

$T^0$ {\em is not finite} because (8.10) no longer holds when $\kbar_2 = 0$.
However, if $T^0$ is regarded as defined by the sequence of diagrams in
Fig.~8.6, then it is clear from the representation of $T^0$ in terms of $T^0_2$
that the infra-red divergences can arise only from the last transverse momentum
integration performed. To discuss further how such divergences might be
eliminated and also what is the significance of the infinities in $T^I_{2,2}$
and $T^I_2$ for I non-zero, we need to consider the general relationship of
reggeon unitarity to the exponentiation of infra-red divergences.

\subhead{8.3 Exponentiation of Divergences and Reggeon Unitarity}

{}From (7.24) we already see that the infra-red divergences of $T^1$ produce a
simple exponentiation in rapidity space. This exponentiation can be
straightforwardly undone using the Fourier transform relation (6.7). This shows
that in the $E$-plane the series (7.24) corresponds to
$$
{{1} \over {tE}} \sum^\infty_{n=0} E^{-n}
\left[ {{g^2} \over {16\pi^2}} ln \left( {{M^2} \over {t}}\right)\right]^n =
{ {1} \over
{t\left[ E - {{g^2} \over {16\pi^2}} ln
\left( {{M^2} \over {t}}\right)\right]}}
\auto
$$

$$\eqalign{
\quad&\centerunder{$\sim$}
{$M^2\rightarrow 0$}~~\quad
{{1} \over {ln M^2}} ~~~\longrightarrow 0\cr }
\auto
$$
In this simple example we see how exponential suppression in rapidity space is
equivalent to inversion of the infra-red divergence in the $E$-plane amplitude
- leading to {\em complete removal of any $E$-plane singularity}. Removal of
the $E$-plane singularity implies, of course, that when the Fourier integral
over $E$ is carried out, the answer will simply be zero.

It is a general property of the reggeon unitarity equations ((5.45) and (5.46)
of I) that the nature of any reggeon singularity is inverted by iteration. (A
similar property holds for a conventional unitarity equation). We also
demonstrated in (6.20)--(6.24) that although the two-reggeon propagator was
eventually eliminated, reggeon diagrams can be utilised to reproduce the
contribution of the corresponding nonsense particle state which produces the
contribution $\Delta(\kbar^2)$ to the reggeon trajectory function. Consider
therefore the unitarity equation for the two-reggeon cut. This gives
$$
{\rm disc}\, T^I_{2,2} = \left[ T^I_{2,2}\right]^+-\left[ T^I_{2,2}\right]^- =
\int d\Omega_2 \delta\Gamma^I_{2,2} \left[ T^I_{2,2}\right]^+
\left[ T^I_{2,2}\right]^-
\auto
$$
where
$$
\delta \Gamma^I_{2,2} = \left( \kbar^2+M^2\right)^{-1}
\left( \kbar^\prime + M^2\right)^{-1} \delta \left[ E-\alpha^\prime
(\kbar^2 + M^2)- \alpha^\prime \left( (\kbar^\prime)^2 + M^2\right) \right]
\auto
$$

If we write formally
$$
\delta \Omega  = \int d\Omega_2 \delta \Gamma^I_{2,2}
\auto
$$
to denote the discontinuity due to the singularity of the two-reggeon
phase-space then (8.19) gives directly
$$
{\rm disc} \left[ T^I_{2,2}\right]^{-1} \sim \delta \Omega
\auto
$$
and so near the two-reggeon branch-point
$$
T^I_{2,2} \sim {{1} \over {C^I + \delta \Omega}}
\auto
$$
where $C^I$ is non-singular at this branch-point. It is straightforward to
extend this argument to give
$$
T^I_2 \sim {{C^I_1} \over {C^I + \delta \Omega}}
\auto
$$
and
$$
T^I \sim C^I_2 + {{\left( C^I_1\right)^2} \over {C^I + \delta \Omega}}
\auto
$$
where $C^I_1$ and $C^I_2$ are also non-singular at the branch-point. It follows
from (8.23)--(8.25) that any infra-red divergence will be inverted in E-plane
amplitudes if can be constructed via reggeon diagrams in such a manner that
the divergence originates within the two reggeon phase-space $\delta\Omega$.
Also since reggeon unitarity implies a similar form to (8.25) for any
reggeon cut it clearly follows that in general an infra-red divergence will
be inverted in E-space (that is exponentiated in rapidity-space) whenever it
originates from a (wrong-signature) nonsense state that can be directly
associated with one or more reggeon states.

However, we argued above that gauge invariance implies that physical reggeon
states (i.e. reggeon states in which signature does not eliminate the reggeon
propagator) do not give divergences. To reconcile the arguments, we note
that the divergence of $K^I$ given by (8.8) can be seen in terms
of (8.23)--(8.25) as follows. First we consider the compatibility of (8.13)
with (8.23). This is straightforward since (8.13) has the formal sum
$$
T^I_{2,2} \sim {{\hat{K}^I} \over {1 + \Gamma K^I}}
\auto
$$
where $\hat{K}^I$ is not integrated over phase-space and so is finite at a
general point. Therefore, formally
$$
T^I_{2,2} \sim {{1} \over { {{1} \over {K^I}} + \delta \Omega} }
\auto
$$
and so if we go to the two-reggeon branch-point with $M^2$ non-zero (8.23)
is valid. However, from (8.26) it is clear that for {\em non-zero} $I$
$$
\eqalign{
T^I_{2,2} \sim T^I_2 \sim T^I\quad&\centerunder{$\sim$}
{$M^2\rightarrow 0$}\quad
{{1} \over {K^I}} \longrightarrow 0\cr
}
\auto
$$
and so all amplitudes with non-zero $I$ vanish as $M^2\to 0$.

To derive (8.28) from reggeon unitarity we have to view the divergences
within $K^I_2$ as originating directly from (wrong-signature) {\em
three-reggeon
contributions} that are cancelled as angular-momentum plane singularities by
nonsense-zeroes, but remain as nonsense-state transverse momentum
singularities. In this case (8.28) can be interpreted as due to
representations of the form of (8.23)--(8.25) but with $\delta\Omega$ due to
the three-reggeon state so that $\delta\Omega \sim K\to \infty$. As a
general property we conclude that whether we regard a divergence as occuring
within a reggeon interaction or as due to a wrong-signature reggeon state it
will be inverted because of reggeon unitarity and will eliminate any
physical reggeon state to which it couples.

There is, nevertheless, a further subtlety in drawing conclusions from
(8.23)-(8.25) which is crucial for the properties of the $I=0$ channel.
First we note that if $T^0$ is defined by the series of diagrams in
Fig.~8.6, then the only angular-momentum plane singularity of each term is
the two-reggeon cut. Consequently $T^0$ can be written as an integral over
the two-reggeon discontinuity, that is, as illustrated in Fig.~8.7,
$$
{\rm disc}\, T^0 = \int d\Omega_2 \delta \Gamma^0_{2,2}
\left[ T^0_2\right]^+ \left[ T^0_2\right]^-
\auto
$$
and since $T^0_2$ is infra-red finite, the divergence of $T^0$ can arise only
when we evaluate the discontinuity. Apparently, the infra-red divergence of
$T^0$ can be viewed as due to the two-reggeon state in general
(specifically, of course, it is due to the one-particle pole in
$\delta\Gamma^0_{2,2}$) rather than the last (or first) loop appearing in
Fig.~8.6. From this last viewpoint, (8.25) appears to imply that the
divergence of $T^0$ should be inverted. However, the argument fails in this
case because because the zero (8.10) in $R^0_{2,2}$ implies that the basic
four-reggeon coupling {\em vanishes} at the divergence point. In the notation
of (8.27), this implies that $K^I$ vanishes and in (8.23)
$$
C^0~~ \gsim~~ \delta \Omega
\auto
$$
The finiteness of $T^0_2$ implies also that $C^0_1 \sim C^0$ and so
$$
T^0~~ \sim C^0_1~~ \gsim~~ \delta\, \Omega
\auto
$$
Consequently, this divergence will be inverted {\em only if there is an $I =
0$ four-reggeon interaction which does not vanish} at the divergence point -
that is an $I = 0$ interaction which {\em does not have the zero} (8.10). In
effect the divergence occurs because the two-reggeon coupling to particle
gluon states does not satisfy the constraints of gauge invariance. To invert
it we must find a four-reggeon interaction which also violates the constraints
of gauge-invariance. To do so we shall need to discuss the properties of
massless quarks in some detail.

\subhead{8.4 General Reggeon Kernels}

It is straightforward to anticipate the generalization of the above discussion
to a general N-reggeon state. We define the kernel for such a state as the full
set of reggeon diagrams describing its propagation that are
{\em irreducible} with respect to the state. This {\em includes the reggeon
self-interaction diagrams} and, as for $K^0_2$, also includes all relevant
momentum conservation $\delta$-functions. For the present we shall effectively
assume that the reggeons involved are reggeized ($\alpha^\prime$ is non-zero)
without the inclusion of the self-interaction diagrams but they could as well
be elementary massless gluons. Within the general triple-regge formalism all
infra-red divergences originate from the particle pole in the reggeon
propagator
(6.2). We always refer to the corresponding massless particle as a gluon and
use the dashed line notation of Figs.~8.1 and 8.2. The structure of
higher-order divergences is then determined by the combination of gluon
divergences and the nonsense-zero of the triple-reggeon vertex. That is, in the
infra-red region the nonsense-zero cancels the gluon pole associated with the
distinct single leg of the vertex whenever the gluon poles associated with the
other two legs both produce divergences. The divergences of some higher-order
diagrams are illustrated in Fig.~8.8.

Denoting the kernel for an N-reggeon state with t-channel isospin as $K^I_N$
we generalize the last sub-section to

$A$ -- $K^0_N$ {\em has a finite limit $\tilde{K}^0_N$ as $M^2\to 0$}

$B$ -- $K^I_N$ {\em is infra-red divergent due to uncanceled self-interaction
divergences (I~>~0).}

Generalizing the above discussion of the two-reggeon state we clearly expect
property $B$ to lead to a generalization of the exponentiation of (7.24) to all
channels with non-zero isospin $I$. We also expect property $A$ to lead to
finite, $I = 0$,  N-reggeon scattering amplitudes $T^0_{N,N}$ and also finite
two-particle $\to N$-reggeon amplitudes $T^0_N$. There will be {\em divergent}
N-reggeon cut contributions (generalizing (8.29)) to $T^0$. This last set of
divergences implies that the massless limit does not exist for particle
scattering amplitudes.

We can summarize the situation by saying that the decoupling (8.10) softens
the infra-red contribution of all gluon reggeon states to the point that there
are no infinities coming {\em directly from reggeon states} in reggeon
scattering amplitudes. For $I$ non-zero, the divergent kernels $K^I_N$
nevertheless remove all such reggeon states - via the corresponding
generalization of (8.26)-(8.28). For $I = 0 $, the finiteness of the $K^O_N$
allows the corresponding reggeon states to survive. As a result all amplitudes,
even those that are finite, have singularities due to
gluon reggeon states at zero momentum transfer. Consequently not only is
{\em there is no sensible massless limit--even in the $I = 0$ channels} for
particle scattering amplitudes, but there are also gluon reggeon states
remaining in all reggeon amplitudes i.e. there is {\em no confinement}.

\subhead{8.5 Scale-Invariance and the Lipatov Pomeron}

Since the kernel $K^0_2$ defined by (8.5)--(8.7) is dimensionless, it follows
that it, and indeed all the $K^0_N$, must be scale-invariant functions
(distributions). If we temporarily ignore the transverse momentum cut-off
$\lambda$ that we have argued we should impose, we can study the Fredholm norm
of the massive kernel $K^0_2$. In fact
$$\eqalignno{
||K^0_2||^2 &= \int \prod_i {{d^2\kbar_i} \over {(\kbar^2_i + M^2)}} K^0_2
(\kbar_1, \kbar_2, \kbar_3, \kbar_4, M^2)\cr
&\sim {{\int} \atop {|\kbar^2_i|>\lambda}} \prod_i
{{d^2\kbar_i} \over {\kbar^2_i}} \tilde{K}^0_2 (\kbar_1, \kbar_2, \kbar_3,
\kbar_4)&\num\label{8.32}\cr
&\equiv \infty&\num\label{8.33}\cr
}
$$
where the infinity comes from the {\em ultra-violet} region
$$
|\kbar_1|^2 \sim |\kbar_2|^2 \sim |\kbar_3|^2 \sim \kbar_4|^2 \longrightarrow
\infty
\auto
$$
and is directly due to the scale-invariance of $K^0_2$. In general
{\em without the transverse momentum cut-off} the $K^0_N$ all have infinite
norm and are {\em non-Fredholm kernels}.

A Fredholm kernel generates only Regge poles in the angular momentum plane,
whereas the ultra-violet divergence of a non-Fredholm kernel, in general,
produces fixed (that is t-independent) singularities. Indeed when the iteration
of Fig.~8.5 is carried out for $K^0_2$, {\em without the cut-off} $\lambda$, it
is well-known that a branch-point - the ``Lipatov Pomeron'' - is
generated\cite{lnl} at
$$
E = -g^2 (2 ln 2)/~\pi^2 ~~~\equiv~~~ j = 1 + g^2 (2 ln 2)/~\pi^2
\auto
$$
The corresponding high-energy behavior
is
$$ \eqalign{
\sigma_T\quad&\centerunder{$\sim$}
{\raisebox{-2mm} {$s\rightarrow \infty$}}\quad
s^{2g^2ln2/\pi^2}}
\auto
$$
and so the Froissart bound is violated.

The fixed-cut is generated by the infinite sum of (5.51) when (6.12) is
inserted
because as $n$ increases the average transverse momentum contributing grows.
Consequently, for the reasons discussed in sub-Section 6.3, the approximation
to the individual production processes given by (6.12) becomes increasingly
worse with increasing $n$ (that is at large transverse momentum the amplitudes
summed are down by powers of the sub-energy from the leading behavior). In
effect an infinite sum of production processes with individually small
amplitudes produces a violation of unitarity.

It is often proposed that (8.36) be used as a starting-point for a study of the
Pomeron in QCD by, for example, coupling the series of reggeon diagrams to
some specifically chosen $I = 0$ external states, such as a heavy quark pair
coupling to a photon\cite{bl}. The couplings for such states will, because
of gauge-invariance, contain zeroes which remove the remaining infra-red
divergences in $T^0$. The fixed-cut (8.35) can be argued\cite{lnl} to become an
accumulation of Regge poles if asymptotic freedom and the running of the
gauge coupling are incorporated . As we discussed in the Introduction, this
might be a legitimate way to study the so-called ``hard Pomeron'' corrections
to short-distance perturbative $QCD$\cite{lnl,glr,bl,ahm}. As a matter of
principle, however, the amplitudes obtained can not represent the physical
or ``soft'' Pomeron in a confining gauge theory. Although they are infra-red
finite, they remain {\em singular at zero transverse momentum because of the
persistence of gluon reggeon intermediate states}. Equivalently, the zero of
(8.11) is sufficient to remove divergences but not to produce amplitudes
which are analytic at zero transverse momentum - as should be the case for a
confining theory with {\em no massless particles}.

There is an even more serious problem with using (8.36) as a starting-point for
studying the Pomeron. The iteration of the higher-order kernels $K^0_N$ will
generate further branch-points in the $E$-plane at $E\sim -g^2N$ which produce
stronger and stronger violations of the Froissart bound. [In SU(3) gauge theory
this includes odd-signature, ``Odderon'', branch-points with\cite{lnl} higher
intercepts than the Pomeron!!] Clearly this singularity structure  bears no
resemblance to the multiperipheral Pomeron discussed in Part I which evolved
phenomenologically over the years and was the basis of the Reggeon  Field
Theory. Indeed our whole development of multi-Regge theory as a technology for
controlling the low transverse momentum unitarity properties of the Pomeron
would be useless if the Pomeron in QCD is really the ultra-violet dominated
multi-singularity object suggested by the present discussion.

         In our opinion the reason for the emergence of this unwelcome
structure
is that the construction ignores the requirement of complementarity that we
must
impose a transverse momentum cut-off to reach a confining gauge theory in the
limit $M^2\to 0$. We have already noted that the reggeon diagram formalism
itself contains this same message in that the multi-Regge production amplitudes
involved are extremely small at large transverse momentum and are down by
powers of the energy with respect to neglected amplitudes. Therefore if no
transverse momentum cut-off is imposed the multi-Regge amplitudes do not
represent a sensible first approximation to the production processes they
describe. We emphasize, therefore, that {\em the ``Lipatov Pomeron'' is
produced} (in a spontaneously-broken gauge theory) {\em by summing a very
large number of very small amplitudes in regions of phase-space where they
make no physical sense as an approximation}. Reggeon diagrams are a sensible
formalism only at small transverse momenta!

As we noted in the last Section, complementarity further implies
that, if we do keep a transverse cut-off, then the states of the theory should
be {\em selected} by the infra-red divergence structure. At this stage our
infra-red analysis actually implies that we have no justification for
considering {\em only} bound-state amplitudes in that gluon (and quark)
particle amplitudes are actually infinite with respect to such amplitudes.
Indeed we do not yet have a sensible limit for reggeon diagram amplitudes in
general. Clearly what we would like to understand is how the low transverse
momentum regions where the reggeon diagram formalism makes sense can dominate
the physics and yet not lead to unphysical divergences. In the next Section
we shall argue that the presence of massless quarks is a vital ingredient for
this to happen.

\newpage

\mainhead{9. MASSLESS FERMIONS IN REGGEON DIAGRAMS}

As we described in Section 4, all fermions (quarks) lie on Regge trajectories
in a spontaneously broken gauge theory. The reggeization of the electron in
(massive) QED has also been demonstrated\cite{mw} up to very high-order in the
leading log approximation and the next-to-leading order modification of the
trajectory function (6.30) has even been calculated\cite{as}.

\subhead{9.1 Quark Reggeons}

Much of the formalism for gluon reggeons described in the last two Sections has
also been developed for quark reggeons. There is an extensive treatment in
\cite{fs1} in particular. The reggeization can again be derived as the outcome
of a multiperipheral equation as illustrated in Fig.~9.1. That is, the
multi-Regge production of quarks and gluons, with both quarks and gluons
reggeized, self-consistently produces the reggeization of the quarks. (Of
course, in massive QED the photons are not reggeized.) The set of gluon and
quark production amplitudes represented in Fig.~9.1 is part of the complete
set of leading-log multi Regge pole amplitudes defined as follows.

Consider a specific non-abelian gauge theory in which all gluons are massive
(from the Higgs mechanism) and all quarks are also (initially) massive. We then
consider the complete set of multi-Regge limits defined by Toller diagrams as
in Part I. The leading-log result for such amplitudes will be zero unless all
t-channels carry either gluon or quark quantum numbers. For the non-zero
amplitudes the result will correspondingly be reggeized gluon or quark
exchange in each such t-channel. A generalization of the multiperipheral
bootstrap illustrated in Fig.~9.1 implies that when these amplitudes are
inserted into the direct-channel unitarity equations, in the form of the
asymptotic dispersion relations described in Part I, then the leading-log
approximation will be self-consistently reproduced for the non-zero amplitudes.

The trajectory function for a reggeized quark differs from that of the electron
only by a group-theoretic factor, that is $\alpha(q^2) = \frac{1}{2} +
\Delta(q^2)$ where
$$
\Delta \left( \qbar^2\right) = \left( \qbarsl -m\right)G {{g^2}
\over {(2\pi)^3}}
\int {{d^2\kbar} \over {\left( \kbarsl - m\right)
\left(\left( \qbar - \kbar\right)^2+ M^2\right)}}
\auto
$$
and for SU(N) gauge theory
$$
G = \left( N^2-1\right)/2N
\auto
$$
Reggeon diagrams containing reggeized quarks will involve the propagator
$$
{\st \Gamma} \left( E,\qbarsl \right)= {{1} \over {(\qbarsl -m)}}\
{{1} \over {\left( E-{{1}\over {2}} -\Delta (\qbar^2) \right)}}
\auto
$$
Note that
$$\eqalign{
\Delta\left( \qbar^2\right)\quad\centerunder{$\longrightarrow$}
{$m\rightarrow 0$}\quad
{{g^2G} \over {(2\pi)^3}}\  \qbarsl\ \int
{{d^2\kbar} \over {\kbarsl \left( (q-k)^2 + M^2\right)}}\cr
}
\auto
$$
which is finite. Therefore quarks remain reggeized as the massless limit is
taken (with all gluons still massive). In general the massless quark limit is
smooth and does not disrupt the basic Regge properties of the theory.

The insertion of the leading-log multi Regge pole amplitudes into the unitarity
equation (via the asymptotic dispersion relations) also generates
next-to-leading log approximations for amplitudes that are zero in the
leading-log approximation. In many cases these amplitudes will contain
Regge cut contributions involving quark-reggeons. For example, quark-gluon
Regge
cuts will be generated in channels involving quark quantum number exchange, as
illustrated in Fig.~9.2, but with {\em opposite} signature to the quark
reggeon.

The exchange of two reggeized quarks (or a quark/anti-quark pair) produces a
Regge cut at
$$
E = 1 + 2 \Delta \left( \qbar^2/4\right)
\auto
$$
Therefore amplitudes involving this exchange will be down by a full power of
the energy involved compared to those involving gluon exchange.
Consequently two quark exchange can, in general, be isolated in a well-defined
way only in those channels where a flavor quantum number precludes gluon
exchange. In such a channel there will be a multiperipheral
equation\cite{jk}, of the form illustrated in Fig.~9.3, with a reggeon
interaction $R^Q$ generated by the two reggeized quark/gluon vertex. This
vertex has the form (in the notation of Fig.~9.4)
$$
\Gamma_\mu = \tilde{G} \left[ \gamma^\mu - (m-\kbarsl_1)
{{P^\mu_B} \over {k_2\cdot P_A}} - (m - \kbarsl_2)
{{P^\mu_A} \over {k_1\cdot P_B}}\right]
\auto
$$
The resulting $R^Q$ has two distinct components
$$
R^Q = R^Q_0 + \tilde{R}^Q
\auto
$$
where $R^Q_0$ results from the product of $\gamma^{\mu}$ factors in (9.6) and
$\tilde{R}^Q$ has a similar structure to the gluon vertex $V(q,k_1,k_2)$, that
is - in the notation of Fig.~9.5
$$
\tilde{R}^Q \sim ~~~~
{{(\kbarsl_1 + m)(\qbarsl - \kbarsl_2 + m) + (\qbarsl - \kbarsl_1 + m)
(\kbarsl_2 + m)} \over {(\kbar_1 - \kbar_2)^2 + M^2}}
\auto
$$

Infra-red divergences due to the denominator in (9.8) are cancelled in the
limit $M^2 \to 0$ by analogous trajectory function divergences {\em if and
only if} the quark/antiquark state carries zero color ($I = 0$). $R^Q_0$ is an
infra-red finite four (reggeized) quark vertex which will play an important
role
in the following. Consider first the series of diagrams produced by iterating
$R^Q$ through the multiperipheral equation of Fig.~9.3. The resulting
transverse
momentum integrals are (just) divergent if we include the logarithmic large
$k^2$ behavior of the trajectory function $\Delta(k^2)$ in the denominator of
the quark reggeon propagator. If this denominator is expanded out and the
contribution of $\Delta(k^2)$ combined with the singular part of the kernel to
produce an $I = 0$ infra-red finite kernel as we did for the gluon interaction,
then this kernel will be dominated by $R^Q_0$ at large transverse momentum.
The iteration of the kernel through the equation of Fig.~9.3 then produces a
series containing {\em ultra-violet} divergent transverse momentum integrals of
the form
$$
\int d^2\kbar ~~\gamma_\mu (\kbarsl)^{-1}\ \gamma_\mu (\kbarsl)^{-1}
\auto
$$
$$
\sim \int {{d^2\kbar} \over {\kbar^2}}
\auto
$$

This divergence potentially produces an additional $\log s$ with each iteration
of the multiperipheral equation (the well-known `double-logs' for
fermion-antifermion channels). The divergence is {\em not} eliminated by taking
only the `leading threshold' behavior in analogy with the example discussed in
Section 5. It {\em is removed} by including the ultra-violet running of an
asymptotically-free gauge coupling. In contrast to the $I = 0$ gluon
channels\cite{bs} the infrared finiteness of quark/antiquark reggeon
diagrams does allow a running gauge coupling to be legitimately introduced
to eliminate the divergence (9.10) and, as we shall discuss further later,
bound-states directly related to the quark/antiquark Regge cut generated. To
avoid the ``non-Regge'' behavior generated by (9.10), at the stage when we
do not have asymptotic freedom, it is clearly imperative that we keep a
transverse momentum cut-off (or equivalently - keep a non-zero
$\alpha^{\prime}$ in the quark reggeon propagator). The need for a
transverse momentum cut-off is therefore apparent at an even earlier stage
than we encountered it in the pure gluon case.

Because of (9.5), quark loops have no influence on the reggeized gluon channels
at either the leading or next-to-leading log level. At sufficiently non-leading
log level they will contribute to reggeized gluon interaction vertices.
However,
such contributions are difficult to isolate unless there are distinctive
quantum numbers involved. At first sight this seems to be a problem for our
general purpose. We wish to include quark loop effects in our analysis, but
this will be difficult if we can not describe and manipulate such vertices
within the framework of ``Analytic Multi-Regge Theory''. That is if they can
not
be written in terms of gauge-invariant transverse momentum (or reggeon)
diagrams
which are built from basic reggeon interaction vertices. This requirement is
necessary, firstly to implement a transverse momentum cut-off consistently, and
secondly so that we can build up the complete set of diagrams using the general
formalism of Part I. Indeed, as we have already implied, we ultimately wish to
argue that Regge region quarks contribute to the infrared structure of
reggeized gluon vertices in a manner which critically modifies the analysis of
the last Section.

Before discussing quark loop contributions to gluon vertices in detail it will
be instructive to briefly review the contribution of electron loops to
high-energy (massive) QED. In this case there are well-defined leading-log
contributions but they do {\em not} satisfy our requirements. We shall find
that
there is a {\em non-abelian} component of a quark loop which can be isolated
as we want and is essential for our purpose.

\subhead{9.2 The Tower Diagrams in QED}

Since the photon has no self-interaction the ``leading-log'' diagrams in the
vacuum exchange channel of QED necessarily involve electron loops. The
`tower-diagrams' in Fig.~9.6 involve photon pairs coupled by the (minimal)
gauge-invariant set of electron loop diagrams shown. Each photon pair exchange
produces a factor of $lns$ (apart from the first pair) together with a
transverse momentum integral. The result is\cite{cw} the set of transverse
momentum diagrams shown in Fig.~9.7, with the electron loops producing the
four-photon interaction
$$
\eqalignno{
\hat{V} (\kbar_1,\kbar_2,\qbar) &= \int \, d^2 \pbar \ \int^1_0
{{dx} \over {x(\pbar + \qbar + \kbar_1)^2 + (1-x)\pbar^2 +m^2}}\cr
&\times \left\{ {{(\pbar^2+m^2) Tr \left[ (\pbarsl-\qbarsl+\kbarsl_1+m)
(-\pbarsl-\qbarsl-\kbarsl_1 + m)\right]}
\over
{x(\pbar-\qbar+\kbar_1)^2 + (1-x)\pbar^2 + m^2}}\right.&\num\label{9.11}\cr
&- Tr \left. {{\left[ (-\pbarsl-\qbarsl-\kbarsl_1 +m)
(-\kbarsl_2+\kbarsl_1 + \pbarsl +m)
(-\qbarsl + \kbarsl_2-\pbarsl +m)(\pbarsl +m)\right]}
\over
{(1-x)(\pbar + \qbar - \kbar_2)^2 +
x (\pbar + \kbar_1 - \kbar_2)^2+m^2}} \right\} \cr
}
$$

{}From our perspective, there are a number of points of principle involved in
the derivation of this interaction. First we note that the $x$-integration
is over the relative magnitude of the light-cone momenta of the electrons in
the loop. Consequently {\em the Regge region for electron exchange does not
contribute distinctively}. There is also a vital subtraction\cite{cw} that
has to be made to obtain (9.11) via the light-cone procedure outlined in
Section 5. The subtraction is necessary if the full four-photon amplitude
produced by the sum of the electron loop diagrams is to satisfy the
Ward Identity that we discussed in the last Section for the four-gluon
amplitude. The subtraction is made by introducing a four-photon vertex defined
as (minus) the four-photon amplitude at zero momentum and is equivalent to
Pauli-Villars regularization of the ultra-violet region of the electron
(Feynman) diagrams. The resulting Ward Identity leads directly to the
following property of (9.11), that is
$$
\hat{V} (\pm \qbar,\kbar_2,\qbar) = \hat{V} (\kbar_1, \pm \qbar, \qbar)=0
\auto
$$
This is the same as the property (8.10) satisfied by the four-gluon vertex
$R^I_{2,2}$. That is $\hat{V}$ shares with $R^I_{2,2}$ the zero structure that
we have identified as playing a major role in the non-inversion (or
non-exponentiation) of remaining infrared divergences. $\hat{V}$ also has
essentially the same scale-invariance properties as the gluon kernel described
in sub-Section 8.5. Indeed it is well-known that the iteration of $\hat{V}$ via
the transverse momentum diagrams of Fig.~9.7 produces a fixed branch-point
in the angular momentum plane very similar to (8.33) and similarly violates the
Froissart bound.

The violation of the Froissart bound by massive QED is probably more
fundamental
than the corresponding non-abelian result. For the non-abelian case we shall
eventually find high-energy behavior consistent with unitarity after we have
understood the origin of confinement in the infrared singularities of the
theory. This possibility is not available for the abelian theory and the
violation of unitarity is probably directly coupled to the inconsistency of the
theory at short distances and to the related non-reggeization of the photon.

\subhead{9.3 Quark Loops in a Non-Abelian Theory}

There will surely be a quark-loop contribution to the four reggeized-gluon
vertex which is very similar to (9.11). However, as we have discussed, it will
be buried in non-leading contributions and not distinctively isolated. If it is
derived from the appropriate Feynman diagrams in analogy with the QED
derivation
of Cheng and Wu\cite{cw} then it will have the properties described above
and so will not significantly modify the properties of $R^0_{2,2}$. We must
empasize, however, the point of principle already alluded to. We have argued
consistently throughout this article that we want to study infrared behavior
in the presence of a transverse momentum cut-off. It is clear that since
(9.12) results from a subtraction procedure at large momentum, it is
potentially in conflict with the imposition of any such cut-off. If we
explicitly needed the non-abelian analog of (9.11) in our analysis then
whether or not there was such a conflict would be a major issue that we
would have to resolve. Fortunately this is not the case, a non-abelian gauge
theory has additional structure which resolves the matter.

In SU(2) gauge theory there is, in particular, a next-to-leading log
contribution to $R^0_{2,2}$ which is well-defined in terms of quark transverse
momentum diagrams. We shall find this contribution first by examining Feynman
diagram contributions that might be involved. We shall then describe how the
vertex can be constructed from the general reggeon diagram construction
procedure outlined at the end of Section 6. This will lead us to understand
some key general properties and principles that are involved.

Consider first the two-reggeon cut in the quark/antiquark exchange channel
of a four gluon reggeon amplitude as illustrated in Fig.~9.8. As a reggeon
singularity the quark/antiquark state carries {\em odd} signature and can carry
both $I = 0~and~I = 1$, even if we require $I = 0$ in the overall $t$-channel.
(A property that potentially distinguishes it from {\em any odd-signature gluon
contribution}). However, the vertex for gluon $\to$ gluon + quark/antiquark
pair through which it couples is necessarily a nonsense vertex (that is odd
signature $\to$ odd signature + odd signature) and so could not appear as a
reggeon vertex in signature-conserving reggeon diagrams describing elastic
scattering. Therefore, Fig.~9.8 is well-defined as a component of a
{\em reggeon} diagram only in multiparticle amplitudes. Nevertheless, in
analogy with our discussion in Section 6 of both the gluon reggeon loop
contribution to reggeization and gluon reggeon exchange in the four reggeon
vertex, we can ask whether Fig.~9.8 can contribute in elastic scattering as a
transverse momentum threshold diagram. That is while a nonsense-zero will
cancel the quark/antiquark reggeon propagator in reggeon diagrams, the
presence of non-abelian group-factors can potentially produce an {\em
``even signature'' nonsense-state threshold} in the four reggeized gluon
vertex. The need for the group factors to contribute determines that only
the $I = 1$ component will be involved.

To find potential Feynman diagrams that might contribute we note that the
lowest-order contribution to the gluon $\to$ gluon + quark/antiquark vertex
is a single quark intermediate state. Therefore the simplest reggeon diagram
that could contribute is that shown in Fig.~9.9. The simplest candidate Feynman
diagram contributing to this reggeon diagram is then the planar diagram of
Fig.~9.10. However, if we consider the spin structure of the transverse
momentum diagram generated we find that the diagram of Fig.~9.10 can not
actually contribute.

In analogy with the numerator contribution (5.26) to Fig.~5.6, the left-side
gluons of Fig.~9.10 will give a $\gamma^-$ (say) coupling to the quark loop,
while those on the right-side will give a $\gamma^+$ coupling. Thus, in the
notation shown, the numerator for the quark loop of Fig.~9.10 will be
$$
Tr \left[ \gamma^- (\qbarsl + \pbarsl +m)\gamma^+(\pbarsl - \kbarsl_2 +m)
\gamma^+(\pbarsl-\qbarsl +m)\gamma^-(\pbarsl - \kbarsl_1 +m)\right]
\auto
$$
$$
=0
\auto
$$
since one of the $\gamma^+$'s (or the $\gamma^-$'s) can clearly be commuted
through a transverse numerator to annihilate against the other.

\subhead{9.4 The Four Gluon Reggeon Vertex}

To obtain a non-zero contribution to the reggeon diagram of Fig.~9.8 we must
consider a non-planar Feynman diagram involving the quark loop of Fig.~9.11.
The alternation of $\gamma^+$ and $\gamma^-$ couplings around the loop will
avoid the vanishing of (9.14). Utilizing the identity
$$
\eqalign{
Tr &\left[ \gamma^- (\st{A}_\perp +m)\gamma^+ (\st{B}_\perp +m)\gamma^-
(\st{C}_\perp +m)\gamma^+ (\st{D}_\perp +m)\right]\cr
&= 8Tr \left[ (-\st{A}_\perp +m)(\st{B}_\perp +m)
(\st{C}_\perp -m(\st{D}_\perp+m)\right]\cr
}
\auto
$$
and assuming (for the moment) that the horizontal quark propagators in
Fig.~9.11 are removed by the process of reduction to a transverse momentum
diagram, we obtain for the simplest contribution to Fig.~9.8 (in the notation
shown)
$$\eqalign{
V_D &(\qbar,\kbar,\kbar^\prime) \sim I(\qbar,\kbar,\kbar^\prime)\cr
&= \int d^2p\ {{Tr\left[ (-\pbarsl + \qbarsl +m)(\pbarsl + \kbarsl +m)
(-\pbarsl + \qbarsl +m)(\pbarsl + \kbarsl^\prime +m)\right]}
\over {\left[ (p+q)^2 +m^2\right]\left[(p-q)^2 +m^2\right]}}
}
\auto
$$
which is divergent. As in the examples discussed in Sections 5 and 6 we
anticipate that the appropriate convergent integral is obtained by keeping only
the leading behavior of the numerator at the threshold produced by the
denominators. At first sight there is an ambiguity as to how to proceed, since
there are a number of identities, analogous to (5.28), that we can exploit. We
can use any of
$$
(-\st{p} + \st{q} + m)(\st{p}+\st{k} +m) = (-\st{p}+\st{q}+m)(\st{k}+\st{q})
+4\left[(p-q)^2 +m^2\right]
\auto
$$
$$
(\st{p}+\st{k}+m)(-\st{p}-\st{q}+m) = (\st{k}-\st{q})(-\st{p}-\st{q}+m)
+4 \left[(p+q)^2+m^2\right]
\auto
$$
$$
(-\st{p}-\st{q}+m)(\st{p}+k^\prime +m) = (-\st{p}-\st{q} +m)
(\st{k}^\prime-\st{q}) +4 \left[(p+q)^2 +m^2\right]
\auto
$$
$$
(\st{p}+\st{k}^\prime +m)(-\st{p}+\st{q} +m) = (\st{k}^\prime +\st{q})
(-\st{p} +\st{q} +m) +4 \left[(p-q)^2 +m^2\right]
\auto
$$

Indeed various combinations of (9.17) -- (9.20) can be used to extract a
convergent integral from (9.16). A symmetric procedure (which we shall
shortly argue gives the correct result) is to apply separately (9.17) and
(9.18) and then average, and similarly with (9.19) and (9.20). The result is
the symmetric and convergent vertex function
$$
V_c(q,k,k^\prime) \sim \int d^2p\
{{Tr\left[ (-\st{p}+\st{q}+m)\st{k} (-\st{p}-\st{q} +m)\st{k}^\prime \right]}
\over {\left[(p+q)^2 +m^2\right]\left[(p-q)^2 +m^2\right]}}
\auto
$$

We notice first that $V_c$ is odd under $k \to -k$ or under
$k^\prime \to -k^\prime$. This reflects the essential odd signature nature of
the reggeon diagram in Fig.~9.8. Indeed, in an abelian gauge theory, the
insertion of this interaction into an even signature amplitude, for example
inserting $V_c$ to replace $\hat{V}$ in the transverse momentum diagrams of
Fig.~9.7, will give zero just because of this odd signature property. If this
were not the case the infra-red divergence structure of $QED$ would be
completely different! There would be a $t$-channel iteration of divergences
in addition to the $s$-channel iteration discussed in Section 7.3. That $V_c$
does not contribute in the high-energy transverse momentum diagrams of $QED$
can, as we noted in the Introduction, be regarded as a confirmation of the
cancellation of all anomalous massless fermion effects in the physical S-Matrix
of $QED$\cite{ce}. However, as we have already noted, in a non-abelian theory
and in SU(2) gauge theory in particular, there will be group factors involved.
Signature now involves a combination of $k \to -k$ (say) together with the
conjugation of quark couplings. The presence of an $I = 1$ component of the
quark-antiquark reggeon state then allows the vertex function $V_c$ to give an
{\em even signature} contribution, as a transverse momentum diagram, to the
even-signature reggeized gluon vertex - even though it {\em does not}
contribute as a reggeon diagram because of nonsense zeroes.

$V_c$ is naturally interpreted as follows. First apply (9.18) and (9.19)
to obtain a vertex function
$$
I^1(q,k,k^\prime) = \int d^2p\
{{Tr[(-\st{p}+\st{q}+m)(\st{k}-\st{q})(-\st{p}-\st{q}+m)(\st{k}^\prime-q)]}
\over {[(p+q)^2+m^2][(p-q)^2+m^2]}}
\auto
$$
which satisfies
$$
I^1 (q,q,k^\prime) = I^1 (q,k,q) = 0
\auto
$$
and so, as we shall discuss below, is naturally associated with a {\em reggeon}
diagram having the form of Fig.~9.11. Using the analogous manipulations, the
`full' vertex $V_c$ can be written as the sum of terms
$$
V_c = I^1 + I^2 + I^3 + I^4
\auto
$$
each of which satisfies an analogous property to (9.25) and is associated with
one of the diagrams illustrated in Fig.~9.12.

\subhead{9.5 Reggeon Diagram Origin of the Vertex $V_c$}

We consider now how the vertex $V_c$ would be constructed via the hexagraph
loop
procedure outlined at the end of Section 6. It is very important that we can,
in principle, construct this vertex entirely from multi-Regge amplitudes whose
form is unambiguously determined from our general formalism. The reggeon
diagram
we need to construct is a one-loop amplitude contributing to the hexagraph of
Fig.~9.13. For this we take a multiple discontinuity as shown and integrate the
two multireggeon diagrams (with signature factors removed by the
discontinuities
taken) over the appropriate two reggeon phase space. We can similarly construct
the corresponding loop contribution to the hexagraph of Fig.~9.14. The two
contributions can be combined to give even and odd-signature amplitudes. The
reggeon propagator in the odd-signature amplitude will be cancelled by a
a nonsense-zero as discussed above, while the $I = 1$ even-signature amplitude
will contain the transverse momentum interaction we want.

The discontinuities shown in Fig.~9.13 do indeed remove the intermediate quark
denominator factors that are absent in (9.21) and the corresponding
signature factors are not reinstated when the complete diagram is assembled.
Thus the construction process removes just the quark particle poles that we
assumed should be absent. If there were no particle poles in the amplitudes
initially, then taking discontinuities would introduce additional zeroes and
we would not obtain the expressions we have given. Therefore it is crucial
that the intermediate quark states can be on-shell physical states.

Note also that the nonsense zeroes shown can produce two effects. One is the
cancellation of the reggeon propagator for the reggeon state which splits
Fig.~9.13 in two. In addition, if evaluated on the Regge poles of the
attached quark lines, and the particle pole of the attached gluon line, they
will produce a zero at vanishing momentum for the gluon line. It is clearly
consistent therefore, that the nonsense-zero structure shows up in lowest
order as a product of linear zeroes in two of the gluon momenta together with
the absence of the reggeon propagator - just the structure we obtained by
associating $I^1$ with Fig.~9.11 interpreted as a reggeon diagram.

There is a further point of principle arising from the reggeon diagram
construction of $V_c$. When the full amplitude associated with Fig.~9.13 is
constructed by adding signature factors it does not contain particle poles
in all gluon reggeon channels. This is discussed at the end of Section 4 of
Part I. As a result the four reggeon interaction we extract {\em does not
contribute to the four-gluon scattering amplitude} when all the gluons are
taken
on shell. This implies that $V_c$ plays no role in the $I = 1$ reggeon
channel involving the gluon reggeization. Although we shall not elaborate
the argument, there is necessarily a cancellation analagous to the
cancellation between the diagrams of Fig.~5.12(b) and Fig.~5.13 discussed at
the end of Section 5.

\subhead{9.6 Properties of $V_c$}

We now describe a number of important properties of $V_c$. First we confirm
that (9.21) is indeed a convergent integral. For large values of $p$ the
integral for $V_c$ gives
$$\eqalignno{
\int {{d^2p} \over {p^4}} &Tr\left[ \st{p}\st{k}\st{p}\st{k}^\prime\right]=
4 \int {{d^2p} \over {p^4}}
\left[ 2(k\cdot p)(k^\prime\cdot p)-k\cdot k^\prime p^2\right]
&\num\label{9.25}\cr
&= k k^\prime \int {{d^2p} \over {p^2}} \int ^{2\pi}_0 d \theta
\left[ 2 \cos \theta \cos (\theta + \phi)
- \cos \phi\right]&\num\label{9.26}\cr
&=0&\num\label{9.27}\cr
}
$$
where $\theta$ and $\phi$ are respectively the angles between $k$ and
$p$, and $k$ and $k^\prime$.

Next we consider the limits $m \to 0$ followed by $q \to 0$, that is
$$
V_c\quad\centerunder {$\longrightarrow$}
{$m\rightarrow 0$}\quad
\int d^2p\ {{Tr\left[ (\st{p}-\st{q}) \st{k} (\st{p} + \st{q})
\st{k}^\prime\right]} \over {(p+q)^2(p-q)^2}}
\auto
$$

$$\eqalign{
\quad\centerunder {$\longrightarrow$}
{$q\rightarrow 0$}\quad
&4 \int\ {{d^2p}\ \over {p^4}} \left[ 2(p\cdot k)(p\cdot k^\prime)-p^2(k\cdot
k^\prime)\right] + 0(q^2)\cr
-&4 \left[ 2(q\cdot k)(q\cdot k^\prime) -q^2 (k \cdot k^\prime )\right]
\int {{d^2p} \over {(p+q)^2 (p-q)^2}}\cr
}
\auto
$$
where we have exploited the fact that integrals that are odd with respect to
$p$ vanish. From (9.27) we see that the first integral in (9.29) also
vanishes. Utilizing
$$
\int {{d^2p} \over {(p+q)^2(p-q)^2}} = {{2\pi} \over {q^2}}
\auto
$$
we obtain
$$\eqalign{
V_c \quad&\centerunder{$\longrightarrow$}{\centerunder
{$m\rightarrow 0$}{$q \rightarrow 0$}}\quad
\tilde{V}_c = 8 \pi \left[ {{2(q\cdot k)(q\cdot k^\prime)} \over {q^2}}
- k \cdot k^\prime\right]\cr
}
\auto
$$

Using this last limiting form for $V_c$ (which clearly originates from
the on-shell quark propagators) it follows that the ``Ward Identity property''
(9.12) is not satisfied. Indeed we obtain the simple result
$$\eqalign{
\tilde{V}_c (q,k,k^\prime)\quad\centerunder{$\longrightarrow$}
{$q\rightarrow \pm k, \pm k^\prime$}\quad
8\pi (2k\cdot k^\prime - k\cdot k^\prime)\cr
}
\auto
$$

$$= 8\pi k\cdot k^\prime
\auto
$$

The basic reason that we do not obtain (9.12) is clear from the reggeon diagram
construction illustrated in Figs.~9.13 and 9.14. The initial gluon is able to
couple to two quark reggeons that can carry the necessary helicities to allow
an
on-shell non-vanishing coupling.

The elementary form of (9.33) suggests that a simple subtraction might
restore the Ward Identity property. A conventional `Pauli-Villars' subtraction
of $V_c$ would involve
$$
\eqalignno{
V_p =\quad\centerunder{$lim$}
{$m\rightarrow \infty$}\quad
V_c (q,k,k^\prime,ml)&= \int d^2p\ {{m^2 Tr\left[\st{k}\st{k}^\prime\right]}
\over {(p^2 + m^2)^2}}
&\num\label{9.34}\cr
&= 4 m^2 k\cdot k^\prime \int^\infty_0 {{d(p^2)} \over {(p^2 - m^2)^2}}
\int^{2\pi}_0 d\theta&\num\label{9.35}\cr
&= 8\pi k\cdot k^\prime&\num\label{9.36}\cr
}
$$
Consequently a subtracted vertex $V_c - V_p$ would, for small $q$ and $m$,
satisfy (9.12).

In fact the Ward-Identities and resultant Pauli-Villars subtraction
procedure to be imposed, at the Feynman diagram level, on quark loops are much
more complicated in a non-abelian theory than in an abelian theory. This is
because the existence of an elementary four-gluon vertex implies that the
subtraction process mixes strongly with the problem of renormalization. It is
therefore very interesting that, in the non-abelian case, we are able to
discuss the subject in a self-contained manner in the infrared region of the
transverse momentum interaction that we have derived. This simplicity
enables us to give a straightforward discussion of the {\em interplay} that
we anticipated earlier, between the transverse momentum cut-off and the
subtraction procedure.

\subhead{9.7 Subtraction and the Transverse momentum Cut-Off}

We have emphasized that we wish to take the infrared limits giving massless
quarks and massless SU(2) gauge fields with a (gauge-invariant) transverse
cut-off present in all transverse momentum and reggeon diagrams. It is
straightforward to impose this cut-off in the definition of $V_c$ via (9.21),
that is we define
$$
V^\lambda_c = \int_{|p|^2<\lambda} d^2p\
{{Tr\left[ (-\st{p} + \st{q} + m) \st{k} (-\st{p} - \st{q} + m)
\st{k}^\prime\right]}
\over {\left[ (p + q)^2 + m^2\right] \left[ (p-q)^2 + m^2\right]}}
\auto
$$
Since (9.28) holds also with $|p|^2<\lambda$ imposed and (9.31) is
modified only to
$$
\int_{|p|^2<\lambda}\ {{d^2p} \over {(p + q)^2(p-q)^2}} = {{2\pi} \over {q^2}}
\left( 1 + 0 \left(q^2/\lambda\right)\right)
\auto
$$
we see that
$$\eqalign{
V^\lambda_c\quad\centerunder{$\longrightarrow$}{\centerunder
{$m\rightarrow 0$}{$q\rightarrow 0$}}\quad
\tilde{V}_c + 0 (q^2/\lambda)\cr
}
\auto
$$
and so the leading infrared behavior of $V^{\lambda}_c$ is actually
independent of $\lambda$.

The Pauli-Villars regularization procedure is, not surprisingly, strongly
sensitive to the presence of $\lambda$. If we define $V^{\lambda}_p$ by
imposing $|p|^2<\lambda$ in (9.34) we obtain
$$\eqalignno{
V^\lambda_p &=\quad\centerunder{$\lim$}
{\raisebox{-1mm} {$m\rightarrow \infty$}}\quad
4 m^2 k\cdot k^\prime\ \int^\lambda_0\ {{dp^2} \over {(p^2 + m^2)^2}}
\int^{2\pi}_0 d\theta &\num\label{9.40}\cr
&= \quad\centerunder{$\lim$}
{\raisebox{-1mm} {$m\rightarrow \infty$}}\quad
8\pi k\cdot k^\prime \left[ {{\lambda} \over {\lambda + m^2}}\right]
&\num\label{9.41}\cr
&= 0 \hspace{1.0in} \lambda \neq \infty&\num\label{9.42}\cr
}
$$

This shows clearly that if we insist on taking the infrared limits with
$\lambda \neq \infty$ then we can not impose (9.12) on $V_c$ by a Pauli-Villars
subtraction procedure. This is what we would expect on the general grounds
that the presence of an ultra-violet cut-off prevents the manipulation of the
quark sea at infinite momentum in order to obtain properties of the gluon
interaction at zero momentum. From this perspective, it would clearly be
undesirable to use a ``Pauli-Villars'' procedure involving the ``on-shell''
contribution of infinitely massive reggeized quarks.

We shall see that the failure to impose (9.12) on $V_c$ ultimately leads to
`desirable' rather than `undesirable' consequences. This may be directly
related to Gribov's argument\cite{gr1} that Ward Identities should not be
imposed on fermion loops at too early a stage in the construction of a gauge
theory. We now go on to discuss the extension of the infrared analysis of
the previous Section in the presence of $V_c$.

\subhead{9.8 Infrared Analysis in the Presence of $V_c$}

Clearly the most important implication of the previous sub-Section is that if
we take the limit of massless quarks in SU(2) gauge theory we will obtain a
well-defined contribution, $V_c$, to the four reggeon vertex $R^0_{2,2}$, which
is independent of the cut-off and which satisfies (9.33). That is $R^0_{2,2}$
now does not vanish when one of the reggeons carries zero transverse momentum .
Consequently we anticipate that the infrared divergence of $T^0$ which,
following (8.29), we attributed to the two-reggeon state, will be inverted by
reggeon unitarity as in (8.23) -- (8.25). Actually there is a minor
complication
here. To be completely sure how $V_c$ contributes to $R^0_{2,2}$, we must
construct the complete set of reggeon diagrams contributing to elastic
scattering of the form illustrated in Fig.~9.15 with the appropriate signature
factors. If we follow the procedure outlined at the end of Section 6 we find
that the surviving infra-red divergences are as illustrated in Fig.~9.16.

The inversion process eliminates not only the infrared
divergences, but also removes the corresponding angular-momentum plane
singularity. As a result all contributions of the two-reggeon state are
eliminated {\em with one exception}. Since $V_c$ does vanish when {\em all} of
its arguments vanish (this can be regarded as a consequence of the overall
scaling property of the vertex) the argument for inversion fails for that part
of phase-space where both reggeons carry a finite fraction of an overall
{\em vanishing} transverse momentum. To give a finite contribution such a
region
must produce a further divergence.

As we noted above, the fundamental reason that $V_c(q,k,k^{\prime})$ does not
vanish with individual gluon transverse momenta, is that a massless physical
gluon has a non-zero amplitude for transition into a massless
quark/antiquark pair. This leads to a non-zero amplitude for the exchange of
a quark/antiquark pair as an interaction between reggeized gluons. Even
color-zero gluon reggeon states are unstable in the presence of
this interaction. Indeed it follows from (5.22)-(5.25) that we could equally
well replace each of the gluons coupling via $V_c$ by a general number of
gluons. This would give contributions to $R^0_{N,N}$ (the $N \to N$ reggeized
gluon vertex) having the general form illustrated in Fig.~9.17. In general such
interactions will prevent the vanishing of $R^0_{N,N}$ when any subset of the
transverse momenta involved vanish. This is sufficient to ensure that all
infra-red divergences of the $N$-reggeon state associated with such
configurations are inverted in $T^0$. Again, the only part of the
$N$-reggeon state which might survive is that where {\em all reggeons carry a
finite fraction of an overall vanishing transverse momentum} and participate in
a further infra-red divergence.

\subhead{9.9 Infrared Scale Invariance Divergences}

We have now eliminated almost all infra-red divergences (and almost all
contributions of reggeized gluons in the process). However, as we now discuss,
there is a particular class of zero transverse momentum gluon configurations
that we have not yet eliminated and which produce infrared divergences which
are directly related to the {\em scale-invariance properties of the color-zero
kernels $\tilde{K}^0_N$}.

We have noted, in (8.32) and (8.33), the ultra-violet divergence of the
massive kernels $K^0_N$ due to the scale-invariance of the massless kernels
$\tilde{K}^0_N$. This scale-invariance property is also a potential source of
infrared divergences in the massless theory since
$$
\int_{|k_i|^2, |k^\prime_j|^2<\lambda}\ \prod_i {{d^2k_i} \over {k^2_i}}
\prod_j {{d^2k^\prime_j} \over {k^2_j}}
\ \tilde{K}_N \left( k_1, \cdots k_N,  k^\prime_1, \cdots k^\prime_N\right)
\auto
$$
$$\eqalignno{
\sim &\int^\lambda_0 {{dk^2} \over {k^2}}\hspace{0.5in}; \hspace{0.5in}
k^2 = k^2_1 + \cdots + k^2_N + \cdots + k^{\prime 2}_N&\num\label{9.44}\cr
\sim~~~ &\infty~~ \sim~~ ln\, M^2~~ (?)&\num\label{9.45}\cr
}
$$
The physical significance of this infrared divergence is entirely different to
that of the ultra-violet divergence, as we now discuss.

We have already noted that the scale-invariance property of $\tilde{K}^0_N$ is
not destroyed by the addition of the $V_c$ interaction. The additional
inversion
of infrared divergences produced by this interaction does, however, ensure that
there are no divergences as any subset of the momenta in (9.43) go to zero. The
{\em overall} divergence as all momenta scale uniformly to zero will not be
eliminated - precisely because it is a direct consequence of scale-invariance.
The question we now consider is whether this behavior is actually associated
with any $log~M^2$ divergences, as the gluon mass $M \to 0$, in physical
amplitudes.

The divergence (9.45) will appear in all color-zero $N$-gluon channels for
elastic scattering amplitudes. However, in this case it can only occur when the
total transverse momentum $q$ vanishes and so can only produce a $log~q^2$
divergence, {\em not a $log~M^2$ divergence}. However, our previous discussion
has already shown that all finite $q^2$ amplitudes are eliminated by inversion
of divergences.

A divergence of the form (9.45) could occur in amplitudes of the form
illustrated in Fig.~9.18. That is an $I = 0$ configuration of gluons is
exchanged accompanying an $I = 0$ quark/antiquark pair. As we noted in
sub-Section 9.1 the quark/antiquark channel is infrared finite as the gluon
mass $M \to 0$. Indeed, because of the ultra-violet divergences associated with
$R^Q_0$, the dynamics of this channel will be dominated by finite $q^2$. The
infrared divergence of the accompanying gluons will not mix with the
quark/antiquark channel and the divergence (9.45) will occur as illustrated in
Fig.~9.18.

Given the discussion of the last Section we anticipate that reggeon unitarity
will invert this last divergence via the sequence of diagrams shown in
Fig.~9.19. As illustrated in Fig.~9.20, the necessary interaction for this
iteration will be provided by a generalization of the gluon vertices of the
form of Fig.~9.17, that is the reggeized gluons interact with the reggeized
quarks by the exchange of quark/antiquark pairs. Interactions of the form of
Fig.~9.20 will indeed invert the divergence in Fig.~9.19 unless the coupling of
the gluons and quarks shown {\em either does not exist, or vanishes at zero
transverse momentum}. To discuss this possibility we must consider in detail
the {\em color charge parity} of multigluon states.

\subhead{9.10 Color Charge Parity and Signature}

We have so far denoted multi-reggeon states only by their total color $I$.
There has also been a simple relationship between $I$ and signature. That is
multi-gluon configurations with color $I$ have signature $(-1)^I$. However, as
we go beyond the next-to-leading log approximation this correspondence is not
obviously maintained. In particular an odd-signature three-reggeon
configuration with $I = 0$ can apparently be formed. If $A^i (i = 1,2,3,)$
denotes the components of the color matrix of the gluon field, then the
{\em Odderon} configuration
$$
0 \equiv Tr \left[ \epsilon_{ijk} A^i A^j A^k\right]
\auto
$$
has $I = 0$. The question is whether there is indeed a three-reggeon
singularity
associated with this configuration. If there is, we know from the general
signature rules for Regge cuts that it must have {\em odd-signature} and so
violate the $(-1)^I$ rule. Note that this is not the three-gluon Odderon
studied in $QCD$\cite{lnl}. As we now describe, unlike the $QCD$ Odderon, the
configuration (9.46) has anomalous {\em color-charge parity}.

For gauge fields with color matrix $A^i_{\alpha,\beta}$ the {\em charge
conjugation} operation $C$ is defined by
$$
C\, A^i_{\alpha\beta}\ C^{-1} = -A^i_{\beta\alpha}
\auto
$$
The additional transposition of the color matrix generalizes the sign change
involved in the charge conjugation of the photon field. The gauge field part of
the lagrangian is invariant if (9.47) is combined with a change of sign of the
gauge coupling. We can also define $C$ for the global SU(2) symmetry of the
spontaneously broken lagrangian (7.4) where it can be extended as a
color-charge
parity operation analogous to conventional G-parity. In this case
the signature-rule $(-1)^I$ simply says that the signature of a reggeon
state is given by its color-charge parity. (We have already noted that the
color factors of the global symmetry do become the gauge symmetry color factors
for Feynman diagrams when the gauge symmetry is restored). We now show that
this signature rule is a simple consequence of the helicity-conservation
property of a gauge theory.

\subhead{9.11 Signature and $T, C, P$ Properties}

In Part I we defined signature analytically by the process of {\em twisting}
hexagraphs (or planar Toller diagrams) and adding or subtracting corresponding
spectral component contributions. An equivalent ``group-theoretic'' definition,
which is particularly simple for elastic scattering, is as follows.

Consider the `direct channel' (defined by (2.46) of Part I) corresponding to a
particular planar Toller diagram for the amplitude under discussion.
Consider the {\em full scattering amplitude} in this channel. We define
signature by {\em adding the $TCP$ transformation $\tau$} to the SO(2,1) little
group defined for each line of the Toller diagram. For the $i$-th line,
$\tau_i$ is defined as making a complete $TCP$ transformation on all particles
attached (through the diagram) to one of the two vertices (say the left) which
the $i$-line connects. $\tau_i$ can then be applied directly to the planar
Toller diagram and a new direct channel obtained. The amplitude with signature
$\tau_i$ = +1 (-1) is defined by adding (subtracting) the {\em full amplitude}
for the new physical process to (from) the original full amplitude. (Since all
amplitudes are invariant under a full $TCP$ transformation of all particles it
does not, of course, matter whether the right or the left vertex of the $i$-th
line is used in this definition.)

Repeating the process for each channel defines amplitudes
$$
M^{(\tau_1,\cdots,\tau_{N-3})}_N (t_1,\cdots, t_{N-3}, g_1, \cdots, g_{N-3})
\auto
$$
Regge singularities with definite signature $\tau$ then appear in a particular
$t_i$-channel only in amplitudes with $\tau_i = \tau$.

For elastic scattering this definition of signature implies we add or subtract
amplitudes as illustrated in Fig.~9.21. For a reggeon state to appear, the
coupling to external states must be correspondingly even or odd under the $TCP$
transformation as illustrated in Fig.~9.22. Since the combination $TP$ simply
transforms an ingoing (outgoing) particle to an outgoing (ingoing) particle
with the same helicity, it follows that if reggeized gluon couplings are
{\em helicity conserving} then the contribution of a reggeized gluon state to
an elastic amplitude {\em at zero momentum transfer} must satisfy
$$
\tau~~=~~TCP~~=~~C
\auto
$$
which is the signature rule we have discussed. The odderon configuration (9.46)
is {\em anomalous} in that it {\em does not satisfy (9.49)}.

The helicity conservation property leading to (9.49) is, as we have emphasized,
present in all leading and next-to-leading log calculations. For quark
scattering this will not be the case beyond this approximation if the quarks
are massive. However, if we take the massless limit as discussed earlier in
this Section, then the helicity conservation property becomes absolute in
perturbation theory. Therefore, a reggeized gluon state with the quantum
numbers of (9.46) can not appear in forward quark scattering amplitudes ($T^0$)
in the limit of zero quark mass.

Consider now whether the Odderon configuration (9.46) can couple in an
amplitude
of the form of Fig.~9.18, that is accompanying a quark/antiquark state. We
anticipate that helicity conservation (at zero quark mass) will prevent this
configuration from coupling to such a state via a reggeon interaction at zero
transverse momentum. Therefore we expect that if an infra-red divergence of the
form (9.46) does occur {\em it will not be simply iterated by an interaction of
the form of Fig.~9.20}. The vital question then becomes whether the Odderon can
couple initially or finally to {\em any} physical state! We shall require the
triangle anomaly analysis of the next Section to answer this.

\subhead{9.12 Confinement and Chiral Symmetry Breaking in SU(2) Gauge Theory}

We can finally provide a glimpse of how a sensible confinement spectrum may
emerge out of our analysis. First we note that a single quark (or antiquark)
state is not an eigenstate of either $C$ or $P$, but quark/antiquark states can
be. Indeed the quark/antiquark reggeon state with trajectory (9.5), which
appears in Figs.~9.18 - 9.20 accompanied by gluons, can carry both $C$ = +1
and $C$ = -1. It must be a `nonsense' state with
$$
j \sim |n_1-n_2|-1 \sim 0
\auto
$$
but both $n_1 \sim -n_2 \sim \frac{1}{2}$ and $n_1 \sim -n_2 \sim -\frac{1}{2}$
are possible. The even and odd combinations of these states will have $P = -1$
and $P = +1$ respectively, giving both ``vector'' and ``axial-vector'' nonsense
states.

We clearly have a ``confinement'' spectrum in the sense that all states with
$I$ non-zero have been exponentiated (inverted) out of the theory by
infra-red divergences. We have briefly noted that the infra-red finite, {\em
odd signature}, $I = 0$, quark/antiquark reggeon states (9.50) can be
converted to ``bound-state'' Regge poles via the ``ultra-violet''
interaction $R^Q_0$. If an anomalous $I = 0$ set of gluons is also present
then an {\em even signature} reggeon state capable of producing a spin-zero
physical particle will be generated. Therefore, if the normal parity
``vector'' quark/antiquark state is somehow ``selected'' by the anomalous gluon
configuration then the complete state (involving the anomalous gluons) will
have abnormal parity. (This selection is just what we shall argue is achieved
by
the anomalous interactions discussed in the next Section.) Therefore it seems
that the scaling infra-red divergence of the Odderon configuration
can create the correct physical spectrum of {\em pseudo-scalar} Goldstone
bosons for a confining, chiral symmetry breaking, solution of the theory.
Before we can study this more explicitly we must understand just how the
anomalous gluon state can couple to quark reggeon states.

\newpage

\mainhead{10. THE TRIPLE-REGGE VERTEX AND THE TRIANGLE ANOMALY}

In the previous Section we raised the possibility that an ``anomalous Odderon''
three-gluon state could produce an infra-red divergence of the form (9.45).
However, as we have discussed, because of helicity-conservation this state will
normally {\em not couple} at zero $Q^2$ - where the infra-red divergence is
generated. ($Q$ now denotes the overall transverse momentum of the Odderon
state). We can restate this property by noting that if (9.49) is not
satisfied then the reggeon coupling (to the external states involved) must be
odd under $TP$. That is, if helicity conservation is maintained, the coupling
must be linear in $Q$ and so must vanish at $Q = 0$ {\em unless it has singular
$Q$-dependence - odd under $Q \to -Q$}. Alternatively (or as we shall find
equivalently) we must find a singularity which violates helicity conservation.
We must consider, therefore, whether such singular dependence can arise from
quark (reggeon) loop interactions?

We have already shown that the quark/antiquark threshold in reggeon diagrams
can
produce {\em finite} (reggeized) gluon interactions at zero momentum transfer.
To obtain more singular dependence we must locate a singularity involving
{\em three} quark propagators in a transverse momentum diagram - that is a
{\em triangle} singularity. To this end we consider whether a configuration
of the form illustrated in Fig.~10.1 could produce a singular coupling for
the anomalous gluon state. To do this we shall have to review some
properties of the general triple-regge vertex which were not covered in
sufficient detail for our present purposes in Part I.

\subhead{10.1 General Structure of the Triple-Regge Vertex}

We consider the Toller diagram shown in Fig.~10.2~ (cf. Fig.~2.2 in Part I).
The
appropriate variables are as illustrated. The corresponding
Sommerfeld-Watson (SW) representation is constructed by writing the sum of
hexagraphs shown in Fig.~10.3. Each hexagraph corresponds to a set of allowable
(asymptotically distinct) triple discontinuities. For the first hexagraph in
Fig.~10.3, the set of triple discontinuities is as shown in Figs.~10.4 and
10.5.
The SW representation is obtained by starting with the partial-wave expansion
$$
A^H_6 (z_1,z_2,z_3,w_2,w_3) = \sum T_{\ell_1}D_{\ell_2}D_{\ell_3}
a_{\underline{\ell}\, \underline{n}}
\auto
$$
$$
= \sum d^{\ell_1}_{0,n_2+n_3} (z_1)\, d^{\ell_2}_{n_2,0} (z_2)\,
d^{\ell_3}_{n_3,0} (z_3)\,
u^{n_2}_2 u^{n_3}_3 a_{\underline{\ell}\, \underline{n}}
\auto
$$
(The notation here is the same as in Section 4 of Part I)

It will be useful to first distinguish the triple-regge and helicity-pole
limits
associated with Fig.~10.2. The triple-regge limit is
$$
z_1,z_2,z_3 \rightarrow \infty
\auto
$$
while there are two distinct helicity-pole limits. The first is
$$
z_1,u_2,u_3 \rightarrow \infty \ ({\rm or}\ u_2, u_3 \rightarrow 0)
\auto
$$
and the second is
$$
z_1,u_2,u_3^{-1} \rightarrow \infty \ ({\rm or}\ u_2, u_3^{-1}
\rightarrow 0)
\auto
$$
The following approximations are (essentially) uniformly valid in all
three limits
$$
s_{23} \sim z_1z_2 (u_2 + 1/u_2)
\auto
$$
$$
s_{234} \sim z_1
\auto
$$
$$
s_{25} \sim s_{16} \sim z_1z_3 (u_3 + 1/u_3)
\auto
$$
$$
-s_{45} \sim s_{35} \sim z_2z_3 (u_2/u_3 + u_3/u_2)
\auto
$$
$$
s_{235} \sim s_{23} + s_{25} + s_{23}
\auto
$$

The triple discontinuity corresponding to Fig.~10.4 is in
$$
s_{23},~~ s_{16}, ~~s_{234}
\auto
$$
{}From (10.6) - (10.8) we see that (asymptotically) powers of these invariants
will contribute to partial waves with
$$
n_2, n_3 \geq 0,\, ~~~\ell_1 \geq n_2 + n_3
\auto
$$
Correspondingly a Froissart-Gribov (FG) continuation can be made to complex
$n_2, n_3$ and $l_1 - n_2 - n_3$. The part of (10.2) satisfying (10.12) can
then
be SW transformed and an asymptotic representation valid in the limits (10.3)
and (10.4) obtained. Since $s_{23}$ and $s_{16}$ are functions of
$\cos \omega_2~(~= (u_1 + \frac{1}{u_1})/2) $ and $\cos \omega_3$ respectively,
the same asymptotic behavior holds in the limit (10.5) and indeed an equivalent
asymptotic representation would be obtained if $n_2,n_3 \geq 0$ in (10.12) is
replaced by any combination of $\pm n_2 \geq 0$ and $\pm n_3 \geq
0$.

The cuts represented by Fig.~10.5(a) are in
$$
s_{23},~~ s_{16},~~ s_{45}
\auto
$$
while the cuts represented by Fig.~10.5(b) are in
$$
s_{23}, ~~s_{16}, ~~s_{235}
\auto
$$
{}From (10.9) and (10.10) we see that asymptotically the $s_{235}$ and $s_{45}$
cuts can be treated as (right and left) cuts in the same variable. (10.13) [or
(10.14)] appear as three asymptotically distinct cuts only in the Regge limit
(10.3) and the helicity-pole limit (10.5). The resulting partial-waves satisfy

$$
\eqalignno{
& n_2 \geq 0,\,~~ n_3 \leq 0,~~ \tilde{\ell}_1 \geq n_2 -n_3\cr
or &{}&\num\label{10.15}\cr
& n_2 \leq 0,\, ~~n_3 \geq 0,~~ \tilde{\ell}_1 \geq n_3 -n_2\cr}
$$
and the SW representation is correspondingly obtained from that part of the sum
in (10.2).

The combination of cuts (10.13) actually appears in each of the first three
hexagraphs in Fig.~10.3 and so is not uniquely associated with a particular
hexagraph. In the dual model studies of the triple-regge vertex by Detar and
Weiss\cite{dw}, four contributions are identified (before signature is
introduced). These are effectively the first three hexagraphs of
Fig.~10.3 and a fourth term identified with triple discontinuities of the
form of Fig.~10.5. The above discussion shows that we can identify the
fourth term as a sum of terms originating from all three hexagraphs but {\em
with opposite signs} to the usual terms {\em for the helicity labels
involved}. In Part I we adopted a (potentially confusing) convention whereby
all helicity labels appearing explicitly would be positive. In fact if we
were to continue to particle poles in the $t_2$ and $t_3$ channels, $n_2$
and $-n_3$ would correspond to $t_1$-channel center of mass helicities.
Consequently $n_2$ and $n_3$ could directly be ``$s$-channel'' helicities
for scattering by the exchange of a $t_1$-channel state and so, in the language
of previous Sections, the partial-waves corresponding to (10.15) involve
helicity-flip processes.

The full SW representation of the triple discontinuities of Fig.~10.5 is
$$
\eqalignno{
A_6 &= \int {{dn_2dn_3} \over {\sin \pi n_2\sin\pi n_3}}\
\int {{d\ell_1 (u_2)^{n_2}(u_3)^{n_3}} \over {\sin\pi(\ell_1-n_2+n_3)}}\
d^{\ell_1}_{0,n_2-n_3} (z_1)\cr
&\times \sum^\infty_{\ell_2-n_2=0}\ d^{\ell_2}_{n_2,0}(z_2)\
\sum^\infty_{\ell_3+n_3=0}\
d^{\ell_3}_{-n_3,0}(z_3)\,
a(\underline{\ell}, \underline{n})&\num\label{10.15}\cr
&+(2 \rightarrow 3,\ 3\rightarrow 1,\ 1\rightarrow 2) +
(2\rightarrow 1,\ 3\rightarrow 2,\ 1\rightarrow 3)\cr
&+ {\rm signatured\ contributions}\cr
&+ {\rm remaining\ summations}\cr
}
$$

In the helicity-pole limit (10.5) Regge poles at $l_1 = \alpha_1$,
$l_2 = \alpha_2$, and $l_3 = \alpha_3$, would give
$$
A_6 \sim u^{\alpha_2}_2\, u_3^{-\alpha_3}\, z_1^{\alpha_1}\,
\beta_{\alpha_1\alpha_2\alpha_3}
\auto
$$
$$
\sim (z_1u_2)^{(\alpha_1+\alpha_2-\alpha_3)}\,
(z_1u_3^{-1})^{(\alpha_1+\alpha_3-\alpha_2)/2}\,
(u_2u_3^{-1})^{(\alpha_2+\alpha_3-\alpha_1)/2}\,
\beta_{\alpha_1\alpha_2\alpha_3}
\auto
$$

$$
\sim (s_{23})^{(\alpha_1+\alpha_2-\alpha_3)/2}
(s_{16})^{(\alpha_1+\alpha_3-\alpha_2)/2}
(s_{35})^{(\alpha_2+\alpha_3-\alpha_1)/2}
\beta_{\alpha_1\alpha_2\alpha_3}
\auto
$$
Note that taking the discontinuity in $s_{23}$ (say) of this expression gives
$$
\centerunder{$\rm{disc}$}
{$s_{23}$}\quad \sim \sin [\pi (\alpha_1+\alpha_2 -\alpha_3)/2]\ A_6
\auto
$$
As in the discussion following I.(4.79) we see that since such a discontinuity
factor will not cancel any of the pole factors in
$\beta_{\alpha_1\alpha_2\alpha_3}$ emerging from the
$\sin \pi n_2 \sin \pi n_3 \sin \pi (l_1 - n_2 - n_3)$ factor in (10.12), the
Steinmann relations imply that these factors cancel in the sum over
permutations in (10.16). As a consequence, the triple-regge contribution
(10.16)
corresponding to triple discontinuities of the form of Fig.~10.5 {\em has no
particle poles in the $t$-channels}. It is a pure {\em multiparticle
multi-regge
amplitude}.

Finally we note that if a triple-regge amplitude contributes to elastic
scattering, as in Fig.~6.8 for example, then since the effective $s_{35}$ is
finite the helicity-pole limit (10.5) is necessarily involved. This is a
straightforward way to see that only the triple discontinuity of Fig.~10.4 and
the FG amplitudes obtained from (10.12) contribute. {\em The triple
discontinuities of Fig.~10.5 and the FG amplitudes obtained from (10.15) do not
contribute}.

\subhead{10.2 The Triangle Singularity in Reggeon Diagrams}

For the reasons discussed above we are interested in determining whether all
three quark propagators could be simultaneously singular in a reggeon diagram
having the structure of Fig.~10.1 and contributing to the Toller diagram of
Fig.~10.2. As described at the end of Section 6 (and in Part I) we construct
reggeon loop diagrams for Fig.~10.2 by building the triple discontinuities
for each hexagraph. If the signature factor of a reggeon propagator within a
loop is to be singular, there must not be a discontinuity taken through the
reggeon. Consequently, if all quark propagators in Fig.~10.1 are to remain
singular the triple discontinuity constructed must be as shown in Fig.~10.6,
that is it {\em must} be of the form of Fig.~10.5(a).

We conclude that if a triangle singularity appears in a triple-regge amplitude
then, from the above analysis, it can not appear in the helicity-pole limit
(10.4) and therefore {\em can not appear within a reggeon diagram for elastic
scattering}. Such an amplitude can appear in the helicity-pole limit (10.5) or
the triple-Regge limit (10.3). The essence of these limits is that the three
invariants of (10.13) must all be independently large. This requires that the
six-particle scattering process be {\em three-dimensional, that is not limited
to a single transverse plane}.

The three-dimensional requirement for the scattering can be seen by studying
directly the simplest possible Feynman diagram that could be involved. That is
we consider inserting the three gluon Odderon configuration into Fig.~9.11 to
obtain the reggeon diagram shown in Fig.~10.7. As discussed in Section 5, if
the
three gluons composing $O$ are emitted in the same transverse plane as the rest
of the scattering their coupling would reduce to a single $\gamma^+$ factor.
Given that $(\gamma^+)^2 = (\gamma^-)^2 = 0$, and that $\gamma^+$ and
$\gamma^-$ commute with transverse propagators, it is clearly not possible to
obtain a non-zero result by coupling $O$ into a quark loop as in Fig.~10.7
if the whole process takes place in a single transverse plane. However, if
the reggeons comprising $O$ are exchanged in a plane orthogonal to the
scattering plane (in which the quarks lie) then the ``$\gamma^+$~'' coupling
will be defined with respect to a different light-cone and we can write
$$
``\gamma^+~{\hbox {''}} = \gamma^0 + \gamma^z
\auto
$$
where now the ``z-axis'' lies in the $k_{\perp}$-plane of the rest of the
scattering. In this case the transverse-momentum diagram emerging from
Fig.~10.7
will be obtained by inserting $\gamma^z$ and a further propagator into (9.23).
This gives (see Fig.~10.7 for notation)
$$
V_A(q,k,k^\prime,Q) \sim \int d^2p\
{{Tr\left[ (-\st{p}+\st{q}+m)\st{k}(-\st{p}-\st{q}+m)\gamma^z
(-\st{p}-\st{q}+\st{Q} +m)\st{k}^\prime\right]} \over
{\left[ (p-q)^2+m^2\right] \left[ (p+q)^2 +m^2\right]
\left[ (p + q -Q)^2 + m^2\right]}}
\auto
$$
where $Q$ is orthogonal to $\gamma^z$.

To understand some of the essential properties of $V_A$ we consider first the
limit $q^2 \to 0$ with $m$ non-zero.
$$
V_A\quad \centerunder{$\sim$}
{$q\rightarrow 0$}\quad
\int d^2p\
{{Tr\left[ (-\st{p}+m)\st{k}
(-\st{p}+m)\gamma^z(-\st{p}+\st{Q}+m)\st{k^\prime}
\right]} \over
{\left[ p^2+m^2\right] \left[ (p-Q)^2 + m^2 \right]}}
\auto
$$
Picking out the ``$m^2$~'' term in the numerator we obtain
$$
V_A\quad \centerunder{$\sim$}
{$m\rightarrow 0$}\quad
m^2 \int
{{d^2p\ Tr \left[ \st{k} \gamma^z \st{Q}\st{k}^\prime \right]} \over
{\left[ p^2+m^2\right]^2 Q^2}}\ + \cdots
\auto
$$
$$
\sim {{\epsilon_{ij} k_i k^\prime_j} \over {Q}}
\auto
$$
which illustrates that (when $q^2 \to 0$) there is a singular component of
the vertex (of just the form we are looking for) which involves helicity
transitions. Clearly this vertex has a structure very close to that of the
familiar triangle anomaly. The reduction to a transverse momentum diagram
produced by the Regge limit has enhanced the triangle singularity. As a
result a helicity transition on the quark line to which the Odderon is
coupled survives the massless limit and produces the same result as if we
had coupled an axial-vector current to the quark-loop rather than the
anomalous Odderon. (In the next Section we shall identify the corresponding
axial-vector current as the ``winding-number operator''.)

An essential feature of $V_A$ that we shall need in the following is that if
we consider the ``scaling limit''
$q, k, k^{\prime} \sim 0,$ involving all the
gluons besides those in the Odderon configuration, followed by $ Q \to 0$, then
as illustrated by (10.25) we can write
$$
V_A \sim {{R(k,k^\prime,q)} \over {Q}}
\auto
$$
where $R(q,k,k^{\prime})$ is non-singular and has the straightforward scaling
dimensions of the four-reggeized gluon vertex $R^0_{2,2}$. By similarly
coupling
the Odderon configuration in to a multigluon quark loop interaction of the form
illustrated in Fig.~9.17 and considering the corresponding scaling limit, we
will obtain a general result of the form of (10.26) with the numerator a
scaling
multi-gluon vertex. Therefore, such vertices allow the O configuration to be
emitted {\em out of the transverse plane} by a general zero-transverse
momentum set of gluons, with a $Q^{-1}$ coupling and {\em without affecting
the scaling interaction of the gluon set}. As we have already discussed
(following (10.20)), because of the structure of the triple-regge vertex
involved, there are necessarily no singular signature factors (particle
poles) associated with the propagation of the O configuration. However, if
the configuration couples between sets of propagating gluons (in distinct
transverse planes), as illustrated in Fig.~10.8, then the combination of
$Q^{-1}$ factors for each coupling will produce the same divergence as if
there were a signature factor present. As we shall discuss again in the next
Section, we believe that the role of the triangle anomaly interaction among
pure gluon configurations is therefore to provide a coupling between overall
infra-red divergences associated with the Odderon configuration in each
distinct transverse plane of a multi-regge diagram.

It will be important for the discussion of Section 12 to consider the
possibility that some of the gluons involved in a multi-gluon coupling via the
anomaly may actually be massive and carrying finite transverse momentum. (For
our present discussion it will be sufficient to consider such gluons as
singlets
under the SU(2) gauge symmetry). To obtain the full triangle singularity, there
must be zero transverse momentum flowing around the quark loop involved.
However, a massive gluon pair can clearly be produced at one vertex as
illustrated in in Fig.~10.9, for example. If we select the part of this vertex
which does not vanish when the transverse momentum of the massive pair vanishes
then (essentially because of the structure of (10.24)) it {\em will vanish} as
the transverse momentum of the massless gluons coupled at the same vertex
vanishes. Consequently while the $Q^{-1}$ divergence produced by the O
configuration coupling can compensate for this vanishing and combine with the
coupling of the other massless gluons to produce overall scaling behavior, it
can not generate a new divergence. Interactions of this form, where multiple
massive gluon states are produced at zero transverse momentum by a massless
configuration involving the Odderon, are a crucial part of the relationship
between the Super-Critical Pomeron and QCD with the gauge symmetry partly
broken.

To complete our description of the role played by triangle anomaly interactions
we now discuss the emission of the Odderon configuration from quark/antiquark
reggeon states. This relates to the anomaly structure of flavor currents and so
we expect it to be crucial in determining whether chiral symmetry breaking is
an outcome of our analysis. It is well-known that the triangle anomaly in
flavor
axial-vector current vertices determines\cite{gth2} that chiral symmetry
breaking must accompany confinement in QCD  (with a sufficiently large number
of flavors).

We have noted that $I = 0$ quark/antiquark reggeon states are infra-red finite
and that if they carry flavor they produce leading reggeon singularities. We
have also noted that the dynamics of reggeon interactions in such channels will
be dominated by the four-quark vertex $R^Q_0$. An Odderon configuration can
also
be emitted {\em out of the transverse plane} by a triangle reggeon interaction
involving $R^Q_0$, as illustrated in Fig.~10.11. The interaction now
has the form
$$
V^Q_A (k,k^\prime,q,Q) \sim \int d^2p\
{{Tr\left[ (-\st{p}+\st{q}+m) K^Q_0 (-\st{p}+\st{q}+m) \gamma^z
(-\st{p}-\st{q}+Q+m) K^Q_0\right]} \over
{\left[ (p-q)^2+m^2\right] \left[ (p+q)^2+m^2\right] \left[
(p+q-Q)^2+m^2\right]
}}
\auto
$$
As we discussed in Section 8 we can separate the quark/antiquark reggeon state
into ``vector'' and ``axial-vector'' components which are respectively odd and
even under parity. We shall assume now that the corresponding anomalous vertex
functions $V^Q_{VV}$ and $V^Q_{AA}$ satisfy similar Ward Identity properties
to current vertex functions. That is
$$
V^Q_{AVV}\ (k,k^\prime,q,Q)\quad \centerunder{$\longrightarrow$}
{$q\rightarrow 0$}\quad 0
\auto
$$
whereas
$$
V^Q_{AAA}\ (k,k^\prime,q,Q)\quad \centerunder{$\st{\longrightarrow}$}
{$q\rightarrow 0$}\quad 0
\auto
$$

These properties can only hold away from the triangle singularity,
that is for $Q \neq 0$. Since both $V^Q_{VV}$ and $V^Q_{AA}$ contain the
infra-red anomaly they will also satisfy (10.26) if we first set $q^2 = 0$.
In the next Sections we effectively use the presence of the triangle
singularity at $Q = 0,~q = 0$ to determine when a coupling to the anomalous
Odderon exists. We then assume, and exploit crucially in the next Section,
that in addition there is also a {\em finite, non-divergent, coupling} to
the quark/antiquark states given by $V^Q_{VV}$ and $V^Q_{AA}$ at
$Q = 0, q \neq 0$, and that (10.28) and (10.29) are satisfied by these
finite couplings.

Clearly we should study whether (10.28) and (10.29) hold in more detail. In
particular we should consider whether a more elaborate treatment of the
transverse momentum cut-off we impose is necessary to achieve (10.28). We would
not expect this to be the case since the integral defining $V^Q_A$ is strongly
convergent in the ultra-violet region, in contrast to the convergence of the
integrals defining $V_c$ in Section 9, where we have argued that Ward Identity
properties are necessarily violated in the presence of the transverse cut-off.

We now move on to discuss how the various infra-red properties we have
discussed
in the last three Sections can be put together to extract a ``hadron''
high-energy S-Matrix from the reggeon diagrams of SU(2) gauge theory.

\newpage

\mainhead{11. THE SU(2) HIGH-ENERGY S-MATRIX AND THE SCHWINGER MODEL}

In this Section we describe how the structure of the anomalous interactions
discussed in the last Section can build up the confining spectrum and
high-energy S-Matrix for SU(2) gauge theory with many fermions. We shall then
go on to briefly review properties of the two-dimensional Schwinger model and
draw on this to understand the physics behind our results. Indeed we are more
confident of the physical description of our results than we are of many of the
details of the (necessarily) elaborate technical derivation that we outline. Of
necessity we must discuss reggeon diagrams for a large class of Toller diagrams
for which we do not strictly have the appropriate rules in hand. Indeed the
anomalous helicity structure of triple-regge vertices exploited in the previous
Section has until now been outside of the framework of reggeon diagram and
Reggeon Field Theory studies. This provides a major reason for the largely
qualitative nature of the description that follows. Another is simply the large
amount of work that is still needed to fully implement the analysis.

We first summarize the separate ingredients which we have established in
previous Sections and then describe their combination into a (potentially)
comprehensive analysis.

\subhead{11.1 Zero Transverse Momentum Gluons Surviving Exponentiation}

We have shown that all configurations of gluon reggeons carrying non-zero color
are removed by the process of inversion of infra-red divergences.
Configurations
with color zero and with some subset carrying zero transverse momentum are
similarly eliminated. Gluon configurations which {\em all} carry zero
transverse momentum can occur accompanying color zero quark reggeon states that
carry the transverse momentum of the state. To avoid removal by inversion of
infra-red divergences (via further reggeon interactions) such configurations
must carry anomalous Odderon quantum numbers. Therefore these are the only zero
transverse momentum gluon configurations which survive in the theory. We have
not explicitly stated it but clearly the concept of an ``anomalous Odderon''
gluon configuration generalises to {\em any odd number} of reggeized gluons
carrying $I = 0$ and $C = +1$. As we discuss below, full gauge invariance
implies that we should expect the sum total of such configurations to be
related to the exponentiation of the winding-number operator.

\subhead{11.2 Anomalous Odderon Couplings}

The most crucial Odderon coupling is the pure glue coupling $V_A$ since this
provides a ``three-Odderon coupling'' at zero transverse momentum as
illustrated
in Fig.~11.1. This coupling will produce an additional divergence for every
Odderon configuration that propagates out of one transverse plane and in to
another. Consequently, as we elaborate below, the most infra-red divergent
multi-Regge amplitudes are those which have (essentially) an Odderon divergence
in each reggeon channel (that is for each line of the associated Toller
diagram) and this is what ultimately defines physical reggeon amplitudes.

In addition to the divergent Odderon configuration, there must also be a
quark/antiquark component to each reggeon state which carries the total
transverse momentum of the state. The quark/antiquark component can emit (or
absorb) an Odderon configuration via an anomalous interaction but since it
carries finite transverse momentum {\em there will be no
additional infra-red divergence} produced by such an emission (or
absorption). We therefore envisage a reggeon state in a general Toller
diagram as emitting an arbitrary (odd) number of Odderon configurations out
of the transverse plane which are absorbed elsewhere in the reggeon diagram
as illustrated in Fig.~11.2. In general any two lines of the Toller diagram
separated by at least one further internal line can be connected. However,
as we discuss below, there are very important constraints placed on such
connections by the discontinuity structure of the hexagraph involved.

Note that a major implication of the necessity to have a quark/antiquark
component in all reggeon states is that there is {\em no Pomeron}. That is
{\em there are no pure gluon reggeon states}. Therefore all cross-sections
will decrease like O(1/S) and we will have a non-interacting theory at
high-energy. In the language of Part I {\em the S-Matrix will be ``trivial''
in an extreme sense.}

\subhead{11.3 The Flavor Anomaly and Chiral Symmetry Breaking}

It is well-known that the presence of a triangle anomaly in vertex functions
of the flavor axial-vector currents determines that in general chiral symmetry
breaking must accompany confinement in a gauge theory\cite{gth2}. As we implied
in the last Section we expect the Odderon coupling to quark/antiquark
reggeon states to reflect this anomaly structure and to determine the
pattern of chiral symmetry breaking as follows.

As we discussed in Section 9, the leading reggeon singularity at low transverse
momentum $q$, in quark/antiquark channels, is the two-reggeon cut (9.5) -
before
the gluon mass is taken to zero. In the massless (gluon) limit the quark
reggeization disappears via the cancellation of infra-red divergences.
Complimentarity should imply that this behaviour is smoothly replaced by a
``bound state'' reggeon singularity formed (predominantly) by the iteration of
the $R^Q_0$ interaction. This singularity will be a Regge pole if we can
introduce asymptotic freedom and the running of the gauge coupling as the
transverse cut-off is removed. From (9.5) it follows that this singularity
will have a $q^2 = 0$ intercept of $j = 0$ as the gluon mass is removed {\em
if the quark mass is zero}. It can be moved from zero if the cut-off is
varied {\em after} the zero gluon mass limit is taken. Therefore (if we
allow the cut-off to vary) we can say only that there will be an {\em odd
signature} reggeon bound state with ``mass close to zero''. As we have noted
this odd signature state will combine with the Odderon state to give an
overall {\em even signature} state capable of producing a physical, spin
zero, particle.

At non-zero $q^2$ the bound state can emit and absorb O configurations
as described above, and illustrated in Fig.~11.2. This produces a
renormalization of the bound state Regge trajectory which, because of
(10.28) and (10.29), distinguishes between the distinct parity states. The
{\em abnormal} parity (``axial'') state will be {\em driven away from zero
mass} by this renormalization whereas, because of (10.28), the {\em normal}
parity (``vector'') state will remain close to zero. The combined reggeon
state - Odderon plus bound state - producing a physical spin zero state will
then have {\em abnormal parity} and the {\em physical particle will be a
pseudoscalar}.

\subhead{11.4 The Multi-Regge S-Matrix}

We now describe how, in principle at least, we anticipate physical amplitudes
emerging from our analysis. As an extension of the complimentarity principle
discussed in Section 7, we assume that the multi-reggeon S-Matrix (associated
with the complete set of Toller diagrams and) describing those Regge poles and
cuts that can be exchanged by physical scattering states should evolve smoothly
as the infra-red limit is taken. For this to be the case it must be that any
infra-red divergences that occur can be absorbed into a redefinition (as a
normalization factor) of the {\em particle} states that are scattering -
leaving a finite multi-Regge amplitude. If we assume that all physical states
are correspondingly renormalized, then it will follow that finite physical
scattering amplitudes are produced only when the (originally) infra-red
divergent reggeon amplitudes are involved.

We shall shortly discuss the construction of physical states in the massless
Schwinger model. This will lead us to suggest that the Odderon divergences we
find are related to the development of an expectation value for the exponential
of the winding number operator. In effect the need to include a ``cloud of
topological gluon configurations'' in our definition of physical states
replaces
the need to include a ``simpler cloud of soft photons'' in our definition of
physical states in $QED$, and is responsible for confinement. For the reasons
discussed in Section 7, we are assuming that complimentarity allows the
detection of this phenomenon in the infra-red divergences of reggeon
diagrams. We clearly can not follow the process of redefinition of the
particle states in detail, but we can discuss the reggeon states. As we have
just described, these will include the bound-state Regge poles that will
produce hadrons and by continuing to the associated particle poles we can,
in principle, extract high-energy hadron scattering amplitudes.

Consider the general divergence structure to be expected in an arbitrary Toller
amplitude. At first sight it might appear that the propagating O configurations
illustrated in Fig.~11.2 can be arbitrarily attached across a general
multi-regge amplitude. It would then appear that {\em divergent} O
configurations could connect, via the triple Odderon coupling of Fig.~11.1,
to any pair propagating between distinct transverse planes. However, as we
have emphasised, our general procedure requires that we construct the
individual hexagraph amplitudes involved via their multiple discontinuities.
This is particularly important, as we found when discussing the
Super-Critical Pomeron in Part I, if we want to ensure that the Regge pole
states we generate produce physical particle poles (and corresponding
particle amplitudes as residues) in the amplitudes we study. In a general
diagram of the form of Fig.~11.2, the discontinuity structure of individual
hexagraphs strongly constrains both the non-divergent O configurations that
can be present and the triple Odderon couplings that can be added.

As an example, if we want a four-particle amplitude to be associated with the
sub-Toller diagram of Fig.~11.3, then the discontinuity shown (or the
corresponding cross-channel discontinuity) must be present. This
discontinuity is present only if O configurations crossing from the bottom
to the top of the diagram (and vice versa) are absent. When this requirement
is combined with the constraint that only odd number O configurations must
be associated with any particular reggeon channel, we essentially obtain the
maximum degree of divergence in diagrams of the form of Fig.~11.4 (together
with the corresponding crossed diagrams) in which a single divergent O
cofiguration can be explicitly identified with the reggeon state in the
discontinuity channel. When this argument is extended to the more complicated
multiple discontinuities of general diagrams, a divergence can always be
identified with each of the internal reggeon lines. For example, divergences
of the sub-triple regge diagram with the discontinuities of Fig.~11.5 can be
identified as in Fig.~11.6. We conclude, therefore, that there is a
divergent zero transverse momentum component, that is a ``reggeon
condensate'', as part of the definition of each reggeon state.

\subhead{11.5 The Winding Number Operator}

To begin our discussion of the physical significance of the Odderon divergence
we remark that the Odderon has the same color structure as the first term in
the
well-known gauge dependent current
$$
K^\mu(x) = {{g^2} \over {8\pi^2}}\epsilon_{\mu\alpha\beta\gamma}
Tr\bigg[-{{2ig} \over {3}}\epsilon_{ijk}A^i_\alpha
A^j_\beta A^k_\gamma + A^i_\alpha \partial_\beta A^i_\gamma \bigg]
\auto
$$
whose divergence is given by
$$
\partial_\mu K^\mu = F\tilde{F}
\auto
$$
If $K_{\mu}$ is integrated over a three-dimensional volume, we obtain a
zero-momentum component which defines the familiar ``winding-number'' for which
instantons produce integer-valued transitions and (the exponential of) which is
{\em gauge invariant}. Note that if a gauge field is pure gauge then {\em only}
the first term in (11.1) contributes to the winding number.

The Odderon is not produced by a local operator but rather is an infra-red
configuration that couples into the infra-red region of quark loops in
essentially the same way as the winding number operator. Clearly the existence
of an Odderon reggeon condensate is very similar to the existence of a winding
number condensate and can be usefully thought of as a ``generalized winding
number condensate. Nevertheless we should emphasize that, as we have introduced
it, a reggeon condensate is something specifically involved in the definition
of a scattering hadron in the Regge region and as such is {\em not a property
of the vacuum}. However, since we essentially have a non-interacting theory at
high-energy there may be {\em no distinction between a property of
hadrons and a property of ``the infinite momentum vacuum''}.

\subhead{11.6 The Schwinger Model}

The Schwinger model, or massless $QED_2$, is explicitly solvable and provides
considerable physical insight into the meaning of an expectation value for a
winding number operator. For this reason it will be helpful to our overall
purpose to provide a very brief synopsis of the solution of the model based on
Manton's treatment\cite{man}. Manton's comments on the relationship between the
Schwinger Model and $QCD$ will also help us elucidate the potential physical
significance of our analysis.

The lagrangian is
$$
L = {{1} \over {2e^2}}\left(\partial_tA_x-\partial_x A_t \right)^2 +
\overline{\psi} i\gamma^{\mu} \left(\partial_\mu + iA_\mu \right)\psi
\auto
$$
It is helpful to initially ``compactify space'' by regarding the theory as
defined on a circle $0~\leq x \leq~2\pi$ and ultimately take the infinite
radius limit. In the Coulomb gauge we can take
$$
\partial_xA_x = 0,~~~~~\ 0 \leq A_x \leq 1
\auto
$$
with $A_x = 0,1$ being gauge equivalent. A winding number $n$ can be defined
for
gauge transformations
$$
g(x) = \exp \left[i\Lambda(x)\right],~~~~\ \Lambda(2\pi) =
\Lambda(0) + 2\pi n
\auto
$$
A topologically non-trivial gauge transformation takes $A_x$ from the interval
(11.4) to an alternative interval.

We define the electric current $j^{\mu} = \overline{\psi}\gamma^\mu \psi$
and the chiral current $j^{\mu}_5 = \overline{\psi}\gamma^\mu\gamma^5 \psi$.
There is an {\em anomaly equation}
$$
\eqalign{
\partial_\mu j^\mu_s &= ~{{1} \over {\pi}} \partial_\mu W^\mu =
{}~{{1} \over {\pi}} \partial_\mu \epsilon^{\mu\nu} A_\nu \cr
&= -\partial_t A_x\ \rm {in\ Coulomb\ gauge}
}
\auto
$$
and so $W^{\mu}$ is the analogue of the winding number current in a
four-dimensional gauge theory. In Coulomb gauge we can take $A_x$ itself to
be the winding number. Clearly it changes by an integer under a gauge
transformation with non-trivial topology.

The Dirac operator appearing in (11.3) is
$$
\Bigg({{-i\partial_x + A_x} \atop {0}} \hspace{0.25in}
{{0} \atop {i\partial_x-A_x}} \Bigg)
\auto
$$
and gives an energy spectrum
$$
\eqalign{
&p+A_x~~~\ \rm{for\ ``left-handed"\ fermions}\cr
-&p-A_x~~~\ \rm{for\ ``right-handed"\ fermions}
}
\auto
$$
with $p$ taking all integer values. The anomaly equation (11.6) is satisfied
non-trivially because the Dirac operator has a ``spectral flow'' of two in that
as we ``orbit'' the configuration space $0\leq ~A_x~ \leq~1$ the spectrum is
permuted. The energy of left-handed fermions increases by one while that of
right-handed fermions decreases by one.

Introducing creation operators via
$$
\psi_j(x) = {{1} \over {\sqrt{2\pi}}} \sum^\infty_{k=-\infty}
a_{j,k}e^{ikx}\hspace{0.5in} j=1,2
\auto
$$
the left and right-handed charge operators must be regularized. A
gauge-invariant procedure for states with a finite number of positive energy
left-handed particles is to define
$$
Q^\lambda_L = \sum_{-\infty} e^{\lambda(k+A_x)} a^+_{1k}a_{1k}
\auto
$$
$$
{{\longrightarrow} \atop {\lambda\rightarrow 0}}\ ~~~~\sum^M_{k=-\infty}\
e^{\lambda(k+A_x)} = {{1}\over {\lambda}} + \left(M + A_x+{{1} \over {2}}
\right) + 0(\lambda)
\auto
$$
if all energy levels $\leq M$ are occupied, with a similar definition for a
state with all right-handed momenta $\leq N$ occupied. Removing the
$\frac{1}{\lambda}$ divergences gives regularized charges for such states
$$
Q^{\rm reg}_L = M + A_x + {{1} \over {2}},~~~\ Q^{\rm reg}_R =
-N-A_x + {{1} \over {2}}
\auto
$$
which are non-integer (``fractional charge'') and depend on $A_x$. The
dependence on $A_x$ is a necessary consequence of the spectral flow, and
leads to the crucial conclusion that {\em states with non-zero electric
charge are unphysical} precisely because the dependence on $A_x$ does not
cancel in physical quantities. In particular, charged states do not have
gauge-invariant momenta.

The zero momentum states of the theory can be found very easily and involve no
fermionic excitations. They are specified by a set of wave-functions
$$
\left\{\psi_p (A_x):\ P\epsilon~ Z,\ 0\leq A_x \leq 1 \right\}
\auto
$$
where $\psi_p$ denotes the amplitude for a state in which left-handed particles
with all momenta $\geq~p$ are present and right-handed particles with all
moment
a
$\leq~p + 1$ are present. Clearly such states have zero charge. The existence
of spectral flow imposes the boundary conditions
$$
\psi_p (1) = \psi_{p+1}(0),~~~~\ \partial_{A_x} \psi_p (1) = \partial_{A_x}
\psi_{p+1} (0)
\auto
$$
If the kinetic energy in (11.3) is regularized as we have regularized the
charges, we obtain the regularized energy
$$
V^{\rm reg}_p (A_x) = \left(p + A_x + {{1}\over{2}} \right)^2
\auto
$$
We now effectively mix the topological sectors of the gauge field by defining
an
energy $V(A_x)$ and a wave-function $\psi (A_x)$ on the whole interval
$ -\infty~<~A_x~<~\infty$ by writing
$$
V(p + A_x) = V_p (A_x),~~~\ \psi (p + A_x) = \psi (A_x)
\auto
$$
If we then use Gauss's Law to eliminate $A_t$ in (11.3) and make the
approximation of ignoring the resultant Coulomb energy we obtain the
Schroedinger equation
$$
i{{\partial \psi}\over{\partial t}} =
\bigg[-{{e^2}\over {4\pi^2}}\ {{d^2}\over {dA^2_x}} +
\left(A_x + {{1}\over{2}} \right)^2 \bigg] \psi
\auto
$$
This is a harmonic oscillator equation with the energy eigenvalues spaced by
$e/\sqrt{\pi}$, giving the energy  spectrum of the free scalar multiparticle
zer
o
momentum states which are known to solve the Schwinger model. The ground-state
has
$$
A_x = -{{1}\over {2}}
\auto
$$
which includes
$$
p = -1, ~~~\ A_x = {{1}\over {2}}
\auto
$$
in which, from (11.12), there are ``no fermions'', but also includes additional
topological sectors with compensating pairs of fermions - because of spectral
flow. The gauge-invariant statement of (11.18) is that the ``Wilson Loop''
operator
$$
\eqalign{
\exp \bigg[i\ \int^{2\pi}_o dx W^t \bigg] &= \exp
\bigg[2\pi  i\ A_x \bigg]\ {\rm in\ Coulomb\ gauge}\cr
&= -1
}
\auto
$$
since this holds in all topological sectors.

Summarizing - {\em the existence of spectral flow and the chiral anomaly leads
to the appearance of the winding number field in the regularized chiral
charges.
The topological classification of gauge fields breaks down and a vacuum
winding-number field emerges which stabilizes a ``confinement spectrum'' of
non-interacting particles}.

\subhead{11.7 Massive $QED_2$}

It is possible to define massive $QED_2$ in which a fermion mass term is added
to (11.3). The theory is rigorously defined by mass perturbation theory around
the massless theory\cite{fr}. After bosonization of the theory the fermion mass
becomes
the coupling constant for the non-interacting massive bosons of the massless
fermion theory. That is the massless Schwinger model provides the free theory
around which the interacting theory is defined perturbatively. The spectrum of
the massive theory is continuously related to that of the massless theory and
therefore is also confining. However, in the massive theory there is {\em no
vacuum field} and instead the theory has {\em interactions} - with the
interaction between charged particles increasing with distance as in
strong-coupling lattice $QCD$. This is the well-known parallel\cite{kog}
between
strong-coupling confinement in $QCD$ and confinement in massive $QED_2$.

The ``chiral limit'' of massive $QED_2$ exists, in a sense, because of the
interplay between massless fermions and gauge field topology. It is presumably
also true that topology is able to govern the spectrum of the massless theory
just because of the lack of interactions among the physical states.

\subhead{11.8 Comparison with Non-Abelian Theories in Four Dimensions}

In a four-dimensional theory the operator which plays a similar role to the
phase-factor (11.20) is the gauge invariant exponential of the winding number
operator referred to above, that is
$$
X = \exp\ \bigg[-{{1}\over{8\pi}} \int d^3 \underline{x}
K^o(x) \bigg]
\auto
$$
where $K^0(x)$ is defined by (11.1). In addition to being referred to as the
winding number operator, $K^0(x)$ is also known as the ``Chern-Simons 3-form''.
Similarly $A_x$ is known as the abelian ``Chern-Simons 1-form''. Topologically
trivial gauge transformations do not change $X$ at all. A gauge transformation
with non-trivial topology produces a phase change of $2\pi$$n$, where $n$ is an
integer, and therefore also leaves $X$ unchanged. In general a closed loop in
Yang-Mills configuration space (produced by some parameter variation) is
defined to be ``noncontractable'' if (and only if) the phase of $X$ changes
continuously by $2\pi$$n$ along it.

An instanton is a gauge field which traverses a noncontractable loop with
Euclidean time as the parameter. Also it is known that the Dirac operator has a
nonvanishing spectral flow along noncontractible loops. In analogy with the
Schwinger model, therefore, the presence of massless fermions changes the
topology of the Yang-Mills gauge field configuration space and it is again
necessary to pass to the simply connected covering space. Therefore we might
anticipate that the U(1) axial anomaly equation for the color singlet
axial-vector current
$j^{\mu}_5$
$$
\partial_\mu j^\mu_5 = -N_f F\tilde{F}~/16\pi^2
\auto
$$
combined with the effects of instantons will similarily lead to a non-trivial
winding number vacuum field which minimizes the sum of the classical potential
energy and the regularized energy of the Dirac sea.

As we discussed at length in Section 3, we expect that when there are only a
small number of massless fermions in the theory, the effects of large gauge
field fluctuations associated with renormalons (including the infra-red growth
of the gauge coupling) will overwhelm instanton contributions - if these are
even well-defined! However, as the theory is saturated with massless fermions
we have argued that instantons and the regularization of the Dirac sea that we
are now discussing become the dominant dynamical problem. Clearly we want to
argue that the reggeon condensate we have discovered is due to just this
phenomenon.

First we note that we begin our analysis with both infra-red and ultra-violet
cut-offs in the theory. In this case perturbation theory is Borel summable and
there are no non-perturbative contributions associated with either instantons
or renormalons. In principle at least, there is no reason why the the
appropriate ``non-perturbative'' contributions can not be generated by the
limiting processes involved in removing the infra-red and ultra-violet
cut-offs. As a simple illustrative example, note that
$$
\exp\ \bigg[1/(g^2 + \mu/\lambda) \bigg]
\auto
$$
has a perturbation expansion for finite $\mu$ and $\lambda$. If the
``infra-red'' cut-off $\mu$ is removed, with $\lambda$ finite, a
``non-perturbative'' result is smoothly obtained. This provides a very
over-simplified version of how we believe the ``non-perturbative'' elements of
a theory can be produced by the process of removing cut-offs. Indeed the
removal
of cut-offs while maintaining unitarity and exploiting ``complimentarity'' as
we
have done {\em could conceivably be the only way to rigorously define
non-perturbative contributions to a theory}.

Secondly we note that while we have described the ``topological'' formulation
of
the Schwinger model in order to draw the parallel with the physics of gauge
theories that we believe our analysis is exposing, the Schwinger model can also
be ``solved''\cite{frd} (or defined) by the process of summing perturbation
theory and removing cut-offs - {\em provided that the anomaly is
appropriately regularized from the outset}. Analogously, we believe that the
interaction of the Odderon configuration with the massless quark
singularities of reggeon diagrams, as we remove the gluon mass, does indeed
anticipate the development of instanton contributions and the problem of the
Dirac sea. In effect we are imposing unitarity on the solution of this
problem. However, that there should be a direct correspondence between all
the infra-red ``perturbative'' phenomena that we have been discussing and
the ``non-perturbative'' instanton interactions of the theory is a very
strong requirement. It could be that, in the case of $QCD$, the complicated
set of fermion interactions produced by the condensate and instanton
interactions of a color-sextet quark sector are (as we suggest in Section 14)
essential to realise this requirement.

We also anticipate that, again as in the Schwinger model, the development of a
``topological'' condensate can be the whole story only if the theory is
essentially non-interacting. As we have described, our high-energy {\em
confining} solution of $SU(2)$ gauge theory is noninteracting because there
is no Pomeron. In effect the anomaly acts as a transverse-momentum infra-red
boundary effect, coupling non-interacting (quark/antiquark) reggeon states in
distinct reggeon channels A conventional description of confinement can
become appropriate as we go to lower energy (if quarks are massive) just as
a conventional description of confinement becomes appropriate when the
fermions of the Schwinger model are given a mass. Note that once the
infra-red limit for gluons has been taken there is no reason why we should
not add quark masses via some form of chiral perturbation theory. Clearly
the high-energy non-interacting S-Matrix will not be affected.

We have still not discussed the removal of the transverse cut-off in reggeon
diagrams. Of course, if we consider only the non-interacting high-energy
S-Matrix it is trivial that it is cut-off independent. Also, there is no
obvious
conflict in assuming that asymptotic freedom can be utilized to remove the
cut-off in diagrams, such as those of Figs.~11.4 and 11.6, which describe
the scattering of mesons via meson exchange. However, since the meson mass
is a cut-off dependent parameter, we might expect problems. As we discuss in
the next two Sections, when the full gauge symmetry group is bigger than
$SU(2)$, it seems that (at least in principle) there are physical parameters
which remain after the cut-off is removed and which determine the chiral and
high-energy limits.

\newpage

\mainhead{12. SPONTANEOUSLY-BROKEN QCD AND THE SUPER-CRITICAL POMERON.}

Having analysed $SU(2)$ gauge theory exhaustively we finally move on to the
physical problem of $SU(3)$ gauge symmetry. We recall from Section 7 that we
are
instructed to first break the symmetry with two color triplet (Higgs) scalar
fields, each of which acquires an expectation value. This breaks the local
gauge symmetry completely but leaves the same global $SU(2)$ symmetry that
we have exploited throughout our discussion of $SU(2)$ gauge theory.
Decoupling one scalar field then converts the global $SU(2)$ symmetry to a
local gauge symmetry. The corresponding infra-red analysis of reggeon
diagrams is very closely related to our previous discussion and will be the
subject of this Section. As we have discussed several times already, we
anticipate that in this case the reggeon (winding number) condensate
described in the last Section will be directly related to the Pomeron
condensate of the Super-Critical Pomeron phase.

\subhead{12.1 Gluon Representation Structure}

The two Higgs field expectation values provide two mass scales which we can
associate with the global $SU(2)$ properties of the octet of $SU(3)$ gluons as
follows. As illustrated in Fig.~12.1, there is

i)~~~~ an $SU(2)$ triplet [$G_3$] with mass $M_1$,

ii)~~~ 2  $SU(2)$ doublets [$G^1_2, G^2_2$] with
mass $\frac{2}{\sqrt{3}}M_2 + O(M_1)$,

iii)~~ an $SU(2)$ singlet [$G_0$] with mass $M_2 + O(M_1)$.
$$
{}~~\auto
$$
In the limit $M_1 \to 0$ the triplet of gluons $G_3$ becomes massless and
$SU(2)$ gauge symmetry is restored. If there is confinement at this stage then
we would expect not only the triplet $G_3$ but also the doublets $G^1_2, G^2_2$
to disappear from the physical spectrum. We shall discuss just how we
incorporate this in our analysis shortly. The most important aspect of the
above gluon spectrum is, however, the existence of the $SU(2)$ singlet $G_0$.
This will provide the reggeized vector gluon trajectory which must exist in the
Super-Critical Pomeron phase and which should be exchange degenerate with the
Pomeron trajectory.

The $SU(3)$ gauge couplings between the different $SU(2)$ gluon representations
$G_i$ are as illustrated in Fig.~12.2. They are simply given by the
``antisymmetric'' $SU(3)$ structure constants $c_{ijk}$ and in an obvious
notation (sufficient for our purposes) the couplings which exist are
$$
r_{333},~~~~~r_{322},~~~~~r_{022}.
\auto
$$
Note that there is no self-coupling for the $G_0$ singlet. For $SU(3)$ and
bigger gauge groups there is a ``symmetric'' $d_{ijk}$ tensor which will play a
role in our discussion shortly. It will be convenient to list the corresponding
couplings in the same notation as (12.2). They are, as illustrated in
Fig.~12.3,
$$
r_{322},~~~~~r_{033},~~~~~r_{000}
\auto
$$
and clearly a self-coupling for $G_0$ is included. The existence of the triple
Pomeron coupling will be tied to the existence of $r_{000}$.

Consider now the properties of the $G_i$ and the $r_{ijk}$ under the {\em color
parity} operation introduced in Section 9. If this is defined as ``G - parity''
with respect to the $SU(2)$ global symmetry then both (12.2) {\em and} (12.3)
are possible couplings for vector states. However, if it is defined from the
gauge field transposition (9.47) extended to $SU(3)$ gauge symmetry then the
couplings (12.3) violate this symmetry. (9.47) is, of course, a symmetry of the
pure $SU(3)$ lagrangian and it is well-known that the $d_{ijk}$ couplings can
not appear in the gauge-field self-interaction. Nevertheless a triple-Pomeron
coupling has to have the symmetry property of the $d_{ijk}$ and so it will be
important to understand how this coupling can arise for an ``unphysical''
reggeon such as the Pomeron. It will also be important for understanding
whether
the Pomeron can produce physical particles.

\subhead{12.2 Trajectory Functions}

When $M_1, M_2~\neq 0$ all gluons reggeize and we can represent the
reggeization
in terms of reggeon diagrams and the couplings of Fig.~12.2 - as shown in
Fig.~12.4. In the infra-red limit $M_1 \to 0$, the $G_0$ trajectory is
infra-red finite simply because the triplet $G_3$ does not couple to $G_0$ in
lowest order. The existence of the massive doublets $G_2$ allows $G_0$ to
remain
reggeized. The doublet trajectory contains an infra-red divergence from the
first of the two reggeon diagrams in Fig.~12.4. In the notation of Section 5
this gives the transverse momentum integral
$$
\eqalign{
J_1 \left(M^2_1,\ M^2_2,\ \underline{q}^2 \right)
&= {{1} \over {(2\pi)^3}}\ \int \
{{d^2\underline{k}}
\over {\bigg[ \left(\underline{q}-\underline{k}\right)^2
-M^2_1 \bigg]
\bigg[ \underline{k}^2-M^2_2 \bigg]}}}
\auto
$$
$$
\eqalign
{= {{1}\over {2\sqrt{Q}}}\ \ell n
\left[{{\left(\underline{q}^2 + \Delta + \sqrt{Q}\right)
\left(\underline{q}^2 - \Delta + \sqrt{Q} \right)}\over
{\left(\underline{q}^2 + \Delta - \sqrt{Q} \right)
\left(\underline{q}^2 - \Delta - \sqrt{Q} \right)}} \right]
}
\auto
$$
with
$$
\eqalign{
Q &= \bigg[\underline{q}^2 - \left(M_1 - M_2 \right)^2 \bigg]
\bigg[\underline{q}^2 - \left(M_1 + M_2 \right)^2 \bigg]\cr
\Delta &= M^2_2 - M^2_1.
}
\auto
$$
We can separate the divergences of $J_1$ as $M_1~and~ M_2 \to 0$ by writing
$$
J_1\left(M^2_1,\ M^2_2,\ \underline{q}^2 \right)\ =\
J^1_1\left(M^2_1,\ M^2_2,\ \underline{q}^2 \right) +
J^2_1\left(M^2_1,\ M^2_2,\ \underline{q}^2 \right)
\auto
$$
where
$$
\eqalign{
J^1_1\ =\ &{{2}\over {\sqrt{Q}}}\ \ell n
\left[{{\underline{q}^2+\Delta + \sqrt{Q}}\over
{\underline{q}^2 + \Delta - \sqrt{Q}}} \right]\cr
&\centerunder{$\sim$}
{$M^2_1\rightarrow 0$} -
{{1}\over {2\left(\underline{q}^2-M^2_2 \right)}}\  \ell n
\left[{{\underline{q}^2}\over {M^2_2}}\right]
}
\auto
$$

If we absorb $J^2_1$ into the reggeon interaction for the doublet there will be
an infra-red cancellation in the color-zero channel and furthermore the
trajectory will then satisfy
$$
\eqalign{
\alpha_2\left(\underline{q}^2 \right)\quad&\centerunder{$\sim$}
{$M^2_1 \rightarrow 0$}\quad
1 - {{3g^2}\over {4\pi^2}}\ \ell n
\left[{{\underline{q}^2}\over {M^2_2}}\right] +\cdots
}
\auto
$$
$$
\eqalign{
&\centerunder{$\sim$}
{$\underline{q}^2 \rightarrow 0$}\quad - \infty
}
\auto
$$
The high-energy behavior associated with the doublet is then negligible for
$q \to 0$ and we shall ignore it in the following. The only contribution of the
doublets is therefore that they provide physical massive states responsible for
the reggeization of the singlet gluon $G_0$. Hopefully, this is an adequate
description of their confined status in the limit $M_1 \to 0$.

\subhead{12.3 Reggeon Diagrams and the Pomeron}

We suppose now that the $SU(2)$ infra-red analysis has been carried out exactly
as discussed in previous Sections. Clearly the Odderon plus quark/antiquark
reggeon states can form in just the same way. The difference will be that we
will now have additional ``unconfined'' gluon and quark states that are $SU(2)$
singlets. Since we ignore the high-energy behavior of the $G_2$ states, the
only
gluon reggeon states we need to consider are those involving the singlet $G_0$.
We have illustrated the reggeization of $G_0$ in Fig.~12.4. The reggeization of
the singlet quarks will involve both $G_0$ and the $G^i_2$.

The first point we note is that the $SU(2)$ anomalous Odderon divergence can
now
accompany $G_0$ states (and singlet quark states) in just the same manner as
for the quark/antiquark states discussed in the previous Section. The first
approximation to meson elastic scattering will be as illustrated in Fig.~12.5.
That is we {\em now have a Pomeron}. The ``Pomeron'' is simply the exchange of
a $G_0$ reggeon in the ``background'' of the Odderon condensate. Note that
$G_0$ will, in first approximation, simply couple to either the quark or
antiquark in the meson. Because the Odderon gluons contribute only at zero
transverse momentum (via the divergence), it follows that the Pomeron Regge
trajectory will simply be given by
$$
\alpha_\spom (t) = \alpha_{G_o} (t)
\auto
$$
that is the ``Pomeron'' trajectory is exchange degenerate with the $G_0$
trajectory.

Since the ($G_0$ plus Odderon) ``Pomeron'' state originally, that is before the
$SU(2)$ limit is taken, involves four (or a higher even number) of reggeized
gluons, we expect it to be purely even signature. However, given the
complicated multiparticle amplitudes in which this configuration couples it
is likely that the usual signature rules are violated and that the
singularity produced is actually exchange degenerate. As a result the
elementary $G_0$ trajectory will not only be exchange degenerate with the
Pomeron trajectory, but the actual physical {\em odd signature} reggeon will
also involve the Odderon condensate. We suspect this is necessary for the
overall consistency of the condensate theory we are trying to formulate.

We clearly have already identified crucial features of the Super-Critical
Pomeron described in Part I. The most detailed feature of the Super-Critical
phase that we would like to identify is the structure of Pomeron vertices
produced by the Pomeron condensate. We will not attempt to identify all of
this structure explicitly since we feel that further study of the anomalous
interactions is still needed. The vacuum production of pairs of Pomerons is
clearly present as a consequence of the pair production of $G_0$'s from
massless
gluon configurations discussed in Section 10 and illustrated in Fig.~10.9.
Note that since the $r_{000}$ coupling does not violate $SU(2)$ color charge
parity it will be generated by quark loops if the quark masses violate $SU(3)$
symmetry. Indeed we have argued that the $SU(2)$ singlet quarks should be kept
massive as the $SU(2)$ symmetry is restored (whereas the doublet masses must
be set to zero {\em before} the symmetry limit is taken). If the transverse
momentum cut-off is removed via asymptotic freedom then {\em both} the mass
of the $G_0$ vector meson amd the masses of the singlet quarks are physical
parameters. These parameters can be used to track both the Pomeron intercept
and the intercepts of multiquark reggeon states that are potential pion
states.

\newpage

\mainhead{13. THE CRITICAL POMERON, SU(3) GAUGE INVARIANCE AND HIGHER GAUGE
GROUPS}

In this Section we will outline, how we believe all the features
of the Critical Pomeron limit are realised as we restore the $SU(3)$ gauge
symmetry (by taking the limit $M_2 \to 0$) within the description of
partially-broken QCD given in the last Section. We shall then go on to discuss
the structure for the Pomeron in $SU(N)$ gauge theories that our analysis
implies. We shall be able to relate this directly to the structure of
transverse
tubes of color flux determined by the center of the gauge group (via 't Hooft
commutation relations\cite{gth} etc.). This will allow us to understand why a
simple Pomeron Regge pole (and hence the Critical Pomeron) is uniquely
related to the $SU(3)$ gauge symmetry of QCD.

\subhead{13.1 The $M_2 \to 0$ Limit and the Transverse Momentum Cut-Off}

As we have already discussed we intend to exploit the (``exchange'') degeneracy
(12.11) between the Super-Critical Pomeron trajectory and the $G_0$ trajectory.
As the renormalized physical vector gluon trajectory is built up we expect this
relation to be preserved by the presence of the reggeon condensate. Following
the discussion of Section 7, we anticipate that if the theory is saturated with
quarks we can remove the transverse momentum cut-off $\lambda$ (by exploiting
asymptotic freedom) already after the $M_1 \to 0$ limit is taken. In
this case $M_2$ can be identified with the physical gluon mass and the limit
$M_2 \to 0$ is a well-defined (cut-off independent) limit which
unambiguously restores $SU(3)$ gauge symmetry and simultaneously gives
$$
\eqalign{
\alpha_\spom (0)\quad&\centerunder{$\longrightarrow$}
{$M^2_2\rightarrow 0$}\quad
\alpha_{G_o} \left(M^2_2 \right) = 1
}
\auto
$$
There could be a question as to whether the result of this limit is pure
$QCD$. That is whether the limit (7.11), which we will again employ, is
sufficient to decouple the remaining Higgs sector when only the dynamical
cut-off provided by asymptotic freedom is present. As we remarked in
Section 7, we can only assume that the power-counting argument given there
suffices - for at least the leading power high-energy behavior.

Since the reggeon condensate and the $G_0$ reggeon are only $SU(2)$ gauge
invariant {\em both must disappear} in the limit $M_2 \to 0$. We have argued
that, before the symmetry is restored, the complicated couplings of the Pomeron
will give an exchange degenerate singularity. However, as the symmetry is
restored these couplings must collapse into (essentially) local symmetric
couplings and so the conventional signature rules should imply that the
odd-signature component does indeed decouple. This clearly
suggests that all of the essential characteristics of the Critical Pomeron
limit will be realised as the $SU(3)$ gauge symmetry is restored.

We can also take the $M_2 \to 0$ limit at a fixed value of $\lambda$ and
many of the arguments for the occurence of the Critical Pomeron will go
through. However, in this case, it is clearly not guaranteed that all
consequences of $SU(3)$ symmetry will follow just from the $M_2 \to 0$ limit.
We know that $\lambda$ is a relevant parameter at the Reggeon Field Theory
critical point. We shall assume, therefore, that for each number of quark
flavors $N_f$ there is a critical value $\lambda_c(N_f)$ of $\lambda$ at
which the critical behavior occurs. It will be consistent with the arguments
of Section 3 and all decoupling arguments if we assume that
$\lambda~<~\lambda_c$ gives the Super-Critical theory, $\lambda~>~\lambda_c$
gives the Sub-Critical theory, $\lambda_c$ increases with $N_f$ and, of
course, $\lambda_c = \infty$ when the theory is saturated.

\subhead{13.2 $C$ - Parity and the Triple Pomeron Vertex}

The Pomeron, as a single $G_0$ reggeized gluon in the condensate background,
has a projection on an $SU(3)$ invariant combination of gluons which is {\em
negative under the $SU(3)$ color charge conjugation} defined by (9.47), and
referred to from now on as $C$-parity. That is, in first approximation,
the $SU(3)$ invariant Pomeron is four gluons with the color structure
$$
\pom~=~ d_{ijk}\ f_{k\ell m}\ A^iA^jA^\ell A^m
\auto
$$
The negative $C$-parity of (13.2) will be particularly important when we come
to discuss the transverse flux tube picture of the Pomeron later in this
Section. As we now discuss $C$-parity is actually broken by Pomeron
interaction vertices although it remains conserved non-locally in Pomeron
graphs. In a sense this already determines that the Pomeron is a non-local
object in confining $SU(3)$ gauge theory.

In the last Section we noted that the $SU(2)$ symmetry limit should be taken
with singlet quarks kept massive. To fully restore the $SU(3)$ symmetry
these masses must also be set to zero. Indeed since they are physical
particle masses they can be used to ensure that (in the $SU(3)$ limit the pions
discussed in previous Sections do indeed have zero mass.  The triangle
diagram shown in Fig.~13.1 will contribute to an $SU(3)$ invariant triple
Pomeron vertex. Before the $M_2 \to 0$ limit is taken the condensate lines
will be attached out of the transverse plane and the triangle diagram will,
following the discussion in Section 10, contain infra-red divergences
involving helicity transitions, but will not violate $SU(2)$ charge parity.
As the the $SU(3)$ limit is taken the helicity transitions can produce
a $C$-parity violating vertex. However, the condensate lines traveling out
of the transverse plane will couple in via similar divergences in another
part of the reggeon diagram. As a result the condensate lines can produce
coupled violations of $C$-parity in distinct parts of a complicated reggeon
diagram while the overall diagram does not violate the symmetry. If $SU(3)$
$C$-parity is indeed only conserved non-locally in Pomeron reggeon diagrams
then both Pomeron self-couplings {\em and Pomeron couplings to external
states} will violate the symmetry. This presumably implies that at the
physical (glueball) particle poles that would be generated at $j = 2, 4,
..$ by the {\em even signature} Pomeron trajectory, the physical couplings
must vanish. That is there will be {\em no glueball states on the Pomeron
trajectory}.

While we have not presented any detailed analysis and have only outlined a
classification of diagrams we believe this last discussion does provide an
understanding of the basic origin of triple Pomeron interactions. Clearly
the existence of interactions is very closely related to the ``non-local''
nature of the particle states and the Pomeron itself. An important feature
of the $M_2 \to 0$ limit is the summation of vacuum production graphs, the
simplest example of which is shown in Fig.~13.2. As illustrated, such graphs
should correspond directly to Super-Critical Pomeron graphs and it should be
straightforward to use Pomeron RFT to carry out the summation.

We certainly hope to establish a complete correspondence between the
Critical Pomeron limit and $SU(3)$ symmetry restoration. However, as we already
remarked in the last Section, the detailed mapping of the reggeon diagrams
of spontaneously-broken $QCD$ onto Super-Critical Pomeron theory clearly
requires a comprehensive treatment of all possible quark loop anomalous
interactions. For the moment we are satisfied that we have established a
sufficiently close correspondence to leave very little doubt that the
mapping can be made.

\subhead{13.3 Symmetry Restoration for Higher Gauge Groups}

Consider now $SU(N)$ gauge theory with $N~>~3$. The discussion of Section 7
implies that we should initially break the gauge symmetry completely by adding
{}~(N - 1)~ scalar fields that are fundamental (that is ``N-tuplet'')
representations of the gauge group. These should then be decoupled
sequentially so that the gauge symmetry is restored through the sequence
$$
SU(1)\quad\centerunder{$\longrightarrow$}
{$M^2_1\rightarrow 0$}\quad
SU(2)\quad\centerunder{$\longrightarrow$}
{$M^2_2\rightarrow 0$}\quad SU(3)\longrightarrow \cdots
\quad\centerunder{$\longrightarrow$}
{$M^2_{N-1}\rightarrow 0$}\quad SU(N)
\auto
$$
While the $SU(3)$ limit $M_2 \to 0$ can again give the Critical Pomeron, there
will also be additional $SU(3)$ color zero trajectories in the theory as
illustrated in Fig.~13.3. For $SU(N)$ there will be an $(N - 3)~X~(N - 3)$
matrix of such trajectories. Consequently as further limits $M_3~\to 0,
M_4~\to 0$ etc. are taken there will potentially be more complicated
critical phenomena involving additional Pomeron Regge trajectories which
actually need not all be even signature. We can understand this phenomenon
from two very different viewpoints - as we describe in the following
sub-sections.

\subhead{13.4 Stability Analysis for Higher Gauge Groups}

We begin by discussing the generalization of the stability analysis of Section
3 to $SU(N)$ gauge theory. If we wish to {\em retain asymptotic freedom for
the complete theory} then we can add no more than $~(N - 2)~$ fundamental
representation scalars\cite{cel}. The Higgs mechanism then potentially
breaks the gauge symmetry down to $SU(2)$. Again, however, the theory must
contain close to the maximum number of fermions allowed by asymptotic freedom,
even if the symmetry breaking is less than the maximum allowed.

Consider, for example, an $SU(4)$ theory with the gauge symmetry broken down to
$SU(3)$ only. In this case just one quartet of scalars is added and (3.7) holds
for the corresponding coupling constant, except that
$$
A = {{1} \over {\pi^2}},~~\ B^\prime = {{-45} \over {32\pi^2}},~~\
C = {{99}\over {256\pi^2}}
\auto
$$
and (for ``color quartet'' quarks)
$$
b_o = {{1}\over {8\pi^2}} \left[{{44}\over{3}} - {{2}\over{3}}\
N_f \ -{{1}\over{6}}\right]
\auto
$$
The stability condition now gives
$$
\left(4\pi - 32\pi^2 b_o \right)^2~ > 16.99
\auto
$$
$$
\sim 45 - 32\pi^2 b_o~ > 40
\auto
$$
$$
\Rightarrow {{5}\over {4}} > 8\pi^2 b_o =
\left[{{44}\over {3}} - {{2N_f}\over {3}} -{{1}\over {6}} \right]
\auto
$$
The dependence on $N_F$ is as follows
$$
{{5}\over {4}} > \bigg[N_f = 21 \bigg] = {{1}\over {2}},~~\ {{5}\over {4}} >
\bigg[N_f = 20 \bigg] = {{7}\over {6}},~~\ {{5}\over {4}}\st{>}
\bigg[N_f = 19 \bigg] = {{11}\over {6}}
\auto
$$
so the first critical point for $SU(4)$ is reached at $N_F = 20$, one less than
the maximum allowed for asymptotic freedom of the pure gauge theory. To
consider
adding more representations of scalars it is necessary to discuss a more
complicated set of stability equations. The conclusion is, as we stated in
general above, that two representations of scalars, but no more, can be added,
allowing the gauge symmetry to be broken to $SU(2)$. The original
studies\cite{cel} did not determine (presumably because it was too
complicated) whether both representations can be added at $N_F = 20$ or only
at $N_F = 21$. Given the nature of the analysis, it is natural to assume
that as $N_F$ increases more stability conditions are satisfied. This would
imply that the gauge symmetry can be broken to $SU(3)$ at $N_F = 20$ and to
$SU(2)$ at $N_F = 21$, so that there are in effect two critical points.
Similarly in $SU(5)$ the symmetry can be broken to $SU(4)$ at four flavors
less than the maximum, and altogether the symmetry can again be broken only
to $SU(2)$. It is natural to assume that there are three critical points
distinguished by the number of flavors, and so on.

As illustrated in Fig.~13.3, when $SU(4)$ gauge symmetry is broken to $SU(2)$
the singlet $G_0$ that we have discussed for $QCD$ is replaced by a quartet of
reggeized massive vectors which are singlets under the $SU(2)$ symmetry. This
quartet will form exchange degenerate vacuum trajectories as a consequence of
the $SU(2)$ reggeon condensate. If we first restore the $SU(3)$ symmetry then
one of the massive gluon trajectories will be part of an $SU(3)$ octet and will
go to zero mass, giving the Critical Pomeron as in $QCD$. Two of the other
massive gluons will combine with the $SU(2)$ doublets $G^i_2$ to form confined
$SU(3)$ triplets. There will remain a massive reggeized vector which is a
singlet under the SU(3) gauge group. Since it is known that a vector boson
trajectory can be left essentially unperturbed by the Critical
Pomeron\cite{abs} it is reasonable to assume that this singlet trajectory
can be taken smoothly to unit intercept. We also have no a priori reason to
expect it to decouple from either the Pomeron or the physical spectrum. We
are therefore led to believe that the critical phenomenon as $SU(2)$ gauge
symmetry is restored to $SU(4)$ involves an odd signature trajectory
contributing to the Pomeron in addition to an even signature trajectory.
This is the case for $N_F = 21$. For $N_F = 20$ the symmetry can only be
broken to $SU(3)$ and so only one gauge vector trajectory can be brought
into the spectrum by the Higgs mechanism. We take this to imply that the
even signature Pomeron is still critical but that in this case the
odd-signature component has intercept less than one.

As we have implied we can generalize this analysis to higher gauge groups by
again beginning with $SU(2)$ gauge symmetry realised via the reggeon
condensate.
The obvious generalization of the above discussion would lead to the conclusion
that when the asymptotic freedom constraint on the number of quarks is
saturated
there will be a critical phenomenon involving many Pomeron trajectories of both
signatures. In particular the above arguments suggest that in $SU(N)$ gauge
theory there will be $(N - 2)$ such trajectories. As the number of quark
flavors
is reduced these trajectories will move successively away from unit intercept
(at each critical value of the number of flavors), with the even signature
Critical Pomeron presumably the last to go. It is clear from the above analysis
that the spectrum of Pomeron trajectories is closely related to the {\em
center}
(or ``diagonal part'') of the gauge group. This picture becomes particularly
attractive when we interpret it in terms of the flux loop structure of gauge
theories {\em when such loops are projected on a transverse plane}. This
structure is also determined by the center of the gauge group.

\subhead{13.5 Transverse Color Flux Tubes}

As we already alluded to in Section 7, it is a common expectation that the
confining solution of a gauge theory can be at least approximated by (if not
exactly described by) a string theory in which the ``string'' is a ``tube'' of
color flux. (Indeed dual models and then string theories were developed
originally as Regge region approximations and it is presumably in this
kinematic regime that a string approximation to QCD has the most validity.) It
is also a very familiar idea that with an ultra-violet cut-off (and on the
lattice in particular) we can approximate the operator that creates a flux-tube
by a line-integral of the form of (7.6). In this case a ``closed-string'' will
be created by the (fixed-time) Wilson loop operator
$$
\phi(\Omega) = Tr \bigg[P\exp - g \int_\Omega dx_\mu A_\mu(x) \bigg]
\auto
$$
where $\Omega$ is the closed-curve around which the string lies. Indeed we
argued in Section 7 that if we exploit complimentarity we anticipate the smooth
formation of ``string states'' from perturbative states. The exchange of
a closed string - {\em which in a hadronic string theory gives the Pomeron} -
will be related to the propagator
$$
\langle \phi (\Omega_1) \phi (\Omega_2)  \rangle
\auto
$$
If we suppose that this propagator contains a Regge pole (``bound-state'')
Pomeron {\em and perturbation theory has some validity} we might expect the
vector nature of the gluon to give a scattering amplitude which has an
approximation of the form
$$
A\left(\langle \phi (\Omega_1) \phi (\Omega_2)\rangle\right) \sim s
\langle \phi (\Omega_{1T}) \phi (\Omega_{2T})\rangle_T
\auto
$$
where $\Omega_{1T}$ and $\Omega_{2T}$ are transverse projections of $\Omega_1$
and $\Omega_2$ and $<~~>_T$ is a transverse propagator evaluated from some
(suitably complicated) set of transverse momentum diagrams. That is we
effectively ignore all the powers of $\ln s$ generated, or equivalently, assume
they simply cancel out (which is almost the case when the Pomeron is Critical).

Next we note that the $C$- parity operation of (9.47) is equivalent to a
transformation within $\phi(\Omega)$
$$
A^i_{\alpha\beta} \longrightarrow A^i_{\beta\alpha},~\ dx_\mu \longrightarrow
-dx_\mu
\auto
$$
and so reverses the direction of the loop integration. Because $\phi(\Omega)$
is
a trace of hermitian (color) matrices this says that the $C$- parity
transformation is equivalent to hermitian conjugation of $\phi(\Omega)$.
Also to define signatured amplitudes we add or subtract the amplitude
obtained by interchanging one incoming and outgoing particle. This produces
the kinematic effect that $s~\to~-s$ and (where the interchanged particles
are attached) produces, via the reversal of quark lines, the reversal of the
orientation in (13.10). It then follows that even and odd-signatured amplitudes
should be obtained from (13.12) by
$$
A^+ \sim s \langle Im \phi (\Omega_{1T}) Im\, \phi
(\Omega_{2T}) \rangle
\auto
$$
$$
A^- \sim s \langle Re\, \phi (\Omega_{1T}) Re\, \phi (\Omega_{2T})\rangle
\auto
$$
Hence the even-signatured amplitude defined from a (Wilson-loop) closed string
involves the imaginary part of the loop operator and hence {\em odd $C$- parity
exchange}. That is the Pomeron we have found perturbatively has {\em the
right $C$- parity to be identified with the even-signatured exchange of a
color flux-tube}.

Once we have made the identification (13.14) and (13.15) we can immediately
understand the complete structure of Pomeron trajectories that we obtained in
the last sub-section in terms of the orientability properties of transverse
line integrals - as determined by the center of the gauge group.

\subhead{13.6 Closed Loop Integrals and the Center of the Gauge Group}

For $SU(2)$ gauge theory the line integral (13.10) is necessarily real because
it is the trace of an $SU(2)$ matrix and all $SU(2)$ matrices have real trace.
Or - what is essentially the same property - $SU(2)$ line integrals can not
be given an orientation because of the equivalence of the $2$ and $2^*$
representations of $SU(2)$. As a result it immediately follows from (13.14)
that there will not be an even signature Pomeron in $SU(2)$ gauge theory.
Since a simple closed string coupling is inevitably even signature (because
of the symmetry under rotation by $\pi$) this implies {\em there is no
Pomeron in $SU(2)$ gauge theory}. A somewhat remarkable independent
confirmation of the perturbative infra-red analysis of previous Sections.

For $SU(3)$ gauge theory it is the fact that line integrals have an
orientation,
as illustrated in Fig.~13.4, which leads to the identification of the even
signature Pomeron with the amplitude for the exchange of a ``closed-string''
configuration with negative $C$ - parity. However, the symmetry of the
closed loop under rotation by $\pi$ again implies that there is no
odd-signature
amplitude. Therefore the $SU(3)$ ``closed string Pomeron'' has exactly the
properties determined by our perturbative analysis.

In SU(N) gauge theory the distinct transverse loop operators that can
be defined, {\em when there are no quarks in the theory}, are determined by
starting from the general commutation relations\cite{gth} for three-dimensional
Wilson loop (electric flux) operators with (magnetic flux) 't Hooft loop
operators when the loops intertwine. The result is that we can obtain
distinct operators as limits into the transverse plane of loops with any
number of windings less than $N$. However, a loop with $N - n$ windings will
give the inverse of the operator with $n$ windings, which from (13.13) is
simply the same operator but with the inverse orientation. (Note that this
discussion is only valid when quarks are not present in the theory.
Nevertheless we might expect that this would be the appropriate place to
start with a flux-tube picture of the Pomeron and that the only relevance of
quarks would be in their contributions to Pomeron interaction vertices.)

For $SU(4)$ gauge theory it is possible to define not only orientation
dependent
single loop operators but also a ``double-loop'' operator $\phi_2(\Omega)$
defined on a transverse ``loop'' which encircles the same spacial loop twice.
As illustrated in Fig.~13.5, this operator is defined via a limit of a three
dimensional loop. $\phi_2(\omega)$ will be {\em orientation independent}, and
therefore real, but the coupling to an open string will not be symmetric under
rotation by $\pi$. As a result this operator will give an {\em odd-signature}
component to the Pomeron via (13.15). For $SU(4)$ gauge theory, therefore, we
also conclude that the structure of the Pomeron is just as we concluded from
the infra-red analysis.

Moving on to $SU(5)$ we can now define an orientation dependent
$\phi_2(\omega)$
and as a result we anticipate an extra {\em even signature} Pomeron trajectory
will contribute to (13.14). For $SU(N)$ we can define $N - 1$ independent loop
operators on a single space loop as illustrated in Fig.~13.6. As a result we
expect $N - 2$ Pomeron trajectories in $SU(N)$ gauge theory, just as we found
in the last sub-section.

We have focussed on the transverse loop operators that can be defined on a
single space loop even though, for example, the $SU(4)$ two-loop operator we
have discussed is smoothly related to a general crossed double loop operator as
illustrated in Fig.~13.7. We believe that we can think of the above discussion
as directly relating to the reggeon diagram infra-red analysis of previous
Sections if we consider the single space loop as the ``circle at infinity'' in
the transverse plane (or equivalently at zero transverse momentum).
Complimentarity is therefore allowing us to build up the possible infra-red
structure of flux-tubes directly by a (very complex) perturbative analysis in
which, of course, the infra-red contribution of the anomaly plays a crucial
role. Clearly the anomaly and the development of the reggeon condensate plays a
bridging role in the coexistence of apparently very different perturbative and
non-perturbative pictures of the Pomeron.

If we broaden the above discussion to include multi-loop configurations (in
transverse space) then the above arguments imply that in general we should
think of connected multi-loop operators as distinct from multiple
disconnected loop operators and as generating new Pomeron trajectories. We can
give a much more physical description of this conclusion if we go back to the
phenomenological formulation of Reggeon Field Theory in terms of multiplicity
fluctuations and the $RFT$ cutting rules described in Section 6 of Part I.

\subhead{13.7 The Center of the Gauge Group, Cutting Rules and Unitarity}

We now consider the structure of the $s$-channel intermediate states making
up the string configurations discussed in the last sub-section. The ``world
sheet'' for the scattering process in which two open string hadrons scatter
by the exchange of a closed string Pomeron is illustrated in Fig.~13.8. The
closed string propagation may, of course, involve splitting and joining an
arbitrary number of times so that cutting the world sheet to expose
intermediate
states gives the familiar result of Fig.~13.9 - that is the multiperipheral
production of multiple open string states.

It is hard to represent adequately in a figure, but we now consider cutting the
corresponding world sheet for propagation of the $SU(4)$ double-loop created by
the $\phi_2(\omega)$ operator. As illustrated in Fig.~13.10, we will apparently
obtain the same result as from cutting the world sheet for the propagation of
two closed strings of the form of Fig.~13.9. Indeed, if we consider the
phenomenogical formulation of $RFT$ discussed in Part I, in the context of a
closed string model of the Pomeron, we would start with the assumption that
single closed string propagation corresponds to single Pomeron exchange and so
gives a good description of average multiplicity events. According to this
formulation, events with twice the average multiplicity should be directly
associated with two Pomeron exchange and therefore with the propagation of two
(distinct) closed strings. However, the cutting of the world sheet for
propagation of the $SU(4)$ double-loop shows that ``double multiplicity''
events can actually be associated with a ``new'' string configuration and hence
a new Pomeron trajectory. If we extend this argument to $SU(N)$ gauge theory it
is straightforward to see that events with up to $N - 2$ times the average
multiplicity events actually generate new Pomeron trajectories.

Note that if we sew two open strings together to form a closed string as
illustrated in Fig.~13.10(a) we obtain an elementary representation of how
$s$-channel unitarity is anticipated to work in the simplest string model of
the Pomeron. In the ``string model'' associated with $SU(4)$ gauge
theory we must allow the center of the gauge group to be represented in the
nature of closed string formation via unitarity, and include the process
illustrated in Fig.~13.10(b). For $SU(N)$ gauge theory, this implies that
the corresponding ``cut'' $RFT$ would have to involve not only $N - 2$
Pomeron trajectories but also a representation of the center $Z_N$. Indeed
it is straightforward to show that the `~+~',~`~--~', and `cut' Pomeron
Greens functions of the cut RFT described in Part I do provide a
representation of $Z_3$ - the center of $SU(3)$.

\subhead{13.8 The Critical Pomeron Phenomena of SU(N) Gauge Theory, Unitarity,
and the Parton Model.}

We close by summarizing what we think is a rather beautiful picture of the
Critical Phenomena which can, in principle, occur in a general gauge theory as
we come close to saturating the theory with quarks. At the first critical
number
of flavors (determined by the lowest number which allows an asymptotically-free
Higgs field to be added to the theory) the simple closed loop Pomeron becomes
Critical - that is has intercept one. At the next critical value the
double-loop
odd-signature Pomeron becomes critical, at the next value the double-loop even
signature Pomeron becomes critical and so on, until all possible multi-loop
Pomeron trajectories are critical - giving a full representation of the center
of the gauge group in the high-energy solution of unitarity.

Whether all of the phenomena we have just described can be consistent with
unitarity is a deep question. We know, as we strongly emphasized in Part I,
that the $SU(3)$ even signature Critical Pomeron satisfies all known unitarity
constraints. Although axiomatic bounds would allow the odd signature ``Critical
Odderon'' which is apparently present in all the phenomena associated with
higher gauge groups, these bounds are derived with very little exploitation of
multiparticle $s$-channel unitarity. The general issue might be related to
another, superficially distinct, but equally deep question. That is whether
such phenomena can be consistent with an asymptotically-free short-distance
description of the theory in terms of a ``parton-model''. Confinement and
chiral symmetry breaking are complicated vacuum properties that are presumably
essential to obtain a unitary description of a strongly-interacting theory in
the infra-red region. However, these vacuum properties can not totally
disappear
at infinite momentum. This can probably be reconciled with the simple vacuum of
the parton model only if, in Feynman's language\cite{rpf}, there is a ``{\em
universal wee-parton distribution} which can act like a vacuum and carry the
chiral symmetry breaking property in particular. But the wee-parton
distribution is only truly universal if the Pomeron is a single Pomeron
Regge pole, with unit intercept\cite{arwpr}, as it is {\em only in
flavor-saturated $QCD$!} This brings us to the general subject of the
relationship of all of our results to the physical world, and in particular
the consistency of the Parton Model with the $N_F$ dependence of the Pomeron
in QCD. This is the subject of the final Section.

\newpage

\mainhead{14. IMPLICATIONS FOR THE REAL WORLD}

In this Section we shall briefly discuss a range of topics involving the
relationship of our results to current theoretical ideas about QCD and the
Standard model as well as to the current experimental status of the Pomeron.
We begin with the implications of our results for $QCD$ with a small number of
flavors and the perhaps surprising consistency with experimental results.

\subhead{14.1 The Pomeron at Present Collider Energies}

We shall be particularly brief here since the topic really deserves a separate
article - which we hope to prepare in the near future. The results of the last
Section imply that for a small number of flavors (say five or six) the Critical
Pomeron will occur at some fixed value $\lambda_c$ of the transverse momentum
cut-off. Increasing the cut-off above this value gives the Sub-Critical Pomeron
while a decrease gives the Super-Critical Pomeron.

It is interesting that, as we emphasized in Part I, the $O(\epsilon)$ Critical
Diffraction peak\cite{abbs} closely resembles that seen at the ISR - that is
over the energy range where the rise of the cross-section first becomes
apparent. (The $0(\epsilon^2)$ result\cite{bbd}, which is not necessarily an
improvement on the $0(\epsilon)$ result, is also relatively close to the
diffraction pattern seen at the ISR). At first sight it is (and has been)
surprising and discouraging that this agreement gets worse rather than
better as the energy increases. However, our current understanding of the
role of the transverse cut-off now makes this explicable. There are a number
of phenomenological models\cite{bkw} which attribute the violations of Regge
behavior seen at the CERN $~Sp\bar{p}S$ and the Fermilab Tevatron -
particularly the ``large $t$ shoulder'' of the diffraction peak - to the
contribution of large $p_\bot$ processes. This appears consistent with the
idea that we could recover the Critical Pomeron by removing the contribution
of all production processes involving $p_\bot > \lambda_c$. We shall return to
the question of whether it is consistent to treat processes with
$p_\bot > \lambda_c$ using perturbative $QCD$ (as has been successful
phenomenologically) after we have discussed the possibility that higher
colored quarks could be responsible for electroweak symmetry breaking.

Another result which follows from our analysis is the approximate validity of
the ``Additive Quark Model'' result for total cross-sections. We have not
discussed just how to consistently implement the transverse cut-off. But,
however this is done, variations in the cut-off should not significantly affect
the magnitude of the forward elastic amplitude. This implies we can obtain a
good approximation to this amplitude by going deep into the Super-Critical
Phase where the Pomeron can be described by a single reggeized gluon
coupling to an individual quark. Clearly we also obtain the extended version
of the additive quark model in which there is a ``Pomeron/photon analogy''.
That is the Pomeron has a simple vector coupling to quarks. The
Pomeron/photon analogy is also very successful phenomenologically\cite{dl}.

\subhead{14.2 Should $QCD$ Be Defined Via the Saturated Massless Version?}

Taken at face value, our results provide a conflict with much of the
conventional wisdom of $QCD$. Apparently, if the theory is not saturated with
quarks, then the cross-section for low momentum transfer processes
ultimately goes to zero asymptotically. This appears to conflict with the
requirement that there be a smooth match with large momentum transfer
processes that are calculable via the parton model and perturbative $QCD$,
since such cross-sections certainly do not go to zero asymptotically. It is
perhaps conceivable that large transverse momentum processes completely
dominate at high-energy with the average transverse momentum being very
large and ever-increasing. However, it should be clear, from Part I in
particular, that we believe this would produce a theory that is inconsistent
with (Reggeon) Unitarity. (This argument is also elaborated in
\cite{arwev}). There is no such conflict if the theory is
saturated with quarks since Critical Pomeron scaling can smoothly connect
with perturbative $QCD$ at large transverse momentum.

The conventional wisdom is certainly that $QCD$ can be defined as a continuum
theory with any (small) number of flavors. Underlying the field of
computational lattice $QCD$ is the assumption that, because of confinement,
the theory is insensitive to boundary conditions and the infinite volume
limit. However, the presence of gauge fields with non-trivial topology
implies that the question of whether or not a non-zero ``$\theta$-parameter''
is introduced by the combination of the infinite volume limit with the
boundary bonditions is a non-trivial problem\cite{ms}. In the massless (chiral)
limit fields with non-trivial topology can produce a coupling between
internally
localised fermion zero-modes and those in the neighbourhood of the boundary.
As a result the correct treatment of boundary conditions and the
infinite-volume limit could be a much deeper issue than is acknowledged in
the conventional wisdom. An infra-red stable effective lagrangian which can
successfully define a version of confining $QCD$ which is both insensitive
to the infra-red volume limit and asymptotically-free has, of course, not
been derived. In fact, there exists a minority opinion\cite{ps} that
infinite volume lattice $QCD$ will, if it exists, not be
asymptotically-free. In a sense this parallels our arguments - that in
general a confining solution of the theory, in the Regge region, does not
match with short-distance perturbation theory.

It may be that the only well-defined infinite-volume limit is (a much more
subtle) one in which quarks are massless so that there is no dependence on
the $\theta$-parameter, the infra-red and ultra-violet behavior of the
gauge coupling is controlled by fixed points, and boundary conditions based
on spectral flow are utilised (as in the solution of the Schwinger model
described in Section 11). Given the solution of this theory (which is, of
course, the ``saturated $QCD$'' that we have been discussing) it could be that
mass effects can be consistently added, via some form of chiral perturbation
theory, in a manner which does not introduce a non-zero $\theta$-parameter.
This would be analagous to the construction of the massive Schwinger model, as
discussed in Section 11. This leads us to ask whether there might not be a
further massive quark sector in $QCD$ and brings us to the next topic.

\subhead{14.3 Color Sextet Quarks, Electroweak Symmetry Breaking and the
$\Theta$- Parameter}

There is clearly very little basis for proposing that ten more flavors of very
massive color triplet quarks exist. However, as we noted in Section 3, two
flavors of color sextet quarks would also produce exactly the effect we want
(that is when added to six flavors of conventional triplet quarks). From the
perspective of this article, therefore, it is a remarkable coincidence that
two flavors of color sextet quarks can provide\cite{wjm} a natural form of
dynamical symmetry-breaking for the electroweak interaction which meshes
perfectly with the observed experimental features. Indeed this provides a
self-contained motivation for introducing the higher color quark sector
which we can briefly outline as follows.

Consider adding to the Standard Model (with no scalar Higgs sector), a massless
flavor doublet $(U,D)$ of color sextet quarks with the {\em usual quark quantum
numbers}, except that the role of quarks and antiquarks is interchanged.
For the $SU(2)\otimes U(1)$ anomaly to be cancelled there must also be other
fermions with electroweak quantum numbers added to the theory\cite{wjm}, but
we shall not consider this here. We consider first the QCD interaction of the
massless sextet quark sector. There is a $U(2)\otimes U(2)$ chiral flavor
symmetry. The chiral dynamics we have discussed in previous Sections will
break the axial symmetries spontaneously and produce four massless
pseudoscalar mesons (Goldstone bosons), which we denote as
$\pi^+_6,\;\pi^-_6,\;\pi^0_6$ and $\eta_6$, in analogy with the usual notation
for mesons composed of $u$ and $d$ color triplet quarks.

As long as all quarks are massless, QCD is necessarily $CP$ conserving in
both the sextet and triplet quark sectors. Therefore, in the massless
theory we can, in analogy with the familiar treatment of flavor isospin
in the triplet quark sector, define sextet quark vector and axial-vector
currents $V^{\tau}_{\mu}$ and $A^{\tau}_{\mu}$ which are ``isotriplets'' under
the unbroken $SU(2)$ vector flavor symmetry and singlet currents
$v_{\mu}$, $a_{\mu}$. The pseudoscalar mesons couple ``longitudinally'' to the
axial currents, that is
$$
<0|A^\tau_{\mu}|\pi^{\tau}_6(q)>~\sim F_{\pi_6}q_{\mu}
\auto
$$
$$
<0|a_{\mu}|\eta_6(q)>~\sim F_{\eta_6}q_{\mu}
\auto
$$
while the vector currents remain conserved. (Note that $a_{\mu}$ should
actually
contain a small admixture of the triplet quark flavor singlet axial current if
it is to generate the U(1) symmetry orthogonal to that broken by the QCD
$U(1)$ anomaly).

We consider next the coupling of the electroweak gauge fields to the sextet
quark sector. The massless $SU(2)$ gauge fields $W^{\tau}_{\mu}$ couple to the
isotriplet sextet quark currents in the standard manner, that is
$$
{\cal L}_I=gW^{\tau\mu}\Bigl(V^{\tau}_{\mu}-A^{\tau}_{\mu}\Bigr)
\auto
$$
It then follows that the $\pi^+_6,\;\pi^-_6$ and $\pi^0_6$ are
``eaten'' by the $SU(2)$ gauge bosons and (after the hypercharge interaction is
included) respectively become the third components of the $W^+,\;W^-$ and
$Z^0$. Consequently, QCD chiral symmetry breaking generates masses for the
$W^+,\;W^-$ and $Z^0$ with $M_W\sim g\;F_{\pi_6}$ where $F_{\pi_6}$ is {\em a
QCD scale}. We anticipate that the relative scales of triplet and sextet
chiral symmetry breaking are determined by the ``Casimir Scaling''
rule\cite{wjm}
$$
C_6\alpha_s(F^2_{\pi_6})~\sim~C_3\alpha_s(F^2_{\pi})
\auto
$$
which is consistent with $F_{\pi_6}\sim 250$ GeV!

We conclude that a sextet sector of $QCD$ produces a special version of
``technicolor'' symmetry breaking in which the electroweak scale is
naturally explained as a second $QCD$ scale. Also since we are completely
restricted to a flavor doublet the form of the symmetry-breaking is
automatically equivalent to that of an $SU(2)$ Higgs sector and so
$$
\rho=~(M^2_W/M^2_Zcos^2\theta_W)~=~1
\auto
$$
as required by experiment. Therefore introducing a sextet quark sector not only
produces a matching of the asymptotic freedom and confinement properties of
$QCD$ via the Critical Pomeron, but also gives a natural solution to the
major problem of today's Standard Model i.e. the nature of electroweak symmetry
breaking. The sextet sector may, as we now discuss, also be deeply tied
up\cite{kkw} with the issue of Strong $CP$ conservation and the very definition
of $QCD$ in the infinite volume limit that we have touched on above.

The $\eta_6$ is not involved in generating mass for the electroweak gauge
bosons, but instead remains as a Goldstone boson associated with a $U(1)$
axial chiral symmetry. It is therefore an axion\cite{wil} in the original sense
of the Peccei-Quinn mechanism\cite{pq} and it remains massless until triplet
quark masses are added to the theory. In the present context, this involves
the addition of triplet/sextet four-fermion couplings (that should ultimately
be traceable to a larger unifying gauge group), which, when combined with
the sextet quark condensate, provide triplet quark masses as illustrated in
Fig.~14.1. That $CP$ remains conserved by QCD triplet quark interactions
follows from the original Peccei-Quinn argument utilising the sextet axial
$U(1)$ symmetry.

It can be argued\cite{kkw} that the $\eta_6$ will aquire an electroweak
scale mass as a result of electroweak scale color instanton interactions and
that it may have actually be seen at LEP and at TRISTAN. This would, of
course, be very exciting confirmation of the existence of the sextet sector.
A further point that we wish to emphasise, however, is that not only may the
sextet sector be the explanation (via the $\eta_6$) of Strong $CP$
conservation in the triplet sector, but it may be a necessary ingredient in
introducing masses consistently into $QCD$. That the $\theta$-parameter is
zero for the low-mass sector of the theory could well be essential for the
existence of a purely {\em even signature} Critical Pomeron and the
consistency that this provides with unitarity in the Regge region. It seems
that this is naturally achieved by introducing two different color quark
sectors with the higher color sector automatically having the higher mass
scale. (Note that this does not prevent the sextet sector from being
$CP$-violating at ``low-energy''\cite{kkw}).

A further property that could be part of the need for two different color quark
sectors is the rather complicated set of fermion vertices that is generated
by instanton interactions. Because of the distinct Casimirs involved, the
singlet current
$$
J^0_{\mu}~=~a^6_{\mu}-5a^3_{\mu}
\auto
$$
is conserved in the presence of instantons (6 and 3 now denote sextet and
triplet currents respectively). Consequently the minimum instanton
interaction involves one quark/antiquark pair of each triplet flavor and
five pairs of each sextet flavor. Combining this interaction with the existence
of both sextet and triplet chiral condensates (and, eventually, four-fermion
vertices coupling triplets and sextets) a wide assortment of fermion
vertices is produced whose magnitude is difficult to estimate. It could be
that this complexity is actually required if the quark loop anomalies we
have discussed are to smoothly reproduce the effects of instanton
interactions in the Regge limit - as we have argued should be the case.

If the sextet sector is indeed hidden at short distances then we can easily
reconcile our results and overall picture with the existing successes of
both perturbative and lattice $QCD$. The decoupling theorem\cite{ac} assures
us that, at least in short-distance expansions, we can integrate out the
higher sextet mass scale and apply $QCD$ perturbation theory at current
momentum scales with only the triplet sector included. Since the infra-red
fixed point value of $\alpha_s$ that we expect to dominate dynamics above
the sextet scale is given by (3.14) there is no problem, at least in
principle, with the idea that integrating out the sextet scale simply
increases $\alpha_s$ from 1/33 to about 1/8. Clearly finite size lattice
calculations (of many quantities) should also remain insensitive to the
higher mass sector. This sector would only be relevant if the full
subtleties of the interplay between the infinite volume, chiral and
continuum limits are discussed.

With respect to the Pomeron phenomenology discussed above, the implications
of the sextet sector would be as follows. Accepting that the Critical
Pomeron can be essentially produced at the ISR by appropriate adjustment of
the transverse momentum cut-off (well) below the sextet scale, we would
anticipate that conventional perturbative $QCD$ can be used above the
cut-off. Thus two-component ``soft'' plus ``hard'' models will adequately
fit the data\cite{bkw} at the ISR and immediately higher energies. However,
since the ``perturbative'' $QCD$ component actually contains sextet quark Regge
contributions which will emerge distinctively at much higher energies it
could well be that the large $t$ shoulder, for example, can both be described
by the ``perturbative'' Pomeron and be associated with diffractive production
of the $\eta_6$\cite{kw}.

\subhead{14.5 Unification and the Super-Critical Pomeron}

Finally we comment on the possible relevance of the Super-Critical Pomeron
and the discussion of higher gauge groups in the last Section, to the
unification of $QCD$ and the electroweak interaction. First we note that the
inclusion of sextet quarks, when combined with the requirement of asymptotic
freedom, very severely restricts the possible unification groups. Asking for
an anomaly-free, asymptotically-free, representation containing the
two flavors of sextets, isolates a very special $SU(5)$
representation\cite{kw1} (and an $SO(10)$ representation that contains the
$SU(5)$ representation). It is very interesting that the $SU(5)$
representation ``saturates'' the asymptotic freedom requirement, just as we
have been discussing for $QCD$.

According to the arguments of the previous Section, the SU(5) theory should
contain a generalised Critical Pomeron phenomenon involving additional even
and odd signature Pomeron trajectories. (Note, however, that the question as
to whether left-handed gauge theories can be treated by the methods we have
developed could be an additional issue.) It is attractive to speculate that
such a phenomenon is indeed inconsistent with unitarity as we discussed and
that the theory is forced to break dynamically via the development of a
reggeon (winding-number) condensate. If the condensate develops with respect
to a (gauge-dependent) subgroup, as we have discussed, it would have the
group structure
$$
d_{ijk}f_{jlm}A^kA^lA^m
\auto
$$
and so, being in the adjoint representation of the gauge group, would break
the symmetry to the desired $SU(3)\otimes SU(2)\otimes U(1)$. In effect this
would be saying that the regularisation of the fermion sea necessarily
breaks the symmetry in order to satisfy unitarity. A generalised version of
the Super-Critical Pomeron, containing the Critical Pomeron, would be the
appropriate description of the vector-boson spectrum. The Critical Pomeron
will produce the sextet and triplet chiral symmetry breaking, as we have
discussed. Whether a consistent low-energy mass spectrum might then emerge
is, of course, a very big question. We end, therfore, with the intriguing
conclusion that studying the ``Pomeron'' in a unified gauge theory may actually
be a direct way to uncover the dynamical spectrum of the theory.

\newpage

\noindent{\bf FIGURE CAPTIONS}

\begin{itemize}

\item[{Fig.~3.1}] Evolution of the $\beta$-function with $N_f$ -
a) $N_f \sim 5,6$, b) $N_f \sim 14,15,$ c) $N_f = 16$

\item[{Fig.~3.2}] An infinite series of vacuum diagrams.

\item[{Fig.~3.3}] Renormalon branch points in the Borel plane.

\item[{Fig.~3.4}] Movement of renormalons as $N_f$ increases.

\item[{Fig.~5.1}] The box-diagram.

\item[{Fig.~5.2}] Cutting the box-diagram.

\item[{Fig.~5.3}] Single photon exchange.

\item[{Fig.~5.4}] N-photon exchange.

\item[{Fig.~5.5}] The transverse momentum diagram from N-photon exchange.

\item[{Fig.~5.6}] A sixth-order ladder diagram in $QED$.

\item[{Fig.~5.7}] Additional sixth-order diagrams needed for the transverse
momentum cancellation.

\item[{Fig.~5.8}] Sub-diagrams responsible for softening the transverse
momentum loop integration.

\item[{Fig.~5.9}] Diagrammatic representation of transverse momentum
integrals.

\item[{Fig.~5.10}] Tree diagrams giving the leading high-energy behavior in
a spontaneously-broken gauge theory.

\item[{Fig.~5.11}] Multi-regge kinematics.

\item[{Fig.~5.12}] Transverse momentum diagrams from the three-particle
s-channel states.

\item[{Fig.~5.13}] s-channel iteration of elastic scattering

\item[{Fig.~6.1}] The two-reggeon diagram.

\item[{Fig.~6.2}] The reggeon diagrams contributing up to tenth order.

\item[{Fig.~6.3}] The propagator and vertex for reggeized gluon RFT.

\item[{Fig.~6.4}] Insertion of multi-regge amplitudes into the unitarity
equation.

\item[{Fig.~6.5}] The gluon-reggeon scattering amplitude.

\item[{Fig.~6.6}] Regge pole and two-reggeon cut trajectories.

\item[{Fig.~6.7}] The three-reggeon vertex.

\item[{Fig.~6.8}] The simplest hexagraph loop diagram with a multiperipheral
cut.

\item[{Fig.~6.9}] Evaluation of the reggeon loop diagram via a
multiperipheral cut.

\item[{Fig.~6.10}] The ``bootstrap'' equation satisfied by the reggeised gluon.

\item[{Fig.~6.11}] The series of diagrams generated by the bootstrap
equation.

\item[{Fig.~6.12}] Hexagraph loop contribution to the triple reggeon vertex.

\item[{Fig.~6.13}] Transverse momentum bubble renormalising the triple-reggeon
vertex.

\item[{Fig.~6.14}] Hexagraph loops that build up higher thresholds in the
triple-reggeon vertex.

\item[{Fig.~6.15}] Higher order diagrams in the even-signature amplitude
involving the three-reggeon vertex.

\item[{Fig.~6.16}] A single loop diagram.

\item[{Fig.~6.17}] Evaluation of the first diagram in Fig.~6.15 using
Fig.~6.16.

\item[{Fig.~7.1}] Lattice gauge theory phase diagram.

\item[{Fig.~7.2}] $QED$ untwisted and twisted box-diagrams.

\item[{Fig.~8.1}] Infra-red divergences of the trajectory function.

\item[{Fig.~8.2}] Infra-red divergences of the four-reggeon interaction.

\item[{Fig.~8.3}] The complete four-reggeon kernel.

\item[{Fig.~8.4}] The vanishing of a general reggeon amplitude at
$\kbar = 0$.

\item[{Fig.~8.5}] The four-reggeon amplitude.

\item[{Fig.~8.6}] The four-particle amplitude.

\item[{Fig.~8.7}] The two-reggeon discontinuity of the four=particle
amplitude.

\item[{Fig.~8.8}] Some higher-order divergences.

\item[{Fig.~9.1}] Bootstrap equation for the reggeization of quarks.

\item[{Fig.~9.2}] A quark/gluon Regge cut.

\item[{Fig.~9.3}] The integral equation for the flavored quark/antiquark
channel.

\item[{Fig.~9.4}] The two quark-reggeon/gluon vertex

\item[{Fig.~9.5}] The quark/antiquark kernel.

\item[{Fig.~9.6}] The ``tower diagrams'' in $QED$.

\item[{Fig.~9.7}] Transverse momentum diagrams from the tower diagrams.

\item[{Fig.~9.8}] A quark/antiquark reggeon diagram.

\item[{Fig.~9.9}] A reggeon diagram potentially contributing to Fig.~9.8.

\item[{Fig.~9.10}] A Feynman diagram contributing to Fig.~9.9.

\item[{Fig.~9.11}] A non-planar Feynman diagram contributing to Fig.~9.8.

\item[{Fig.~9.12}] The sum of reggeon diagrams giving $V_c$.

\item[{Fig.~9.13}] The hexagraph corresponding to the one-loop diagram.

\item[{Fig.~9.14}] The crossed hexagraph necessary to define signatured
amplitudes.

\item[{Fig.~9.15}] Elastic scattering diagrams involving $V_c$.

\item[{Fig.~9.16}] Infra-red divergences in the diagrams of Fig.~9.15.

\item[{Fig.~9.17}] Contribution of a quark-reggeon loop to a general
reggeized gluon vertex.

\item[{Fig.~9.18}] A reggeon diagram potentially containing a gluon scaling
infra-red divergence.

\item[{Fig.~9.19}] Iteration of the scaling infra-red divergence.

\item[{Fig.~9.20}] The vertex appearing in Fig.~9.19.

\item[{Fig.~9.21}] Group-theoretic signature for elastic scattering
amplitudes.

\item[{Fig.~9.22}] Signatured multi-reggeon couplings.

\item[{Fig.~10.1}] The anomalous odderon gluon configuration coupling via a
quark reggeon diagram.

\item[{Fig.~10.2}] The triple-regge Toller diagram.

\item[{Fig.~10.3}] Hexagraphs for the triple-regge Toller diagram.

\item[{Fig.~10.4}] A hexagraph triple discontinuity.

\item[{Fig.~10.5}] Further triple discontinuities of the same heaxagraph.

\item[{Fig.~10.6}] A triple discontinuity of Fig.~10.1.

\item[{Fig.~10.7}] The odderon coupling to the quark loop of Fig.~9.11.

\item[{Fig.~10.8}] The odderon coupling reggeons propagating in different
transverse planes.

\item[{Fig.~10.9}] The coupling of a massive gluon pair via the triangle.

\item[{Fig.~10.10}] Emission of the O configuration from a quark reggeon
state.

\item[{Fig.~11.1}] The three-Odderon coupling.

\item[{Fig.~11.2}] Odderon configurations attached across a general Toller
diagram.

\item[{Fig.~11.3}] The essential discontinuity for a four-particle
amplitude.

\item[{Fig.~11.4}] The divergent O configuration associated with the
discontinuity channel of Fig.~11.3.

\item[{Fig.~11.5}] Triple-regge discontinuities.

\item[{Fig.~11.6}] Divergent O configurations associated with the
discontinuity channels of Fig.~11.5.

\item[{Fig.~12.1}] The breakdown of the $SU(3)$ color matrix for gluons into
$SU(2)$ representations.

\item[{Fig.~12.2}] The $SU(3)$ gauge couplings ($f_{ijk}$) of the $SU(2)$
repesentations.

\item[{Fig.~12.3}] The ``$d_{ijk}$'' couplings.

\item[{Fig.~12.4}] Reggeization whem all gluons are massive.

\item[{Fig.~12.5}] The Pomeron in partially-broken $QCD$.

\item[{Fig.~13.1}] A contribution to the triple-Pomeron vertex.

\item[{Fig.~13.2}] A vacuum production graph.

\item[{Fig.~13.3}] The $SU(N)$ color matrix.

\item[{Fig.~13.4}] The even-signature Pomeron flux loop.

\item[{Fig.~13.5}] The double-loop in $SU(4)$.

\item[{Fig.~13.6}] The increasing complexity of transverse flux loops as the
center of the gauge group increases.

\item[{Fig.~13.7}] The relationship of a general crossed loop to the double
loop of Fig.~13.5.

\item[{Fig.~13.8}] The world sheet for scattering of open strings via closed
string exchange.

\item[{Fig.~13.9}] Cutting a closed string.

\item[{Fig.~13.10}] Cutting a double-loop string.

\item[{Fig.~13.11}] Unitarity for (a) the simple closed string, (b) for the
double loop of $SU(4)$

\item[{Fig.~14.1}] Quark mass generation via the sextet quark condensate and
a four-fermion interaction.

\end{itemize}

\end{document}